\documentclass[11pt]{article}
\pdfoutput=1  
\usepackage[linktocpage]{hyperref}
\usepackage{latexsym,amsmath,amssymb,graphicx}
\usepackage{slashed, color}
\usepackage{multirow}
\usepackage{tikz}
\usepackage{caption}
\usepackage{subcaption}
\usepackage{adjustbox}
\usepackage{amsmath}
\allowdisplaybreaks[2]
\usepackage{color}
\usepackage[multiple]{footmisc}
\usepackage{textcomp}
\usepackage{amssymb}
\usetikzlibrary{positioning}
\usepackage{amscd}
\usepackage{amsthm}
\usepackage{array}
\usepackage{mathrsfs}
\usepackage{mathtools}
\usepackage[sorting=none,backend=biber,sortcites,citestyle=numeric-comp]{biblatex}
\DeclareFieldFormat[misc]{title}{%
  {\mkbibquote{#1}}
}
\usepackage{hyperref}                                   %
\renewcommand{\thefootnote}{\fnsymbol{footnote}}
\numberwithin{equation}{section}

\DeclareFontFamily{U}{MnSymbolC}{}
\DeclareSymbolFont{MnSyC}{U}{MnSymbolC}{m}{n}
\DeclareFontShape{U}{MnSymbolC}{m}{n}{
    <-6>  MnSymbolC5
   <6-7>  MnSymbolC6
   <7-8>  MnSymbolC7
   <8-9>  MnSymbolC8
   <9-10> MnSymbolC9
  <10-12> MnSymbolC10
  <12->   MnSymbolC12}{}
\DeclareMathSymbol{\intprod}{\mathbin}{MnSyC}{'270}

\newcommand{\C}{\mathbb{C}}

\newcommand{\Z}{\mathbb{Z}}
\newcommand{\R}{\mathbb{R}}

\newcommand{\longto}{\longrightarrow}

\let\nc\newcommand
\let\renc\renewcommand
\nc{\wbar}{\overline}
\let\td\tilde
\let\wtd\widetilde
\let\wht\widehat
\let\mcl\mathcal

\nc{\ab}{{\bar{a}}} \nc{\at}{\tilde{a}} \nc{\ah}{\hat{a}}
\nc{\bb}{{\bar{b}}} \nc{\bt}{\tilde{b}} \nc{\bh}{\hat{b}}
\nc{\cb}{{\bar{c}}} \nc{\ct}{\tilde{c}} 
\nc{\db}{{\bar{d}}} \nc{\dt}{\tilde{d}} \renc{\dh}{\hat{d}}
\nc{\eb}{{\bar{e}}} \nc{\et}{\tilde{e}} \nc{\eh}{\hat{e}}
\nc{\fb}{{\bar{f}}} \nc{\ft}{\tilde{f}} \nc{\fh}{\hat{f}}
\nc{\gb}{{\bar{g}}} \nc{\gt}{\tilde{g}} \nc{\gh}{\hat{g}}
\nc{\hb}{{\bar{h}}} \nc{\hh}{\hat{h}} 
\nc{\ib}{{\bar{\imath}}} \nc{\ih}{\hat{\imath}} 
\nc{\jb}{{\bar{\jmath}}} \nc{\jt}{\tilde{\jmath}} \nc{\jh}{\hat{\jmath}}
\nc{\kb}{{\bar{k}}} \nc{\kt}{\tilde{k}} \nc{\kh}{\hat{k}}
\nc{\lb}{{\bar{l}}} \nc{\lt}{\tilde{l}} \nc{\lh}{\hat{l}}
\nc{\mb}{{\bar{m}}} \nc{\mt}{\tilde{m}} \nc{\mh}{\hat{m}}
\nc{\nb}{{\bar{n}}} \nc{\nt}{\tilde{n}} \nc{\nh}{\hat{n}}
\nc{\ob}{{\bar{o}}} \nc{\ot}{\tilde{o}} \nc{\oh}{\hat{o}}
\nc{\pb}{{\bar{p}}} \nc{\pt}{\tilde{p}} \nc{\ph}{\hat{p}}
\nc{\qb}{{\bar{q}}} \nc{\qt}{\tilde{q}} \nc{\qh}{\hat{q}}
\nc{\rb}{{\bar{r}}} \nc{\rt}{\tilde{r}} \nc{\rh}{\hat{r}}
\nc{\st}{\tilde{s}} \nc{\sh}{\hat{s}} 
\nc{\tb}{{\bar{t}}} \renc{\th}{\hat{t}} 
\nc{\ub}{{\bar{u}}} \nc{\ut}{\tilde{u}} \nc{\uh}{\hat{u}}
\nc{\vb}{{\bar{v}}} \nc{\vt}{\tilde{v}} \nc{\vh}{\hat{v}}
\nc{\wb}{{\bar{w}}} \nc{\wt}{\tilde{w}} \nc{\wh}{\hat{w}}
\nc{\xb}{{\bar{x}}} \nc{\xt}{\tilde{x}} \nc{\xh}{\hat{x}}
\nc{\yb}{{\bar{y}}} \nc{\yt}{\tilde{y}} \nc{\yh}{\hat{y}}
\nc{\zb}{{\bar{z}}} \nc{\zt}{\tilde{z}} \nc{\zh}{\hat{z}}

\nc{\Ab}{\wbar{A}} \nc{\At}{\wtd{A}} \nc{\Ah}{\wht{A}}
\nc{\Bb}{\wbar{B}} \nc{\Bt}{\wtd{B}} \nc{\Bh}{\wht{B}}
\nc{\Cb}{\wbar{C}} \nc{\Ct}{\wtd{C}} \nc{\Ch}{\wht{C}}
\nc{\Db}{\wbar{D}} \nc{\Dt}{\wtd{D}} \nc{\Dh}{\wht{D}}
\nc{\Eb}{\wbar{E}} \nc{\Et}{\wtd{E}} \nc{\Eh}{\wht{E}}
\nc{\Fb}{\wbar{F}} \nc{\Ft}{\wtd{F}} \nc{\Fh}{\wht{F}}
\nc{\Gb}{\wbar{G}} \nc{\Gt}{\wtd{G}} \nc{\Gh}{\wht{G}}
\nc{\Hb}{\wbar{H}} \nc{\Ht}{\wtd{H}} \nc{\Hh}{\wht{H}}
\nc{\Ib}{\wbar{I}} \nc{\It}{\wtd{I}} \nc{\Ih}{\wht{I}}
\nc{\Jb}{\wbar{J}} \nc{\Jt}{\wtd{J}} \nc{\Jh}{\wht{J}}
\nc{\Kb}{\wbar{K}} \nc{\Kt}{\wtd{K}} \nc{\Kh}{\wht{K}}
\nc{\Lb}{\wbar{L}} \nc{\Lt}{\wtd{L}} \nc{\Lh}{\wht{L}}
\nc{\Mb}{\wbar{M}} \nc{\Mt}{\wtd{M}} \nc{\Mh}{\wht{M}}
\nc{\Nb}{\wbar{N}} \nc{\Nt}{\wtd{N}} \nc{\Nh}{\wht{N}}
\nc{\Ob}{\wbar{O}} \nc{\Ot}{\wtd{O}} \nc{\Oh}{\wht{O}}
\nc{\Pb}{\wbar{P}} \nc{\Pt}{\wtd{P}} \nc{\Ph}{\wht{P}}
\nc{\Qb}{\wbar{Q}} \nc{\Qt}{\wtd{Q}} \nc{\Qh}{\wht{Q}}
\nc{\Rb}{\wbar{R}} \nc{\Rt}{\wtd{R}} \nc{\Rh}{\wht{R}}
\nc{\Sb}{\wbar{S}} \nc{\St}{\wtd{S}} \nc{\Sh}{\wht{S}}
\nc{\Tb}{\wbar{T}} \nc{\Tt}{\wtd{T}} \nc{\Th}{\wht{T}}
\nc{\Ub}{\wbar{U}} \nc{\Ut}{\wtd{U}} \nc{\Uh}{\wht{U}}
\nc{\Vb}{\wbar{V}} \nc{\Vt}{\wtd{V}} \nc{\Vh}{\wht{V}}
\nc{\Wb}{\wbar{W}} \nc{\Wt}{\wtd{W}} \nc{\Wh}{\wht{W}}
\nc{\Xb}{\wbar{X}} \nc{\Xt}{\wtd{X}} \nc{\Xh}{\wht{X}}
\nc{\Yb}{\wbar{Y}} \nc{\Yt}{\wtd{Y}} \nc{\Yh}{\wht{Y}}
\nc{\Zb}{\wbar{Z}} \nc{\Zt}{\wtd{Z}} \nc{\Zh}{\wht{Z}}

\nc{\CA}{\mcl{A}} \nc{\CAb}{\wbar{\CA}} \nc{\CAt}{\wtd{\CA}} \nc{\CAh}{\wht{\CA}}
\nc{\CB}{\mcl{B}} \nc{\CBb}{\wbar{\CB}} \nc{\CBt}{\wtd{\CB}} \nc{\CBh}{\wht{\CB}}
\nc{\CC}{\mcl{C}} \nc{\CCb}{\wbar{\CC}} \nc{\CCt}{\wtd{\CC}} \nc{\CCh}{\wht{\CC}}
\nc{\cDt}{\wtd{\cC}} \nc{\cDh}{\wht{\cD}}
\nc{\CE}{\mcl{E}} \nc{\CEb}{\wbar{\CE}} \nc{\CEt}{\wtd{\CE}} \nc{\CEh}{\wht{\CE}}
\nc{\CF}{\mcl{F}} \nc{\CFb}{\wbar{\CF}} \nc{\CFt}{\wtd{\CF}} \nc{\CFh}{\wht{\CF}}
\nc{\CG}{\mcl{G}} \nc{\CGb}{\wbar{\CG}} \nc{\CGt}{\wtd{\CG}} \nc{\CGh}{\wht{\CG}}
\nc{\CH}{\mcl{H}} \nc{\CHb}{\wbar{\CH}} \nc{\CHt}{\wtd{\CH}} \nc{\CHh}{\wht{\CH}}
\nc{\CI}{\mcl{I}} \nc{\CIb}{\wbar{\CI}} \nc{\CIt}{\wtd{\CI}} \nc{\CIh}{\wht{\CI}}
\nc{\CJ}{\mcl{J}} \nc{\CJb}{\wbar{\CJ}} \nc{\CJt}{\wtd{\CJ}} \nc{\CJh}{\wht{\CJ}}
\nc{\CK}{\mcl{K}} \nc{\CKb}{\wbar{\CK}} \nc{\CKt}{\wtd{\CK}} \nc{\CKh}{\wht{\CK}}
\nc{\CL}{\mcl{L}} \nc{\CLb}{\wbar{\CL}} \nc{\CLt}{\wtd{\CL}} \nc{\CLh}{\wht{\CL}}
\nc{\CM}{\mcl{M}} \nc{\CMb}{\wbar{\CM}} \nc{\CMt}{\wtd{\CM}} \nc{\CMh}{\wht{\CM}}
\nc{\CN}{\mcl{N}} \nc{\CNb}{\wbar{\CN}} \nc{\CNt}{\wtd{\CN}} \nc{\CNh}{\wht{\CN}}
\nc{\CO}{\mcl{O}} \nc{\COb}{\wbar{\CO}} \nc{\COt}{\wtd{\CO}} \nc{\COh}{\wht{\CO}}
\nc{\CQ}{\mcl{Q}} \nc{\CQb}{\wbar{\CQ}} \nc{\CQt}{\wtd{\CQ}} \nc{\CQh}{\wht{\CQ}}
\nc{\CR}{\mcl{R}} \nc{\CRb}{\wbar{\CR}} \nc{\CRt}{\wtd{\CR}} \nc{\CRh}{\wht{\CR}}
\nc{\CS}{\mcl{S}} \nc{\CSb}{\wbar{\CS}} \nc{\CSt}{\wtd{\CS}} \nc{\CSh}{\wht{\CS}}
\nc{\CT}{\mcl{T}} \nc{\CTb}{\wbar{\CT}} \nc{\CTt}{\wtd{\CT}} \nc{\CTh}{\wht{\CT}}
\nc{\CU}{\mcl{U}} \nc{\CUb}{\wbar{\CU}} \nc{\CUt}{\wtd{\CU}} \nc{\CUh}{\wht{\CU}}
\nc{\CV}{\mcl{V}} \nc{\CVb}{\wbar{\CV}} \nc{\CVt}{\wtd{\CV}} \nc{\CVh}{\wht{\CV}}
\nc{\CW}{\mcl{W}} \nc{\CWb}{\wbar{\CW}} \nc{\CWt}{\wtd{\CW}} \nc{\CWh}{\wht{\CW}}
\nc{\CX}{\mcl{X}} \nc{\CXb}{\wbar{\CX}} \nc{\CXt}{\wtd{\CX}} \nc{\CXh}{\wht{\CX}}
\nc{\CY}{\mcl{Y}} \nc{\CYb}{\wbar{\CY}} \nc{\CYt}{\wtd{\CY}} \nc{\CYh}{\wht{\CY}}
\nc{\CZ}{\mcl{Z}} \nc{\CZb}{\wbar{\CZ}} \nc{\CZt}{\wtd{\CZ}} \nc{\CZh}{\wht{\CZ}}

\let\eps\epsilon
\let\ups\upsilon
\let\veps\varepsilon
\let\vtht\vartheta
\let\vsgm\varsigma
\let\vphi\varphi
\let\vrho\varrho

\nc{\alphab}{\bar{\alpha}} \nc{\alphat}{\td{\alpha}} \nc{\alphah}{\hat{\alpha}}
\nc{\betab}{\bar{\beta}}   \nc{\betat}{\td{\beta}}   \nc{\betah}{\hat{\beta}}
\nc{\gammab}{\bar{\gamma}} \nc{\gammat}{\td{\gamma}} \nc{\gammah}{\hat{\gamma}}
\nc{\deltab}{\bar{\delta}} \nc{\deltat}{\td{\delta}} \nc{\deltah}{\hat{\delta}}
\nc{\epsilonb}{\bar{\eps}} \nc{\epsilont}{\td{\eps}} \nc{\epsilonh}{\hat{\eps}}
\nc{\vepsb}{\bar{\veps}}   \nc{\vepst}{\td{\veps}}   \nc{\vepsh}{\hat{\veps}}
\nc{\zetab}{\bar{\zeta}}   \nc{\zetat}{\td{\zeta}}   \nc{\zetah}{\hat{\zeta}}
\nc{\etab}{\bar{\eta}}
\nc{\etah}{\hat{\eta}}
\nc{\thetab}{\bar{\theta}} \nc{\thetat}{\td{\theta}} \nc{\thetah}{\hat{\theta}}
\nc{\vthetab}{\bar{\vtht}} \nc{\vthetat}{\td{\vtht}} \nc{\vthetah}{\hat{\vtht}}
\nc{\lambdat}{\td{\lambda}} \nc{\lambdah}{\hat{\lambda}}
\nc{\iotab}{\bar{\iota}}   \nc{\iotat}{\td{\iota}}   \nc{\iotah}{\hat{\iota}}
\nc{\kappab}{\bar{\kappa}} \nc{\kappat}{\td{\kappa}} \nc{\kappah}{\hat{\kappa}}
\nc{\lmdb}{\bar{\lmd}}     \nc{\lmdt}{\td{\lmd}}     \nc{\lmdh}{\hat{\lmd}}
\nc{\mub}{\bar{\mu}}       \nc{\mut}{\td{\mu}}       \nc{\muh}{\hat{\mu}}
\nc{\nub}{\bar{\nu}}       \nc{\nut}{\td{\nu}}       \nc{\nuh}{\hat{\nu}}
\nc{\xib}{\bar{\xi}}       \nc{\xit}{\td{\xi}}       \nc{\xih}{\hat{\xi}}
\nc{\pib}{\bar{\pi}}       \nc{\pit}{\td{\pi}}       \nc{\pih}{\hat{\pi}}
\nc{\vpib}{\bar{\vpi}}     \nc{\vpit}{\td{\vpi}}     \nc{\vpih}{\hat{\vpi}}
\nc{\rhob}{\bar{\rho}}     \nc{\rhot}{\td{\rho}}     \nc{\rhoh}{\hat{\rho}}
\nc{\vrhob}{\bar{\vrho}}   \nc{\vrhot}{\td{\vrho}}   \nc{\vrhoh}{\hat{\vrho}}
\nc{\sigmab}{\bar{\sigma}} \nc{\sigmat}{\td{\sigma}} \nc{\sigmah}{\hat{\sigma}}
\nc{\vsigmab}{\bar{\vsgm}} \nc{\vsigmat}{\td{\vsgm}} \nc{\vsigmah}{\hat{\vsgm}}
\nc{\taub}{\bar{\tau}}     \nc{\taut}{\td{\tau}}     \nc{\tauh}{\hat{\tau}}
\nc{\upsb}{\bar{\ups}} \nc{\upst}{\td{\ups}} \nc{\upsh}{\hat{\ups}}
\nc{\phib}{\bar{\phi}}     \nc{\phit}{\td{\phi}}     \nc{\phih}{\hat{\phi}}
\nc{\varphib}{\bar{\vphi}}   \nc{\varphit}{\td{\vphi}}   \nc{\varphih}{\hat{\vphi}}
\nc{\chib}{\bar{\chi}}
\nc{\chih}{\hat{\chi}}
\nc{\psib}{\bar{\psi}}
\nc{\psih}{\hat{\psi}}
\nc{\omegab}{\bar{\omega}} \nc{\omegat}{\td{\omega}} \nc{\omegah}{\hat{\omega}}

\nc{\Gammab}{\wbar{\Gamma}}     \nc{\Gammat}{\wtd{\Gamma}}     \nc{\Gammah}{\wht{\Gamma}}
\nc{\Deltab}{\wbar{\Delta}}     \nc{\Deltat}{\wtd{\Delta}}     \nc{\Deltah}{\wht{\Delta}}
\nc{\Thetab}{\wbar{\Theta}}     \nc{\Thetat}{\wtd{\Theta}}     \nc{\Thetah}{\wht{\Theta}}
\nc{\Lambdab}{\wbar{\Lambda}}   \nc{\Lambdat}{\wtd{\Lambda}}   \nc{\Lambdah}{\wht{\Lambda}}
\nc{\Xib}{\wbar{\Xi}}           \nc{\Xit}{\wtd{\Xi}}           \nc{\Xih}{\wht{\Xi}}
\nc{\Pib}{\wbar{\Pi}}           \nc{\Pit}{\wtd{\Pi}}           \nc{\Pih}{\wht{\Pi}}
\nc{\Sigmab}{\wbar{\Sigma}}     \nc{\Sigmat}{\wtd{\Sigma}}     \nc{\Sigmah}{\wht{\Sigma}}
\nc{\Upsilonb}{\wbar{\Upsilon}} \nc{\Upsilont}{\wtd{\Upsilon}} \nc{\Upsilonh}{\wht{\Upsilon}}
\nc{\Phib}{\wbar{\Phi}}         \nc{\Phit}{\wtd{\Phi}}         \nc{\Phih}{\wht{\Phi}}
\nc{\Psib}{\wbar{\Psi}}         \nc{\Psit}{\wtd{\Psi}}         \nc{\Psih}{\wht{\Psi}}
\nc{\Omegab}{\wbar{\Omega}}     \nc{\Omegat}{\wtd{\Omega}}     \nc{\Omegah}{\wht{\Omega}}

\def\bear{\begin{eqnarray}}
\def\eear{\end{eqnarray}}

\newcommand{\Tr}{{\textrm{Tr}\;}}

\newcommand{\cD}{{\cal D}}

\newcommand{\hlf}{\frac{1}{2}}
\newcommand{\gij}{g_{i\jb}}

\let\OLDthebibliography\thebibliography
\renewcommand\thebibliography[1]{
  \OLDthebibliography{#1}
  \setlength{\parskip}{5pt}
  \setlength{\itemsep}{0pt plus 0.3ex}
}
\usepackage{titlesec}
\titleformat*{\section}{\bfseries\large}
\begin{document}
\addtolength{\baselineskip}{1.5mm}

\thispagestyle{empty}
\vbox{}
\vspace{3.0cm}

\begin{center}
\centerline{\LARGE{Topological 5d ${\cal N}=2$ Gauge Theory: Novel Floer Homologies,}}
\bigskip
\centerline{\LARGE{their Dualities, and an $A_\infty$-category of Three-Manifolds}}

\vspace{3.0cm}

{Arif Er\footnote{E-mail: arif.er@u.nus.edu}, Zhi-Cong~Ong\footnote{E-mail: zc\textunderscore ong@nus.edu.sg} and Meng-Chwan~Tan\footnote{E-mail: mctan@nus.edu.sg}}
\\[2mm]
{\it Department of Physics\\
National University of Singapore \\
2 Science Drive 3, Singapore 117551} \\[1mm]
\end{center}

\vspace{1.5cm}

\centerline{\bf Abstract}\smallskip \noindent

We show how one can define novel gauge-theoretic Floer homologies of four, three, and two-manifolds from the physics of a certain topologically-twisted 5d ${\cal N}=2$ gauge theory via its supersymmetric quantum mechanics interpretation.
They are associated with Vafa-Witten, Hitchin, and $G_\C$-BF configurations on the four, three, and two-manifolds, respectively.
We also show how one can define novel symplectic Floer homologies of Hitchin spaces, which in turn will allow us to derive novel Atiyah-Floer correspondences that relate our gauge-theoretic Floer homologies to symplectic intersection Floer homologies of Higgs bundles.
Furthermore, topological invariance and 5d ``S-duality'' suggest a web of relations and a Langlands duality amongst these novel Floer homologies and their loop/toroidal group generalizations.
Last but not least, via a 2d gauged Landau-Ginzburg model interpretation of the 5d theory, we derive, from the soliton string theory that it defines and the 5d partition function, a Fukaya-Seidel type $A_\infty$-category of Hitchin configurations on three-manifolds -- thereby categorifying the aforementioned Floer homology of three-manifolds -- and its novel Atiyah-Floer type correspondence.
Our work therefore furnishes purely physical proofs and generalizations of the mathematical conjectures by Haydys~\cite{haydys2010fukaya}, Abouzaid-Manolescu~\cite{abouzaid2017sheaftheor}, and Bousseau~\cite{bousseau-2024-holom-floer}, and more.

\newpage

\renewcommand{\thefootnote}{\arabic{footnote}}
\setcounter{footnote}{0}

\tableofcontents

\section{Introduction, Summary, and Acknowledgements}
\vspace{0.4cm}
\setlength{\parskip}{5pt}

\bigskip\noindent\textit{Introduction}
\vspace*{0.5em}

In a visionary mathematical paper~\cite{haydys2010fukaya} by Haydys, he conjectured, with support from some preliminary computations, novel invariants assigned to five, four, and three-manifolds from gauge theory.
Specifically, he defined a certain five-dimensional equation from which he could associate an integer to a five-manifold, Floer type homology groups to a four-manifold, and a Fukaya–Seidel type category to a three-manifold.
This five-dimensional equation later appeared in~\cite{witten2011fivebranes}, too, albeit in a purely physical setting where the five-manifold was of the form $M_4 \times \R$, whence it was thereafter named the Haydys-Witten equation.

In this paper, we aim, amongst other things, to physically define the aforementioned gauge invariants assigned to four and three-manifolds, and derive their connection to symplectic invariants.
To this end, we will study, on various possibly decomposable five-manifolds in different topological limits, the topologically-twisted 5d $\mathcal{N} = 2$ gauge theory defined in~\cite{anderson2013five} whose BPS equations are exactly the Haydys-Witten equations.
As an offshoot, we would be able to derive a web of mathematically-novel relations between these and a variety of other invariants, their Atiyah-Floer correspondences, and more.

The computational techniques we employ are mainly those of standard Kaluza-Klein reduction,  topological reduction as pioneered in~\cite{bershadsky1995topolreduc}, recasting gauge theories as supersymmetric quantum mechanics as pioneered in~\cite{blau1993topological}, and the physical interpretation of Floer homology groups in terms of supersymmetric quantum mechanics in infinite-dimensional space as elucidated in~\cite{ong2022vafa}.

Let us now give a brief plan and summary of the paper.

\bigskip\noindent\textit{A Brief Plan and Summary of the Paper}
\vspace*{0.5em}

In $\S$\ref{section: hw theory general}, we discuss general aspects of a topologically-twisted  5d ${\mathcal {N}} = 2$ theory on $M_5=M_4 \times \R$ resulting from the Haydys-Witten (HW) twist, where the gauge group $G$ is taken to be a real, simple, and compact Lie group.

In $\S$\ref{section: hw floer homology}, we recast the 5d-HW theory as a 1d supersymmetric quantum mechanics (SQM) in the space  $\mathfrak{A}_4$ of irreducible $(A,B)$ fields on $M_4$ with action \eqref{hw sqm}, where $A \in \Omega^1(M_4)$ and $B \in \Omega^{2,+}(M_4)$ are a gauge connection and self-dual two-form, respectively. This will in turn allow us to express the partition function as \eqref{5d partition function floer}:
\begin{equation}
  \label{summary: 5d partition function floer}
  \boxed{
    \mathcal{Z}_{\text{HW},M_4 \times \R}(G)
    =  \sum_k  {\cal F}^{G}_{\text{HW}}(\Psi_{M_4}^k)
    = \sum_k  \text{HF}^{\text{HW}}_{d_k}(M_4, G)
    = \mathcal{Z}^{\text{Floer}}_{\text{HW},M_4}(G)
  }
\end{equation}
where $\text{HF}^{\text{HW}}_{d_k}(M_4, G)$ is a \emph{novel} Haydys-Witten Floer homology class assigned to $M_4$ of degree $d_k$, defined by Floer differentials described by the gradient flow equations \eqref{flow on m5 = m4 x R 2/V}:
\begin{equation}
  \label{summary: flow on m5 = m4 x R 2/V}
  \boxed{
    \begin{aligned}
      \frac{d{A}^a}{dt}
      &= -g^{ab}_{{\mathfrak A}_4}\frac{\partial V_4(A, B)}{\partial A^{b}}
      \\
      \frac{d{B}^a}{dt}
      &= -g^{ab}_{{\mathfrak A}_4}\frac{\partial V_4(A, B)}{\partial B^{b}}
    \end{aligned}
  }
\end{equation}
and Morse functional \eqref{morse potential hw}:
\begin{equation}
  \label{summary: morse potential hw}
  \boxed{
    V_4(A, B)
    = \int_{M_4} \, \text{Tr}\, \Big( - F^+\wedge \star B + \frac{1}{3}B \wedge (\star_3 (\mathscr{B} \wedge_3 \mathscr{B})) \Big)
  }
\end{equation}
The chains of the HW Floer complex are generated by fixed critical points of $V_4$, which correspond to \emph{time-invariant Vafa-Witten (VW) configurations on $M_4$} given by time-independent solutions to the 4d equations \eqref{VW configuration}:
\begin{equation}
  \label{VW configuration - 0}
  \boxed{
    \begin{aligned}
      F^{+} - \star_3(\mathscr{B} \wedge_3 \mathscr{B})
      &= 0
      \\
      D \star B
      &= 0
    \end{aligned}
  }
\end{equation}
Note that $\text{HF}^{\text{HW}}_{d_k}(M_4, G)$ was conjectured to exist by Haydys in~\cite[\S5]{haydys2010fukaya}. We have therefore furnished a purely physical proof of his mathematical conjecture.

In $\S$\ref{section: complex flow on m3 x R}, we let
$M_4= M_3 \times S^1$, and perform a Kaluza-Klein (KK) dimensional reduction of 5d-HW theory by shrinking $S^1$ to be infinitesimally small. We obtain the corresponding 1d SQM theory in the space $\mathfrak{A}_3$ of irreducible fields $(A, B) \in \Omega^1(M_3)$ and $C \in \Omega^0(M_3)$ on $M_3$ with action \eqref{vw sqm final}, that is equivalent to the resulting 4d theory on $M_3 \times \mathbb{R}$. As before, this will allow us to express the partition function as \eqref{hitchin floer classes}:
\begin{equation}
  \label{summay: hitchin floer classes}
  \boxed{
    \mathcal{Z}_{\text{HW},M_3 \times \R}(G)
    = \sum_l {\cal F}^{G}_{\text{HW}_4}(\Psi_{M_3}^l)
    = \sum_l \text{HF}_{d_l}^{\text{HW}_4}(M_3, G)
    = \mathcal{Z}^{\text{Floer}}_{\text{HW}_4,M_3}(G)
  }
\end{equation}
where $\text{HF}_{d_l}^{\text{HW}_4}(M_3, G)$ is a \emph{novel} 4d-Haydys-Witten Floer homology class assigned to $M_3$ of degree $d_l$, defined by Floer differentials described by the gradient flow equation \eqref{flow on m4 = m3 x R}:
\begin{equation}\label{summary:flow on m4 = m3 x R}
  \boxed{
    \begin{aligned}
      \frac{d{A}^a}{dt}
      &= -g^{ab}_{{\mathfrak A}_3}\frac{\partial V_3(A, B, C)}{\partial A^{b}}
      \\
      \frac{d{B}^a}{dt}
      &= -g^{ab}_{{\mathfrak A}_3}\frac{\partial V_3(A, B, C)}{\partial B^{b}}
      \\
      \frac{d{C}^a}{dt}
      &= -g^{ab}_{{\mathfrak A}_3}\frac{\partial V_3(A, B, C)}{\partial C^{b}}
    \end{aligned}
  }
\end{equation}
and Morse functional \eqref{v3 functional}:
\begin{equation}\label{summary: v3 functional}
  \boxed{
    V_3(A, B, C)
    = \int_{M_3} \text{Tr}\, \bigg(
    - F \wedge B
    + \frac{1}{3} B \wedge B \wedge B
    - C \wedge D \star B\bigg)
  }
\end{equation}
The chains of the 4d-HW Floer complex are generated by fixed critical points of $V_3$, which correspond to \emph{time-invariant Hitchin configurations on $M_3$}  given by time-independent solutions of the 3d equations \eqref{H configuration}:
\begin{equation}
  \label{summary:H configuration}
  \boxed{
    \begin{aligned}
      F - B \wedge B
      &= 0
      \\
      DB
      &= 0
      \\
      D \star B
      &= 0
    \end{aligned}
  }
\end{equation}

In $\S$\ref{section: m2 x R}, we further specialize to the case where $M_3=M_2 \times S^1$, and perform a second KK dimensional reduction of 5d-HW theory by shrinking this $S^1$ to be infinitesimally small. We obtain the corresponding 1d SQM theory in the space $\mathfrak{A}_2$ of irreducible $(\mathcal{A}, Z, \bar{\mathcal{A}}, \bar{Z})$ fields on $M_2$ with action \eqref{sqm action m2 x R}, where $\mathcal{A}, \bar{\mathcal{A}} \in \Omega^1(M_2)$ and $Z, \bar{Z} \in \Omega^0(M_2)$ are (conjugate) connections and scalar fields of a complexified gauge group $G_\C$, respectively, that is equivalent to the resulting 3d theory on $M_2 \times \mathbb{R}$. Again, this will allow us to express the partition function as \eqref{BF floer classes}:
\begin{equation}
  \label{summary: BF floer classes}
  \boxed{
    \mathcal{Z}_{\text{HW},M_2 \times \R}(G)
    = \sum_p {\cal F}^{G}_{\text{HW}_3}(\Psi_{M_2}^p)
    = \sum_p \text{HF}_{d_p}^{\text{HW}_3}(M_2, G_\C)
    = \mathcal{Z}^{\text{Floer}}_{\text{HW}_3,M_2}(G_\C)
  }
\end{equation}
where $\text{HF}_{d_p}^{\text{HW}_3}(M_2, G_\C) $ is a \emph{novel} 3d-Haydys-Witten Floer homology class assigned to $M_2$ of degree $d_p$, defined by Floer differentials described by the gradient flow equations \eqref{m2 x R flow}:
\begin{equation}
  \label{summary: m2 x R flow}
  \boxed{
    \begin{aligned}
      \frac{d\mathcal{A}^{a}}{dt}
      &= -g^{ab}_{\mathfrak A_2} \left(\frac{\partial V_2(\mathcal{A}, Z, \bar {\mathcal{A}}, \bar Z)}{\partial \mathcal{A}^{b}}\right)^*
      &\qquad
        \frac{d \bar{\mathcal{A}}^{a}}{dt}
      &= -g^{ab}_{\mathfrak A_2} \left(\frac{\partial V_2(\mathcal{A}, Z, \bar {\mathcal{A}}, \bar Z)}{\partial \bar{\mathcal{A}}^{b}}\right)^*
      \\
      \frac{d{Z}^{a}}{dt}
      &= - g^{ab}_{\mathfrak A_2} \left(\frac{\partial V_2(\mathcal{A}, Z, \bar {\mathcal{A}}, \bar Z)}{\partial Z^{b}}\right)^*
      &\qquad
        \frac{d \bar{Z}^{a}}{dt}
      &= - g^{ab}_{\mathfrak A_2} \left(\frac{\partial V_2(\mathcal{A}, Z, \bar {\mathcal{A}}, \bar Z)}{\partial \bar{Z}^{b}}\right)^*
    \end{aligned}
  }
\end{equation}
and Morse functional \eqref{potential funcitonal on space of fields on m2}:
\begin{equation}
  \label{summary: potential funcitonal on space of fields on m2}
  \boxed{
    V_2(\mathcal{A}, Z, \bar {\mathcal{A}}, \bar Z)
    = i \int_{M_2 } \,\text{Tr}\, \Big( Z \wedge \mathcal{F} - \bar{Z} \wedge \bar{\mathcal{F}} \Big)
  }
\end{equation}
The chains of the 3d-HW Floer complex are generated by fixed critical points of $V_2$, which correspond to \emph{time-invariant $G_\C$-BF configurations on $M_2$} given by time-independent solutions of the 2d equations \eqref{BF configurations}:
\begin{equation}\label{BF configurations - 0}
  \boxed{
    \begin{aligned}
      \mathcal{F}
      &= 0 = \bar{\mathcal{F}}
      \\
      \mathcal{D} Z
      &= 0 = \bar{\mathcal{D}} \bar{Z}
    \end{aligned}
  }
\end{equation}

In $\S$\ref{section: symplectic floer homology}, we specialize to the case where $M_5 = \Sigma \times M_2 \times \mathbb{R}$, and perform a topological reduction of 5d-HW theory along a curved Riemann surface $M_2$ using the Bershadsky-Johansen-Sadov-Vafa (BJSV) reduction method from \cite{bershadsky1995topolreduc}, to obtain a 3d ${\mathcal {N}}=4$ topological sigma model on $\Sigma \times \R$. First, by considering $\Sigma = S^1 \times S^1$, we recast the 3d sigma model as a 1d SQM theory in the double loop space $L^2\mathcal{M}^{G_\C}_{\text{flat}}(M_2)$ of the symplectic moduli space $\mathcal{M}^{G_\C}_{\text{flat}}(M_2)$ of flat $G_\C$ connections on $M_2$ with action \eqref{sqm action 3d sigma}.
This will allow us to express the partition function as \eqref{L^2M floer classes}:
\begin{equation}
  \label{summary: L^2M floer classes}
  \boxed{
    \mathcal{Z}_{\text{HW},S^1 \times S^1 \times \mathbb{R}}(G)
    = \sum_q  {\cal F}^{q}_{L^2\mathcal{M}^{G_\C}_{\text{flat}}(M_2)}
    = \sum_q \text{HSF}^{\text{hol}}_{d_q}({L^2\mathcal{M}^{G_\C}_{\text{flat}}(M_2)})
    = \mathcal{Z}^{\text{SympFloer}}_{L^2\mathcal{M}^{G_\C}_{\text{flat}}(M_2)}
  }
\end{equation}
where $\text{HSF}^{\text{hol}}_{d_q}(L^2\mathcal{M}^{G_\C}_{\text{flat}}(M_2))$ is a \emph{novel} symplectic Floer homology class in $L^2\mathcal{M}^{G_{\C}}_{\text{flat}}(M_2)$ of degree $d_q$ counted by Floer differentials described by the gradient flow equations \eqref{3d sigma flow}:
\begin{equation}
  \label{summary:3d sigma flow}
  \boxed{
    \begin{aligned}
      \frac{dZ^a}{dt}
      &= - g^{a \bb}_{L^2{\mathcal{M}^{G_\C}_{\text{flat}}}} \left(\frac{\partial V_\Sigma(Z, \bar{Z})}{\partial Z^b}\right)^*
      &\qquad
        \frac{d\bar{Z}^a}{dt}
      &= - g^{a \bb}_{L^2\mathcal{M}^{G_\C}_{\text{flat}}} \left(\frac{\partial V_\Sigma(Z, \bar{Z})}{\partial \bar{Z}^b}\right)^*
      \\
      \frac{dZ^\ab}{dt}
      &= - g^{\ab b}_{L^2{\mathcal{M}^{G_\C}_{\text{flat}}}} \left(\frac{\partial V_\Sigma(Z, \bar{Z})}{\partial Z^\bb}\right)^*
      &\qquad
        \frac{d\bar{Z}^\ab}{dt}
      &= - g^{\ab b}_{L^2{\mathcal{M}^{G_\C}_{\text{flat}}}} \left(\frac{\partial V_\Sigma(Z, \bar{Z})}{\partial \bar{Z}^\bb}\right)^*
    \end{aligned}
  }
\end{equation}
where $(Z^a, Z^\ab, \bar{Z}^a, \bar{Z}^\ab)$ are the coordinates of $L^2\mathcal{M}^{G_{\C}}_{\text{flat}}(M_2)$, and
\begin{equation}
  \label{summary: potential functions for 3d sigma}
  \boxed{
    V_\Sigma(Z, \bar{Z})
    = \int_{S^1 \times S^1} dr ds \,
    g_{i\jb} \Big( Z^i \partial_s Z^\jb - i Z^i \partial_r \bar{Z}^\jb - \bar{Z}^i \partial_s \bar{Z}^\jb + i \bar{Z}^i \partial_r Z^\jb \Big)
  }
\end{equation}
is the Morse functional in \eqref{potential functions for 3d sigma}. The chains of the symplectic Floer complex are generated by fixed critical points of $V_\Sigma$, which correspond to \emph{time-invariant maps from $S^1 \times S^1$ to $\mathcal{M}^{G_\C}_{\text{flat}}(M_2)$} given by the time-independent solutions of the 2d equations \eqref{critical points of 3d sigma potential} on $S^1 \times S^1$:
\begin{equation}
  \label{summary: critical points of 3d sigma potential}
  \boxed{
    \begin{aligned}
      i \partial_r Z^i + \partial_s \bar{Z}^i
      &= 0 = i \partial_r \bar{Z}^i + \partial_s Z^i
      \\
      i \partial_r \bar{Z}^\ib - \partial_s {Z}^\ib
      &= 0 = i \partial_r Z^\ib - \partial_s \bar{Z}^\ib
    \end{aligned}
  }
\end{equation}

Second, by considering $\Sigma = I \times S^1$, the 3d sigma model is first recast as a 2d A-model on $I \times \R$, and then further recast as a 1d SQM theory in the interval space $\mathcal{M}(\mathscr{L}_0, \mathscr{L}_1)_{L\mathcal{M}^{G_\C, M_2}_{\text{flat}}}$ of smooth trajectories between A-branes starting at $\mathscr{L}_0$ and ending on $\mathscr{L}_1$ in the loop space $L\mathcal{M}^{G_\C}_{\text{flat}}(M_2)$ of maps from $S^1$ to $\mathcal{M}^{G_\C}_{\text{flat}}(M_2)$ with action \eqref{sqm action on interval space to loop space of higgs}. This will allow us to express the partition function as \eqref{interval to loop space floer classes}:
\begin{equation}
  \label{summary: interval to loop space floer classes}
  \boxed{
    \mathcal{Z}_{\text{HW},\R \times I \rightarrow S^1}(G)
    = \sum_u  {\cal F}^{u}_{L\mathcal{M}^{G_\C, M_2}_{\text{flat}}}
    = \sum_u \text{HSF}^{\text{Int}}_{d_u}({L\mathcal{M}^{G_\C, M_2}_{\text{flat}}}, \mathscr{L}_0, \mathscr{L}_1)
    = \mathcal{Z}^{\text{IntSympFloer}}_{L\mathcal{M}^{G_\C, M_2}_{\text{flat}}}
  }
\end{equation}
where $\text{HSF}^{\text{Int}}_{d_u}({L\mathcal{M}^{G_\C, M_2}_{\text{flat}}}, \mathscr{L}_0, \mathscr{L}_1)$ is a symplectic intersection Floer homology class generated by the intersection points of $\mathscr{L}_0$ and $\mathscr{L}_1$ in $L\mathcal{M}^{G_\C, M_2}_{\text{flat}}$, of degree $d_u$ counted by the Floer differential realized as the flow lines of the SQM whose gradient flow equations are the expression within the squared terms in \eqref{sqm action on interval space to loop space of higgs} set to zero.

Third, by considering $\Sigma = I \times \R$, the 3d sigma model is first recast as a 2d A$_\theta$-model on $I \times \R$, and then further recast as a 1d SQM theory in the interval space $\mathcal{M}(\mathscr{P}_0, \mathscr{P}_1)_{\mathcal{M}(\R, \mathcal{M}^{G,M_2}_{\text{H}, \theta})}$ of smooth trajectories between A$_\theta$-branes starting at $\mathscr{P}_0(\theta)$ and ending on $\mathscr{P}_1(\theta)$ in the path space $\mathcal{M}(\R, \mathcal{M}^{G,M_2}_{\text{H}, \theta})$ of maps from $\R$ to $\mathcal{M}^{G,M_2}_{\text{H},\theta}$, the symplectic $\theta$-Hitchin moduli space of $G$ on $M_2$, with action \eqref{sqm action on interval space to path space of higgs}. This will allow us to express the partition function as \eqref{interval to path space floer classes}:
\begin{equation}
  \label{summary: interval to path space floer classes}
  \boxed{
    \mathcal{Z}_{\text{HW},\R \times I \rightarrow \R}(G)
    = \sum_v  {\cal F}^{v}_{\mathcal{M}(\R, \mathcal{M}^{G,M_2}_{\text{H}, \theta})}
    = \sum_v \text{HSF}^{\text{Int}}_{d_v}(\mathcal{M}\big(\R, \mathcal{M}^{G,M_2}_{\text{H}, \theta}\big) , \mathscr{P}_0, \mathscr{P}_1)
    = \mathcal{Z}^{\text{IntSympFloer}}_{\mathcal{M}\big(\R, \mathcal{M}^{G,M_2}_{\text{H}, \theta}\big)}
  }
\end{equation}
where $\text{HSF}^{\text{Int}}_{d_v}(\mathcal{M}\big(\R, \mathcal{M}^{G,M_2}_{\text{H}, \theta}\big) , \mathscr{P}_0, \mathscr{P}_1)$ is a symplectic intersection Floer homology class generated by the intersection points of $\mathscr{P}_0(\theta)$ and $\mathscr{P}_1(\theta)$ in $\mathcal{M}\big(\R, \mathcal{M}^{G,M_2}_{\text{H}, \theta}\big)$, of degree $d_v$ counted by the Floer differential realized as the flow lines of the SQM whose gradient flow equations are the expression within the squared terms in \eqref{sqm action on interval space to path space of higgs} set to zero.

In $\S$\ref{section: hw atiya-floer correspondence}, we consider 5d-HW theory on $M_5 = M_3 \times M_1 \times \R$, and Heegaard split $M_5$ at $M_3$ with Heegaard surface $\Sigma$. When $M_1 = S^1$, via the topological invariance of HW theory and the BJSV reduction procedure, we will obtain a \emph{novel} HW Atiyah-Floer correspondence in \eqref{Atiyah-Floer HW - S1}:
\begin{equation}
  \label{summary: Atiyah-Floer HW - S1}
  \boxed{
    \text{HF}^{\text{HW}}_*(M_3 \times S^1, G) \cong \text{HSF}^{\text{Int}}_*({L\mathcal{M}^{G,\Sigma}_{\text{Higgs}}}, \mathscr{L}_0, \mathscr{L}_1)
  }
\end{equation}
where $\mathcal{M}^{G,\Sigma}_{\text{Higgs}}$ is the symplectic moduli space of Higgs bundles of $G$ on $\Sigma$!
In turn, this will also lead us to a 4d-HW Atiyah-Floer correspondence in \eqref{Atiyah-Floer HW_4}:
\begin{equation}
  \label{summary: Atiyah-Floer HW_4}
  \boxed{
    \text{HF}^{\text{HW}_4}_*(M_3, G) \cong \text{HSF}^{\text{Int}}_*({\mathcal{M}^{G,\Sigma}_{\text{Higgs}}}, {L}_0, {L}_1)
  }
\end{equation}
where $\text{HSF}^{\text{Int}}_*({\mathcal{M}^{G,\Sigma}_{\text{Higgs}}}, {L}_0, {L}_1)$ is a symplectic intersection Floer homology class generated by the intersection points of A-branes $L_0$ and $L_1$ in $\mathcal{M}^{G, \Sigma}_{\text{Higgs}}$!
It has a duality to a VW Atiyah-Floer correspondence derived in an earlier work \cite[eqn.(5.12)]{ong2022vafa} via a topologically-trivial ``rotation'' of the space orthogonal to $M_3$, as expressed in \eqref{4d-HW/VW AF Duality}:
\begin{equation}
  \label{summary: 4d-HW/VW AF Duality}
  \boxed{
    \text{HF}^{\text{HW}_4}_*(M_3, G) \cong \text{HSF}^{\text{Int}}_*({\mathcal{M}^{G,\Sigma}_{\text{Higgs}}}, {L}_0, {L}_1)
    \xleftrightarrow[]{\text{5d ``rotation''}}
    \text{HF}^{\text{inst}}_{*}(M_3, G_\C) \cong  \text{HSF}^{\text{Int}}_{*}({\mathcal{M}^{G_\C,\Sigma}_{\text{flat}}}, {L}_0', {L}_1')
  }
\end{equation}
where $\text{HF}^{\text{inst}}_{*}(M_3, G_\C)$ is a $G_\C$-instanton Floer homology class assigned to $M_3$.


In $\S$\ref{section: s-duality and web of relations}, we first elucidate the implications of the topological invariance of HW theory on the Floer homologies obtained hitherto. The results are given in \eqref{HW = HW_4 = HW_3}--\eqref{de-rham BF correspondence} and summarized as (I)
\begin{equation}
  \label{summary:HW = HW_4 = HW_3}
  \boxed{
    \sum_k  \text{HF}^{\text{HW}}_{d_k}(M_4, G)
    \xleftrightarrow[]{M_4 = M_3 \times {\hat S}^1}
    \sum_l \text{HF}_{d_l}^{\text{HW}_4}(M_3, G)
    \xleftrightarrow[]{M_3 = M_2 \times {\hat S}^1}
    \sum_p \text{HF}_{d_p}^{\text{HW}_3}(M_2, G_\C)
  }
\end{equation}
and
\begin{equation}\label{summary:HW = HSF}
  \boxed{
    \sum_k \text{HF}^{\text{HW}}_{d_k}(M_4, G)
    \xleftrightarrow[]{M_4 = {\hat M}_2 \times S^1 \times S^1}
    \sum_q \text{HSF}^{\text{hol}}_{d_q}({L^2\mathcal{M}^{G_\C}_{\text{flat}}(M_2)})
  }
\end{equation}
where $S^1$ is a circle of fixed radius; ${\hat S}^1$ is a circle of variable radius; and $\hat{M}_2$ is a Riemann surface of genus $g \geq 2$ of variable size. (II)
\begin{equation}
  \label{summary:Atiyah-Floer partition equality S1 - web}
  \boxed{
    \sum_k  \text{HF}^{\text{HW}}_{d_k}(M_4, G)
    \xleftrightarrow[]{M_4 = M_3 \times S^1}
    \sum_u \text{HSF}^{\text{Int}}_{d_u}({L\mathcal{M}^{G,\Sigma}_{\text{Higgs}}}, \mathscr{L}_0, \mathscr{L}_1)
  }
\end{equation}
and
\begin{equation}
  \label{summary:Atiyah-Floer partition equality HW_4 - web}
  \boxed{
    \sum_l  \text{HF}^{\text{HW}_4}_{d_l}(M_3, G) = \sum_w \text{HSF}^{\text{Int}}_{d_w}({\mathcal{M}^{G,\Sigma}_{\text{Higgs}}}, {L}_0, {L}_1)
  }
\end{equation}
(III)
\begin{equation}
  \label{summary:4d-HW/VW AF Duality Partition Functions}
  \boxed{
    \begin{aligned}
      \sum_l \text{HF}^{\text{HW}_4}_{d_l}(M_3, G)
      &= \sum_w \text{HSF}^{\text{Int}}_{d_w}({\mathcal{M}^{G,\Sigma}_{\text{Higgs}}}, {L}_0, {L}_1)
      \\
      & \displaystyle\left\updownarrow\vphantom{\int}\right. \text{5d ``rotation}"
      \\
      \sum_m \text{HF}^{\text{inst}}_{d_m}(M_3, G_\C)
      &= \sum_n \text{HSF}^{\text{Int}}_{d_n}({\mathcal{M}^{G_\C,\Sigma}_{\text{flat}}}, {L}_0', {L}_1')
    \end{aligned}
  }
\end{equation}
(IV)
\begin{equation}
  \label{summary:de-rham BF correspondence}
  \boxed{
    \sum_p \text{HF}_{d_p}^{\text{HW}_3}(M_2, G_\C) = \sum_p {\text H}^0_{\text{dR}}\big(\mathcal{M}^{G_\C}_{\text{BF}}(M_2)\big)
  }
\end{equation}
where ${\text H}^0_{\text{dR}}\big(\mathcal{M}^{G_\C}_{\text{BF}}(M_2)\big)$ is a de Rham class of zero-forms in the moduli space $\mathcal{M}^{G_\C}_{\text{BF}}(M_2)$ of $G_\C$-BF configurations on $M_2$, and the sum $p$ is a sum over all \emph{time-invariant} such configurations.

Second, we elucidate the implications of the 5d ``S-duality'' of HW theory, and the results are given in \eqref{S-dual 4d HW LG Floer}, \eqref{S-dual 3d HW LG Floer} and \eqref{de-rham flat GC correspondence} as
\begin{equation}
  \label{summary: Langlands-dual 4d HW LG Floer}
  \boxed{
    \sum_l \text{HF}_{{d_l}}^{\text{HW}_4}(M_3, L{G}) \xleftrightarrow[]{\text{5d ``S-duality''}}    \sum_{\tilde l} \text{HF}_{{d_{\tilde l}}}^{\text{HW}_4}(M_3, (LG)^\vee)
  }
\end{equation}
\begin{equation}
  \label{summary: Langlands-dual 3d HW LG Floer}
  \boxed{
    \sum_q \text{HF}_{{d_q}}^{\text{HW}_3}(M_2, LL{G}_\C)_{\scriptscriptstyle {Z, \bar{Z} = 0}} \xleftrightarrow[]{\text{5d ``S-duality''}}    \sum_{\tilde q} \text{HF}_{{d_{\tilde q}}}^{\text{HW}_3}(M_2, L(LG_\C)^\vee)_{\scriptscriptstyle {Z, \bar{Z} = 0}}
  }
\end{equation}
where $G$/$G_\C$ are nonsimply-laced gauge groups with loop groups $(LG)^\vee$/ $(LG_\C)^\vee$ being Langlands dual to $LG$/$LG_\C$ (at the level of their loop algebras); and
\begin{equation}
  \label{summary: de-rham flat GC correspondence}
  \boxed{
    \sum_q \text{HF}_{d_q}^{\text{HW}_3}(M_2, LLG_\C)_{\scriptscriptstyle Z, \bar{Z} = 0}
    = \sum_q {\text{H}}^0_{\text{dR}}\big(\mathcal{M}^{LLG_\C}_{\text{flat}}(M_2)\big)
  }
\end{equation}
where ${\text{H}}^0_{\text{dR}}\big(\mathcal{M}^{LLG_\C}_{\text{flat}}(M_2)\big)$ is a de Rham class of zero-forms in the moduli space $\mathcal{M}^{LLG_\C}_{\text{flat}}(M_2)$ of flat toroidal $G_\C$ connections on $M_2$, and the sum $q$ is a sum over all \emph{time-invariant} such connections.

Lastly, we summarize the web of relations amongst all the Floer homologies, and this is shown in Fig.~\ref{summary:fig:web of relations combined}, where the sizes of $\hat S^1$, $\hat M_2$, and $\Sigma$ can be varied. In particular, from the web, we have, amongst other additional relations,
\begin{equation}
  \label{summary: Langlands-dual Atiyah-Floer LG}
  \boxed{
    \sum_{\tilde l} \text{HF}_{{d_{\tilde l}}}^{\text{HW}_4}(M_3, (LG)^\vee)
    \xleftrightarrow[\text{$M_3 = M_3' \bigcup_{\Sigma} M_3''$}]{ \text{$G$ nonsimply-laced}}
    \sum_u \text{HSF}^{\text{Int}}_{d_u}({L\mathcal{M}^{G,\Sigma}_{\text{Higgs}}}, \mathscr{L}_0, \mathscr{L}_1)
  }
\end{equation}
which is a Langlands duality in a 4d-HW Atiyah-Floer correspondence for loop gauge groups!

\usetikzlibrary{arrows,automata,positioning,calc}
\begin{figure}
    \centering
    \begin{adjustbox}{max totalsize={\textwidth}{\textheight},center}
        \begin{tikzpicture}[%
            auto,%
            block/.style={draw, rectangle},%
            every edge/.style={draw, <->, dashed},%
            relation/.style={scale=0.8, sloped, anchor=center, align=center},%
            vertRelation/.style={scale=0.8, anchor=center, align=center},%
            shorten >=4pt,%
            shorten <=4pt,%
            ]
            \def \verRel {2.2} 
            \def \horRel {1.8} 
            \node[block, ultra thick] (HW5) {$ \sum_k \text{HF}^{\text{HW}}_{d_k}(M_4, G)$};
            \node[above={\verRel} of HW5] (aHW5) {};
            \node[below={\verRel} of HW5] (bHW5) {};
            \node[block, below={\verRel} of HW5] (HW4) {$\sum_l \text{HF}_{d_l}^{\text{HW}_4}(M_3, G)$};
            -
            \node[block, left={\horRel} of HW4] (HSF-LM) {$\sum_u \text{HSF}^{\text{Int}}_{d_u}({L\mathcal{M}^{G,\Sigma}_{\text{Higgs}}}, \mathscr{L}_0, \mathscr{L}_1)$};
            \node[block, below={\verRel} of HW4] (HW3) {$\sum_p \text{HF}_{d_p}^{\text{HW}_3}(M_2, G_\C)$};
            -
            \node[block, right={\horRel} of HW3] (HSF-L2M) {$\sum_q \text{HSF}^{\text{hol}}_{d_q}({L^2\mathcal{M}^{G_\C}_{\text{flat}}(M_2)})$};
            -
            \node[block, left={\horRel} of HW3] (HSF-M) {$\sum_w \text{HSF}^{\text{Int}}_{d_w}({\mathcal{M}^{G,\Sigma}_{\text{Higgs}}}, {L}_0, {L}_1)$};
            \node[block, below={\verRel} of HW3] (VW4) {$\sum_m \text{HF}^{\text{inst}}_{d_m}(M_3, G_\C)$};
            -
            \node[block, below left={sqrt(\verRel*\verRel + \horRel*\horRel)} of VW4] (VW-HSF) {$\sum_n \text{HSF}^{\text{Int}}_{d_n}({\mathcal{M}^{G_\C,\Sigma}_{\text{flat}}}, {L}_0', {L}_1')$};
            \node[block, below={\verRel} of HSF-L2M] (HW1) {$\sum_p {\text H}^0_{\text{dR}}\big(\mathcal{M}^{G_\C}_{\text{BF}}(M_2)\big)$};
            \node[block, left={\horRel} of aHW5] (HW4-LG) {$\sum_l \text{HF}_{{d_l}}^{\text{HW}_4}(M_3, L{G})$};
            -
            \node[block, above={\verRel} of HW4-LG] (HW4-LGv) {$\sum_{\tilde l} \text{HF}_{{d_{\tilde l}}}^{\text{HW}_4}(M_3, (LG)^\vee)$};
            \node[block, right={\horRel} of aHW5] (HW3-L2G) {$\sum_q \text{HF}_{{d_q}}^{\text{HW}_3}(M_2, LL{G}_\C)_{\scriptscriptstyle {Z, \bar{Z} = 0}}$};
            -
            \node[block, above={\verRel} of HW3-L2G)] (HW3-L2Gv) {$\sum_{\tilde q} \text{HF}_{{d_{\tilde q}}}^{\text{HW}_3}(M_2, L(LG_\C)^\vee)_{\scriptscriptstyle {Z, \bar{Z} = 0}}$};
            -
            \node[block, right={1.5*\horRel} of HW5] (HW1-L2G) {$\sum_q {\text H}^0_{\text{dR}}\big(\mathcal{M}^{LLG_\C}_{\text{flat}}(M_2)\big)$};
            -
            \node[block, above={3.5*\verRel} of HW1-L2G] (HW1-L2Gv) {$\sum_{\qt} {\text H}^0_{\text{dR}}\big(\mathcal{M}^{L{(LG_\C)^\vee}}_{\text{flat}}(M_2)\big)$};
            \draw
            (HW5) edge node[vertRelation, right] {$M_4 = M_3 \times \hat{S}^1$} (HW4)
            (HW4) edge node[vertRelation, right] {$M_3 = M_2 \times \hat{S}^1$} (HW3)
            (HW5.south east) edge node[relation, above] {$M_4 = \hat{M}_2 \times S^1 \times S^1$} (HSF-L2M)
            (HW5.south west) edge[solid] node[relation, above] {$M_3 = M_3' \bigcup_\Sigma M_3''$} node[relation, below] {$M_4 = M_3 \times S^1$} (HSF-LM)
            (HW4.south west) edge[solid] node [relation, name=HW4-HSF, above] {$M_3 = M_3' \bigcup_\Sigma M_3''$} (HSF-M)
            (VW4.south west) edge[solid] node [relation, name=VW4-VW-HSF, below] {$M_3 = M_3' \bigcup_\Sigma M_3''$} (VW-HSF)
            ($(HW4-HSF) + (0.5em, -0.5em)$) edge[solid] node[vertRelation, right] {5d ``rotation''} (VW4-VW-HSF)
            (HW3.south east) edge node[relation, above] {$M_2 = \hat{M}_2$} (HW1)
            (HW5.north west) edge[solid] node[relation, below] {$M_4 = M_3 \times {S}^1$} (HW4-LG)
            (HW4-LG) edge[solid] node[vertRelation, right] {5d ``S-duality''\\$G$ nonsimply-laced} (HW4-LGv)
            (HW5.north east) edge[solid] node[relation, below] {$M_4 = M_2 \times S^1 \times S^1$} (HW3-L2G)
            (HW3-L2G) edge[solid] node[vertRelation, right] {5d ``S-duality''\\$G_\C$ nonsimply-laced} (HW3-L2Gv)
            (HW4-LG) edge[solid] node[vertRelation, left] {$M_3 = M_3' \bigcup_{\Sigma} M_3''$} (HSF-LM)
            (HW1-L2G) edge node[relation, above] {$M_2 = \hat{M_2}$} (HW3-L2G)
            (HW1-L2G) edge[solid] (HSF-L2M)
            (HW4-LGv.south west) edge[solid, bend right] node[relation, above] {$G$ nonsimply-laced} node[relation, below] {$M_3 = M_3' \bigcup_{\Sigma} M_3''$} (HSF-LM.north west)
            (HW3-L2Gv) edge node[relation, above] {$M_2 = \hat{M}_2$} (HW1-L2Gv)
            (HW1-L2Gv) edge[solid, bend left] node[relation, above] {$G_\C$ nonsimply-laced} (HW1-L2G.north east)
            ;
        \end{tikzpicture}
    \end{adjustbox}
    \caption{A web of relations amongst the Floer homologies.}
    \label{summary:fig:web of relations combined}
\end{figure}
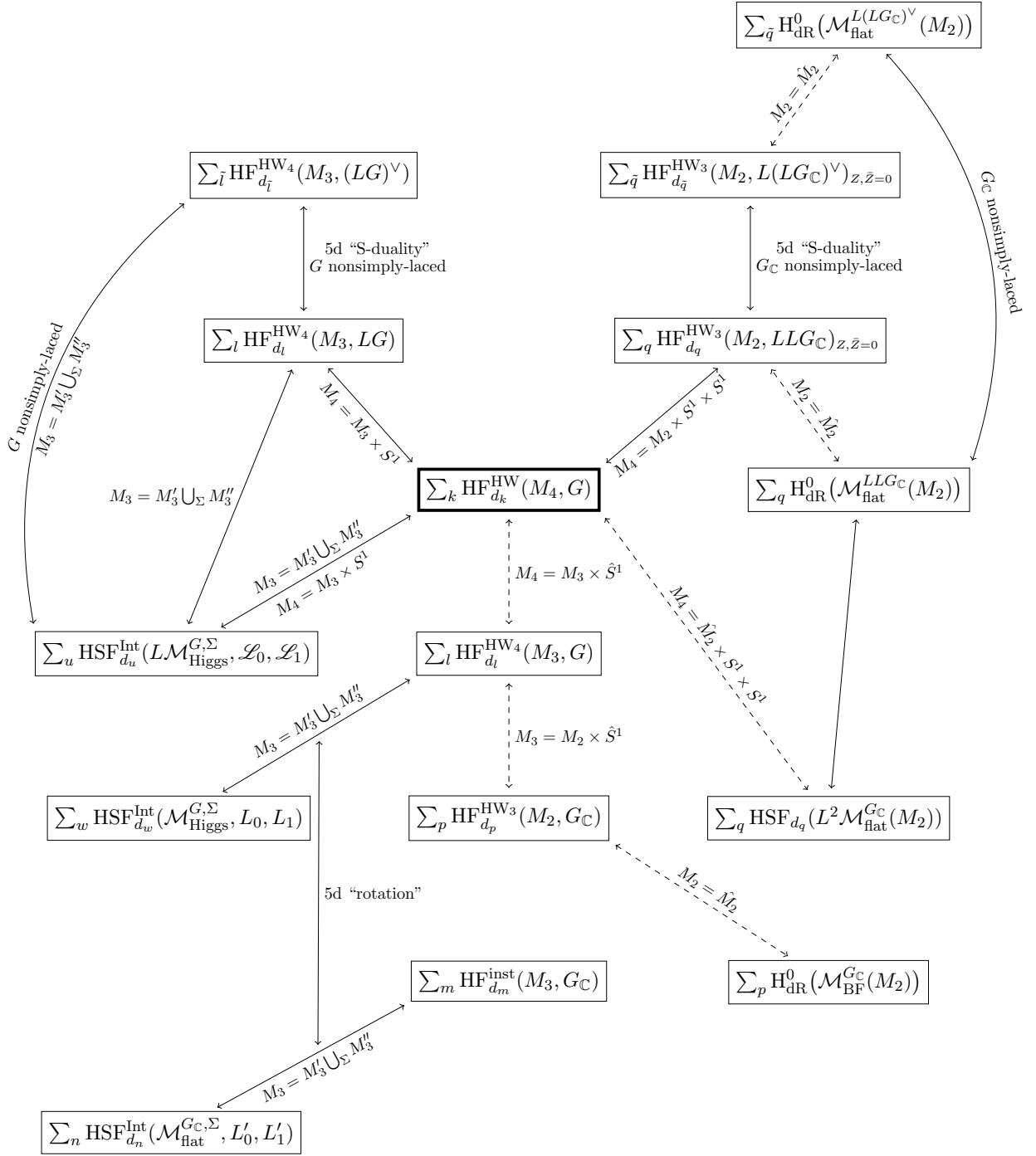

In $\S$\ref{section: Fukaya-Seidel}, we consider the case where $M_5 = M_3 \times \R^2$, and recast 5d-HW theory as a 2d gauged Landau-Ginzburg (LG) model on $\R^2$ with target $\mathscr A_3$, the space of connections of a complexified gauge group $G_\C$ on $M_3$. In turn, this 2d gauged LG model can be recast as a 1d SQM theory in the path space $\mathcal{M}(\R, \mathscr A_3)$. From the SQM and its critical points that can be interpreted as solitons in the 2d gauged LG model, we obtain~\eqref{Haydys Hom = CF}:
\begin{equation}
  \label{summary: Haydys Hom = CF}
  \boxed{
    \text{Hom}(\mathcal{A}_I, \mathcal{A}_J)_\pm  \Longleftrightarrow \text{HF}^{G}_{d_p}(p^{IJ}_\pm)
  }
\end{equation}
where $\text{HF}^G_{d_p}(p_\pm^{IJ})$ is a Floer homology class of degree $d_p$ generated by $p_\pm^{IJ}$, the intersection points of left- and right-thimbles, which represent the LG solitons that can be described as the morphisms $\text{Hom}(\mathcal{A}_I, \mathcal{A}_J)_\pm$ whose endpoints $\mathcal{A}_I, \mathcal{A}_J$ correspond to $G$ Hitchin configurations on $M_3$. Furthermore, via this equivalent description of 5d-HW theory as a 2d gauged LG model with target $\mathscr{A}_3$, we can interpret the normalized 5d partition function as a sum over tree-level scattering amplitudes of LG soliton strings given by the maps of morphisms in \eqref{Haydys mu maps}:
\begin{equation}
  \label{summary: Haydys mu maps}
  \boxed{
    \mu^d: \bigotimes_{i = 1}^d
    \text{Hom}(\mathcal{A}_{I_i}, \mathcal{A}_{I_{i + 1}})_-
    \longto
    \text{Hom}(\mathcal{A}_{I_1}, \mathcal{A}_{I_{d + 1}})_+
  }
\end{equation}
where $\text{Hom}(\mathcal{A}_*, \mathcal{A}_*)_-$ and $\text{Hom}(\mathcal{A}_*, \mathcal{A}_*)_+$ represent incoming and outgoing scattering soliton strings, as shown in Fig. \ref{fig:mud maps}.

Note that \eqref{summary: Haydys Hom = CF} and \eqref{summary: Haydys mu maps} underlie a Fukaya-Seidel type $A_\infty$-category of $G$ Hitchin configurations on $M_3$ that was conjectured to exist by Haydys in \cite[$\S$5]{haydys2010fukaya}, and later rigorously constructed by Wang in~\cite{wang2022complgradien}. As such, we have a purely physical proof of Haydys' mathematical conjecture and realization of Wang's mathematical construction. Also, Abouzaid-Manolescu's conjecture in~\cite[$\S$9.2]{abouzaid2017sheaftheor} implies that $G_\C$-instanton Floer homology of $M_3$, which is generated by $G$ Hitchin configurations on $M_3$ (at $\theta = \pi/2$), can be categorified to give an $A_\infty$-category of $M_3$. Thus, we also have a purely physical proof and generalization of their mathematical conjecture.

Applying the results of $\S$\ref{section: hw atiya-floer correspondence} with $M_1 = \R$, we have the one-to-one correspondence
\begin{equation}
  \label{summary:Haydys Hom = CF Atiyah-Floer}
  \boxed{
    \text{Hom}(\mathcal{A}_I(\theta), \mathcal{A}_J(\theta))_\pm
    \Longleftrightarrow
    \text{HSF}^{\text{Int}}_{d_v}(\mathcal{M}\big(\R, \mathcal{M}^{G,\Sigma}_{\text{H}, \theta}\big) , \mathscr{P}_0, \mathscr{P}_1)
  }
\end{equation}
in \eqref{Haydys Hom = CF Atiyah-Floer}, which is an Atiyah-Floer type correspondence for the Fukaya-Seidel $A_\infty$-category of $G$ Hitchin configurations on $M_3$!

Via the Atiyah-Floer correspondences of $\S$\ref{section: hw atiya-floer correspondence} for general $\theta$, we would arrive at the one-to-one correspondence
\begin{equation}
  \label{summary: FS-cat hom as Hom-cat}
  \boxed{
    \text{Hom}(\mathcal{A}_I(\theta), \mathcal{A}_J(\theta))_\pm
    \Longleftrightarrow
    \text{Hom}\left(
      \text{Hom}(L_{I, 0}(\theta), L_{I, 1}(\theta)),
      \text{Hom}(L_{J, 0}(\theta), L_{J, 1}(\theta))
    \right)_\pm
  }
\end{equation}
in~\eqref{FS-cat hom as Hom-cat}, which implies a correspondence between a Hom-category of morphisms between Lagrangian branes of $\mathcal{M}^{G, \Sigma}_{\text{H}, \theta}$, and a Fukaya-Seidel type $A_\infty$-category of $G$ Hitchin configurations on $M_3$. In particular, at $\theta = \pi/2$, we have a correspondence between (i) a Hom-category of Lagrangian submanifolds of $\mathcal{M}^{G_\C, \Sigma}_{\text{flat}}$, (ii) a Fukaya-Seidel type $A_\infty$-category of flat $G_\C$ connections on $M_3$, and (iii) a $G_\C$-instanton Floer homology of $M_3$. This coincides exactly with Bousseau's conjecture in~\cite[$\S$2.8]{bousseau-2024-holom-floer}. Therefore, we also have a purely physical proof and generalization of his mathematical conjecture!

Furthermore, via~\eqref{summary:Haydys Hom = CF Atiyah-Floer} and~\eqref{summary: FS-cat hom as Hom-cat}, we would arrive at the one-to-one correspondence
\begin{equation}
  \label{summary: CF Atiyah-Floer = Hom-cat}
  \boxed{
    \text{HSF}^{\text{Int}}_{d_v}(\mathcal{M}\big(\R, \mathcal{M}^{G,\Sigma}_{\text{H}, \theta}\big) , \mathscr{P}_0, \mathscr{P}_1)
    \Longleftrightarrow
    \text{Hom}\left(
      \text{Hom}(L_{I, 0}(\theta), L_{I, 1}(\theta)),
      \text{Hom}(L_{J, 0}(\theta), L_{J, 1}(\theta))
    \right)_\pm
  }
\end{equation}
in~\eqref{CF Atiyah-Floer = Hom-cat}, between an intersection Floer homology and a Hom-category of morphisms!

\bigskip\noindent\textit{Acknowledgements}
\vspace*{0.5em}

We would like to thank M.~Ashwinkumar, A.~Haydys, and R.P.~Thomas for useful discussions. We would also like to thank the ATMP referee for suggesting improvements to our paper. This work is supported in part by the MOE AcRF Tier 1 grant R-144-000-470-114.

\mbox{}\par\nobreak
\noindent

\section{A Topological 5d \texorpdfstring{$\mathcal{N}=2$}{N=2} Gauge Theory on \texorpdfstring{$M_4 \times \mathbb{R}$}{M4 x R}}
\label{section: hw theory general}

In this section, we will study a certain topologically-twisted 5d ${\mathcal {N}} = 2$ theory on $M_4 \times \R$ with gauge group a real, simple, and compact Lie group $G$, where the BPS equations that its path integral localizes onto are the Haydys-Witten equations~\cite{haydys2010fukaya,witten2011fivebranes}, and $M_4$ a closed and compact four-manifold.
In turn, this will allow us to get our desired results in later sections.

\subsection{Twisting along \texorpdfstring{$M_4$}{M4}}
\label{subsection twist on M4}

The five-manifold of our 5d theory is $M_4 \times \R$, where we have a holonomy group $SO(4)_E$ for $M_4$. Given that
\begin{equation}
    SO(4)_E \cong SU(2)_l\otimes SU(2)_r \, ,
\end{equation}
and the fact that $Spin(5)_R$, the $R$-symmetry group of this theory, contains $Spin(4)_R\cong SU(2)_a\otimes SU(2)_b$ as a maximal subgroup, the relevant topological twisting amounts to replacing $SU(2)_r$ with $SU(2)_{r'}$, the diagonal subgroup of $SU(2)_r \otimes SU(2)_{a}$. As a $U(1)_R$ $R$-charge can be assigned to the fields under the Cartan subgroup of the $SU(2)_b$ $R$-symmetry, upon twisting, the symmetry group of the theory will become
\begin{equation}
    SU(2)_l \otimes SU(2)_r \otimes SU(2)_a\otimes U(1)_R \longrightarrow   SU(2)_l \otimes SU(2)_r' \otimes U(1)_R \, .
\end{equation}
Noting that $\textbf{1}\otimes \textbf{2}=\textbf{2}$ and $\textbf{2}\otimes \textbf{2}=\textbf{3}\oplus \textbf{1}$, one can compute  that the field content of the 5d $\mathcal{N}=2$ SYM theory will be modified as \cite{anderson2013five}
\begin{equation}\label{field content}
    \begin{aligned}
        A_{\mu}\: (\textbf{2,2,1})^0 &\longrightarrow A_{\mu}\: (\textbf{2,2})^0 \: ,\\
            A_t\: (\textbf{1,1,1})^0 &\longrightarrow A_t\: (\textbf{1,1})^0 \: ,\\
          \phi_{i}\: (\textbf{1,1,1})^2\oplus(\textbf{1,1,1})^{-2}\oplus(\textbf{1,1,3})^0  &\longrightarrow  \sigma\:(\textbf{1,1})^2 \oplus \bar{\sigma}\:(\textbf{1,1})^{-2} \oplus B_{\mu\nu}\:(\textbf{1,3})^0 \: ,\\
        \lambda_{\alpha}\:(\textbf{2,1,2})^1 \oplus (\textbf{2,1,2})^{-1}  &\longrightarrow \psi_{\mu}\:(\textbf{2,2})^1 \oplus \tilde{\psi}_{\mu}\:(\textbf{2,2})^{-1} \: , \\
        \bar{\lambda}_{\dot{\alpha}}\:(\textbf{1,2,2})^{-1}\oplus (\textbf{1,2,2})^{1} &\longrightarrow \chi_{\mu\nu}\:(\textbf{1,3})^{-1}\:\oplus \eta\:(\textbf{1,1})^{-1} \oplus \tilde{\chi}_{\mu\nu}\:(\textbf{1,3})^{1}\:\oplus \tilde{\eta}\:(\textbf{1,1})^{1} \: .
    \end{aligned}
\end{equation}
In \eqref{field content}, $\mu,\nu = 0,1,2,3$ are indices along $M_4$, and $A_t$ is the component of the 5d gauge field along $\mathbb{R}$. The indices $\alpha, \dot{\alpha}$ are, respectively, the $SU(2)_l$ and $SU(2)_r$ indices along $M_4$. The index $i=1,2,\dots5$ label the five scalar fields in 5d $\mathcal{N}=2$ SYM theory. The dimensions of the fields under the various $SU(2)$'s are denoted by boldfaced numbers while the $U(1)_R$ $R$-charges are denoted by the numbers in the superscript.

The resulting field content of this twist leaves, for the bosonic fields, the gauge field $A_{\mu} \in \Omega^1 (M_4, ad(G)) \otimes \Omega^0 (\R, ad(G))$ and $A_t\in \Omega^0(M_4, ad(G)) \otimes \Omega^1 (\R, ad(G))$ unchanged, while the five original scalar fields $\phi_i$ in the untwisted theory are now two scalars $\sigma, \bar{\sigma}\in \Omega^0 (M_4, ad(G)) \otimes \Omega^0 (\R, ad(G))$, and the remaining three now form a self-dual two-form $B_{\mu\nu}\in \Omega^{2,+}(M_4, ad(G)) \otimes \Omega^0 (\R, ad(G))$. The fermionic fields are now one-forms $\psi_{\mu}, \tilde{\psi}_{\mu}\in \Omega^1(M_4, ad(G)) \otimes \Omega^0 (\R, ad(G))$, self-dual two-forms $\chi_{\mu\nu}, \tilde{\chi}_{\mu\nu}\in\Omega^{2, +}(M_4, ad(G)) \otimes \Omega^0 (\R, ad(G))$, and scalars $\eta \in \Omega^0 (M_4, ad(G)) \otimes \Omega^1 (\R, ad(G))$ and $\tilde{\eta}\in \Omega^0 (M_4, ad(G)) \otimes \Omega^0 (\R, ad(G))$. Here, $ad(G)$ is the adjoint bundle of the underlying principal $G$-bundle.

Notice that only the spins along $M_4$ of the fields are modified, since the twisting was performed only along $M_4$. Nonetheless, in doing so, we still obtain a topological theory along the whole of $M_5$. This is to be expected, since $\mathbb{R}$ is a flat direction. Furthermore, the supersymmetry generators transform in the same representation as the fermion fields, and since we now have two scalar fermion fields, we will also have two nilpotent scalar supersymmetry generators, $\mathcal{Q}\: (\textbf{1,1})^1$ and $\bar{\mathcal{Q}}\: (\textbf{1,1})^{-1}$.

The supersymmetry transformations of the twisted theory under $\mathcal{Q}, \bar{\mathcal{Q}}$ are presented in \cite{anderson2013five}. For our purposes in this paper, we will consider only one of the two supercharges, i.e., $\mathcal Q$.
The transformations of the fields under $\mathcal{Q}$ are 
\begin{equation}
  \label{5d susy variations}
  \begin{aligned}
    \delta A_{\mu}
    & =  i \psi_{\mu}
      \, ,  \\
    \delta A_t
    & =  \eta
      \, ,\\
    \delta B_{\mu\nu}
    &=  2 {\chi}_{\mu\nu}
      \, ,\\
    \delta \sigma
    &= 0
      \, ,\\
    \delta \bar{\sigma}
    &= -i \tilde{\eta}
      \, ,\\
    \delta \psi_{\mu}
    &= - 2 D_{\mu}\sigma
      \, ,\\
    \delta \tilde{\psi}_{\mu}
    &= i (F_{t\mu} + D^{\nu}B_{\nu\mu})
      \, ,\\
    \delta \chi_{\mu\nu}
    &= - i [B_{\mu\nu}, \sigma]
      \, , \\
    \delta \tilde{\chi}_{\mu\nu}
    &= - \left(
      F^+_{\mu\nu}
      - \frac{1}{4}[B_{\mu\rho}, B_{\nu}^{\rho}]
      - \frac{1}{2} D_t B_{\mu\nu}
      \right)
      \, , \\
    \delta \eta
    &= -2i D_t \sigma
      \, , \\
    \delta \tilde{\eta}
    &= - 2 [\sigma, \bar{\sigma}]
      \, .
  \end{aligned}
\end{equation}
Here, $F^+_{\mu\nu}=\frac{1}{2}(F_{\mu\nu}+\frac{1}{2}\epsilon_{\mu\nu\rho\lambda}F^{\rho\lambda})$ is the self-dual part of the gauge field strength $F_{\mu\nu}$, $B_{\mu\nu}=\frac{1}{2}\epsilon_{\mu\nu\rho\lambda}B^{\rho\lambda}$, and $[B_{\mu\rho}, B^{\rho}_{\nu}] \equiv [B_{\mu\rho}, B_{\nu\sigma}] g^{\rho\sigma}$.
The supersymmetry variation $\mathcal{Q}$ is nilpotent up to gauge transformations, where
\begin{equation}
\label{eq:delta squared}
  \delta^2 A_{\mu} \propto D_{\mu}\sigma \, ,
  \qquad
  \delta^2 A_t \propto D_t \sigma\,,
  \qquad
  \delta^2 \Phi \propto [\phi, \sigma]\,.
\end{equation}
Here $\Phi$ represents all other fields that are not $A_{\mu}$ or $A_t$.
Since $\sigma$ is a generator of gauge transformations, we shall only consider the case where it has no zero-modes such that the gauge connections are irreducible whence the relevant moduli space is well-behaved.

\subsection{The Topological Action and the Haydys-Witten Equations}

The bosonic part of the action (involving only $A$ and $B$) can be written as
\begin{equation}\label{boson action in s and k}
    S_{\text{bo}} = \frac{1}{2e^2} \int_{M_4 \times \mathbb{R}} d^5 x \, \text{Tr} \big( |k_{\mu} |^2+| s_{\mu\nu}|^2  \big) \, ,
\end{equation}
where
\begin{equation}\label{s and k}
    \begin{aligned}
       k_{\mu}  &= F_{t\mu} +  D^{\nu}B_{\nu\mu} \,,\\
        s_{\mu\nu} &= F^{+}_{\mu\nu} - \frac{1}{4}[B_{\mu\rho}, B^{\rho}_{\nu}] - \frac{1}{2}D_{t}B_{\mu\nu} \,.
    \end{aligned}
\end{equation}
The $\mathcal Q$-exact topological action is \cite{anderson2013five}
\begin{equation}
  \label{5d action}
  \begin{aligned}
    S_{\text{HW}}
    = \frac{1}{e^2}
    & \int_{M_4 \times \R} dt d^4x \, \Tr \bigg(
      \frac{1}{2} \left| F_{t\mu} + D^{\nu} B_{\nu\mu}) \right|^2
      + \frac{1}{2} \left|
        F^+_{\mu\nu}
        - \frac{1}{4} [B_{\mu\rho}, B^{\rho}_{\nu}]
        - \frac{1}{2} D_t B_{\mu\nu}
      \right|^2
    \\
    & + 2 D_{\mu} \sigma D^{\mu} \bar{\sigma}
      + 2 D_t \sigma D^t \bar{\sigma}
      - 2 [\sigma, \bar{\sigma}]^2
      + \frac{1}{2} [B_{\mu\nu}, \sigma] [B^{\mu\nu}, \bar{\sigma}]
    \\
    & + \frac{1}{2} B^{\mu\nu} \{\tilde{\eta}, \chi_{\mu\nu}\}
      - \frac{1}{2} B^{\mu\nu} \{\eta, \tilde{\chi}_{\mu\nu}\}
      - B^{\mu\nu} \{\psi_{\mu}, \tilde{\psi}_{\nu}\}
      - B^{\mu\nu} \{\tilde{\chi}_{\mu\rho}, \chi^{\rho}_{\nu}\}
    \\
    & - i \tilde{\eta} D_{\mu} \psi^{\mu}
      - i \eta D_{\mu} \tilde{\psi}^{\mu}
      - 2 i \tilde{\psi}_{\mu} D_{\nu} \chi^{\mu\nu}
      - 2 i \psi_{\mu} D_{\nu} \tilde{\chi}^{\mu\nu}
      - \tilde{\eta} D_t \eta
      - \tilde{\psi}_{\mu} D_t \psi^{\mu}
      - \tilde{\chi}_{\mu\nu} D_t \chi^{\mu\nu}
    \\
    & - i \sigma \{\tilde{\eta}, \tilde{\eta}\}
      - i \bar{\sigma} \{\eta, \eta\}
      + i \sigma \{\tilde{\psi}_{\mu}, \tilde{\psi}^{\mu}\}
      + i \bar{\sigma} \{\psi_{\mu}, \psi^{\mu}\}
      - i \sigma \{\tilde{\chi}_{\mu\nu}, \tilde{\chi}^{\mu\nu}\}
      - i \bar{\sigma} \{\chi_{\mu\nu}, \chi^{\mu\nu}\}
      \bigg)
      \, .
    \end{aligned}
\end{equation}
By setting the variation of the fermions in \eqref{5d susy variations} to zero, we obtain the BPS equations of the 5d theory as\footnote{Recall from our explanation below \eqref{eq:delta squared} that we are only considering the case where the bosonic field $\sigma$ (and $\bar{\sigma}$) has no zero-modes, so we can take it to be zero in the variations of the fermions.}
\begin{equation}\label{5d bps hw}
    \begin{aligned}
        F_{t\mu} +  D^{\nu}B_{\nu\mu} &= 0 \, ,\\
        F^{+}_{\mu\nu} - \frac{1}{4}[B_{\mu\rho}, B^{\rho}_{\nu}] - \frac{1}{2}D_{t}B_{\mu\nu} &= 0 \, .
    \end{aligned}
\end{equation}
These are the Haydys-Witten (HW) equations \cite{haydys2010fukaya,witten2011fivebranes}, and configurations of $(A_t, A_{\mu}, B_{\mu\nu})$ satisfying \eqref{5d bps hw} constitute a moduli space $\mathcal{M}_{\text{HW}}$ that the path integral of the 5d theory would localize onto, where $S_{\text{bo}}$ and thus  $S_{\text{HW}}$ is indeed minimized.

We shall henceforth denote the 5d theory with action $S_\text{HW}$ as HW theory.

\section{A Haydys-Witten Floer Homology of Four-Manifolds}
\label{section: hw floer homology}

In this section, we shall define, purely physically, a novel HW Floer homology of $M_4$ via the $\mathcal Q$-cohomology of HW theory, through a supersymmetric quantum mechanics (SQM) interpretation of the 5d gauge theory.

\subsection{5d Gauge Theory as SQM}
\label{subsection: 5d theory/5d to sqm}

We would like to re-express the $\mathcal{N}=2$ gauge theory on $M_5=M_4 \times \mathbb{R}$ as an SQM model in $\mathfrak{A}_4$, the space of irreducible $(A_{\mu},B_{\mu\nu})$ fields on $M_4$.\footnote{%
  Since we will ultimately consider only gauge-inequivalent configurations, $\mathfrak{A}_4$ is more precisely the space of irreducible $(A_{\mu},B_{\mu\nu})$ fields on $M_4$ modulo gauge equivalence.
  Similar such spaces to appear in later sections should also be understood as spaces of fields modulo gauge equivalence.
  \label{ft:modulo gauge inequivalence}
}
To this end, we shall employ the methods pioneered in \cite{blau1993topological}.

We begin by subtracting from the action \eqref{5d action}, the $\mathcal{Q}$-exact terms $\delta (\eta D_t \bar{\sigma})$, $\delta(B_{\mu\nu}\tilde{\eta}B^{\mu\nu})$ and $\delta (\tilde{\eta}[\sigma, \bar{\sigma}])$.
As we are ultimately interested in the $\mathcal Q$-cohomology spectrum of HW theory, this subtraction will have no relevant consequence.  Using the self-duality properties of $B_{\mu\nu}$, terms such as $B_{\mu\nu}\{\eta, \tilde{\chi}^{\mu\nu}\}$ are topological, and can also be ignored in \eqref{5d action}, thereby leaving us with
\begin{equation}
  \label{5d action terms removed for sqm}
  \begin{aligned}
    S_{\text{HW}}
    = \frac{1}{e^2}
    \int_{M_4 \times \R} dt d^4x \,
    & \Tr \bigg(
      \frac{1}{2} \left| F_{t\mu} + D^{\nu} B_{\nu\mu} \right|^2
     + \frac{1}{2} \left|
        F^+_{\mu\nu}
        - \frac{1}{4} [B_{\mu\rho}, B^{\rho}_{\nu}]
        - \frac{1}{2} D_t B_{\mu\nu}
      \right|^2
    \\
    & - i \tilde{\eta} D_{\mu} \psi^{\mu}
      - i \eta D_{\mu} \tilde{\psi}^{\mu}
      - 2 i \tilde{\psi}_{\mu} D_{\nu} \chi^{\mu\nu}
      - 2 i \psi_{\mu} D_{\nu} \tilde{\chi}^{\mu\nu}
      - \tilde{\psi}_{\mu} D_t \psi^{\mu}
      - \tilde{\chi}_{\mu\nu} D_t \chi^{\mu\nu}
    \\
    & + 2 D_{\mu} \sigma D^{\mu} \bar{\sigma}
      + i \sigma \{\tilde{\psi}_{\mu}, \tilde{\psi}^{\mu}\}
      - i \sigma \{\tilde{\chi}_{\mu\nu}, \tilde{\chi}^{\mu\nu}\}
      + i \bar{\sigma} \{\psi_{\mu}, \psi^{\mu}\}
      - i \bar{\sigma} \{\chi_{\mu\nu}, \chi^{\mu\nu}\}
      \bigg)
      \, .
  \end{aligned}
\end{equation}

The bosonic terms of the equivalent SQM action can be obtained by expanding out the first line in \eqref{5d action terms removed for sqm} and collecting only those terms that do not contain $A_t$ (as it can and will be integrated out). This gives
\begin{equation}\label{hw sqm bosonic terms}
  S_{\text{HW}} =
  \frac{1}{2e^2} \int_{\mathbb{R}} dt \, \text{Tr} \, \int_{M_4} d^4 x
  \left(
    \Big| \dot{A}_{\mu} + D^{\nu}B_{\nu\mu} \Big|^2
    + \frac{1}{4} \left|
      \dot{B}_{\mu\nu}
      - \left(
        2F^{+}_{\mu\nu} - \frac{1}{2}[B_{\mu\rho}, B^{\rho}_{\nu}]
      \right)
    \right|^2
    +\dots
  \right)
  \, ,
\end{equation}
where ``$\dots$'' contains the fermionic terms in the action,
$\dot{A}_{\mu}$ represents the time-derivative of $A_{\mu}$, and similarly for $\dot{B}_{\mu\nu}$.

The fermionic terms of the equivalent SQM action can be obtained by first integrating  $\eta, \tilde{\eta}$ out from \eqref{5d action terms removed for sqm} (since it is linear in $\eta, \tilde{\eta}$), imposing the condition
\begin{equation}\label{Dpsi=Dpsi=0}
    D_{\mu}\psi^{\mu}=D_{\mu}\tilde{\psi}^{\mu}=0 \, .
\end{equation}
Furthermore, via integration by parts, \eqref{Dpsi=Dpsi=0} also removes from \eqref{5d action terms removed for sqm}, the terms $i\tilde{\psi}_{\mu}D_{\nu}\chi^{\mu\nu}$ and $i \psi_{\mu}D_{\nu}\tilde{\chi}^{\mu\nu}$.

Using the equations of motion of $\sigma, \bar{\sigma}$ and plugging them back into \eqref{5d action terms removed for sqm} yields a term
\begin{equation}\label{5d scalar fields}
    \begin{aligned}
        \text{Tr}\int d^4 x \, \Big(\{\tilde{\psi}^{\mu}(x), \tilde{\psi}_{\mu}(x)\}-\{\tilde{\chi}_{\mu\nu}(x), \tilde{\chi}^{\mu\nu}(x)\}\Big)G(x-y)\Big(\{\psi^{\mu}(y), \psi_{\mu}(y)\}-\{\chi_{\mu\nu}(y), \chi^{\mu\nu}(y)\}\Big)\,,\\
    \end{aligned}
\end{equation}
where $G$ is the Green's function for the scalar Laplacian (on $M_4$).

Using the equations of motion for $A_t$ gives a term similar to \eqref{5d scalar fields}, as well as the term
\begin{equation}\label{A_t eom}
    \text{Tr}\int d^4 x  \,  \Big(\{ \tilde{\psi}_{\mu}(x),\psi^{\mu}(x) \}- \{\tilde{\chi}_{\mu\nu}(x), \chi^{\mu\nu}(x) \}\Big) G(x-y)\Big(D_{\mu}\dot{A}^{\mu}(y) + [\dot{B}^{\mu\nu}(y),B_{\mu\nu}(y)] \Big)\,.
\end{equation}
With regard to the equivalent SQM action, \eqref{5d scalar fields} will contribute towards the four-fermi curvature term, while \eqref{A_t eom} will contribute towards the Christoffel connection in the kinetic terms for the fermions.

After suitable rescalings, the equivalent SQM action can then be obtained from \eqref{hw sqm bosonic terms} as
\begin{equation}\label{hw sqm}
\begin{aligned}
    S_{\text{SQM,HW}} =  \frac{1}{e^2} \int_{\mathbb{R}}dt\,
    & \bigg(
    \Big|\dot{A}^a + g^{ab}_{\mathfrak{A}_4}\frac{\partial V_4}{\partial A^b}\Big|^2
    +\Big|\dot{B}^a + g^{ab}_{\mathfrak{A}_4}\frac{\partial V_4}{\partial B^b}\Big|^2
    + g_{\mathfrak{A}_4ab}\Big(\tilde{\psi}^a\nabla_t \psi^b + \tilde{\chi}^a\nabla_t \chi^b \Big)\\
    &+ R^{ab}_{cd}\,\Big(\tilde{\psi}_a\tilde{\psi}_b\psi^c\psi^d
    -\tilde{\psi}_a\tilde{\psi}_b\chi^c\chi^d-\tilde{\chi}_a\tilde{\chi}_b\psi^c\psi^d
      +\tilde{\chi}_a\tilde{\chi}_b\chi^c\chi^d\Big)
      \bigg) \, ,
\end{aligned}
\end{equation}
where $(A^a, B^a)$ and $a,b,c\dots$ are coordinates and indices of the target $\mathfrak{A}_4$, respectively; $\psi^a$, $\chi^a$ $(\tilde{\psi}^a$, $\tilde{\chi}^a)$ are tangent (co-tangent) vectors to  $A^a$, $B^a$ in $\mathfrak{A}_4$, respectively; $\nabla_t\psi^a = \partial_t\psi^a + \Gamma^a_{bc}\big(\dot{x}^b+\dot{y}^b  \big)\psi^c $; $g_{\mathfrak{A}_4}$ and $R$ are the metric and Riemann tensor on $\mathfrak{A}_4$, respectively; and $V_4(A, B)$ is the potential function.

\subsection{An HW Floer Homology of \texorpdfstring{$M_4$}{M4}}\label{subsection: HW floer on M4}

In a TQFT, the Hamiltonian $H$ vanishes in the $\mathcal{Q}$-cohomology, whence this means that for any state $|\mathcal{O}\rangle$ that is nonvanishing in the  $\mathcal{Q}$-cohomology, we have
\begin{equation}
    H |\mathcal{O}\rangle = \{ \mathcal{Q}, \cdots \} |\mathcal{O}\rangle = \mathcal{Q}(\cdots |\mathcal{O}\rangle ) =\mathcal{Q} |\mathcal{O}'\rangle = \{ \mathcal{Q}, \mathcal{O}' \} |0 \rangle = |\{\mathcal{Q}, \mathcal{O}' \} \rangle \sim 0 \, .
\end{equation}
In other words, the $|\mathcal{O}\rangle$'s which span the relevant $\mathcal Q$-cohomology of states in HW theory are actually ground states that are therefore time-invariant. In particular, for HW theory on $M_5 = M_4 \times \mathbb R$ with $\mathbb R$ as the time coordinate, its relevant spectrum of states is associated only with $M_4$.

With $M_5 = M_4 \times \mathbb R$, $M_4$ is the far boundary of the five-manifold and one needs to specify ``boundary conditions'' on $M_4$ to compute the path integral. We can do this by first defining a restriction of the fields to $M_4$, which we shall denote as $\Psi_{M_4}$, and then specifying boundary values for these restrictions.  Doing this is equivalent to inserting in the path integral, an operator  functional $F_4(\Psi_{M_4})$ that is nonvanishing in the ${\cal Q}$-cohomology (so that the path integral will continue to be topological). This means that the corresponding partition function of HW theory\footnote{The HW equations are elliptic, so the virtual dimension of its moduli space $\mathcal M_{\text{HW}}$ will be zero~\cite{haydys2010fukaya}, whence just as in Vafa-Witten theory, it is a balanced TQFT, and one can define the partition function. \label{balanced TQFT}} can be computed as~\cite[eqn.~(4.12)]{witten1988topological}
\begin{equation}\label{floerfunctional}
    \langle 1 \rangle_{F_4(\Psi_{M_4})} = \int_{\mathcal{M}_{\text{HW}}} F_4(\Psi_{M_4}) \,    e^{-S_{\text{HW}}}\,.
\end{equation}

Since we demonstrated in the previous subsection that HW theory on $M_4 \times \mathbb R$ can be expressed as an SQM model in $\mathfrak{A}_4$, we can write the partition function as
\begin{equation}\label{5d partition floer functional}
    {\mathcal{Z}_{\text{HW},M_4 \times \R}(G) = \langle 1 \rangle_{F_4(\Psi_{M_4})}  = \sum_k  {\cal F}^{G}_{\text{HW}}(\Psi_{M_4}^k)}\,.
\end{equation}
Here,  ${\cal F}^{G}_{\text{HW}}(\Psi_{M_4}^k)$ in the $\mathcal Q$-cohomology is the $k^{\text{th}}$ contribution to the partition function  that depends on the expression of ${F_4(\Psi_{M_4})}$ in the bosonic fields on $M_4$ evaluated over the corresponding solutions of the HW equations \eqref{5d bps hw} restricted to $M_4$, and the summation in `$k$' is over all presumably isolated and non-degenerate configurations on $M_4$ in $\mathfrak{A}_4$ that the equivalent SQM localizes onto.\footnote{This presumption will be justified shortly.}

Let us now ascertain what the ${\cal F}^{G}_{\text{HW}}(\Psi_{M_4}^k)$'s correspond to. To this end, let us first determine the configurations that the SQM localizes onto. These are configurations that minimize the SQM action \eqref{hw sqm}, i.e., they set the (zero-modes of the) expression within the squared terms therein to zero. They are therefore given by
\begin{equation}\label{flow on m5 = m4 x R 2/V}
\boxed{   \begin{aligned}
        \frac{d{A}^a}{dt} &= -g^{ab}_{{\mathfrak A}_4}\frac{\partial V_4(A, B)}{\partial A^{b}}\\
        \frac{d{B}^a}{dt} &= -g^{ab}_{{\mathfrak A}_4}\frac{\partial V_4(A, B)}{\partial B^{b}}
    \end{aligned} }
\end{equation}
where the squaring argument~\cite{blau1993topological} means that they have to obey the vanishing of both the LHS and RHS \emph{simultaneously}. In other words, the configurations that the SQM localizes onto are fixed (i.e., time-invariant) critical points of the potential $V_4$ in ${\mathfrak A}_4$.

What then should the explicit form of $V_4$ be, one might ask. To determine this, note that the squared terms in~\eqref{hw sqm} originate from the squared terms in~\eqref{hw sqm bosonic terms}. Indeed, setting the expression within the squared terms in~\eqref{hw sqm bosonic terms} to zero minimizes the underlying 5d action, and this is consistent with setting the expression within the squared terms in~\eqref{hw sqm} to zero to minimize the equivalent SQM action. As such, one can deduce the explicit form of $V_4$ by comparing \eqref{flow on m5 = m4 x R 2/V} with the vanishing of the expression within the squared terms in \eqref{hw sqm bosonic terms}. Specifically, this would give us
\begin{equation} \label{flow on m5 = m4 x R}
    \begin{aligned}
        \frac{d{A}_{\mu}}{dt} &= - D^{\nu}B_{\nu\mu}\,,\\
        \frac{d{B}_{\mu\nu}}{dt} &=  2F^{+}_{\mu\nu} - \frac{1}{2}[B_{\mu\rho}, B^{\rho}_{\nu}]\,,
    \end{aligned}
\end{equation}
and by comparing this with \eqref{flow on m5 = m4 x R 2/V}, we find that
\begin{equation}\label{morse potential hw}
  \boxed{   V_4(A, B) = \int_{M_4} \, \text{Tr}\, \Big( - F^+\wedge \star B + \frac{1}{3}B \wedge (\star_3 (\mathscr{B} \wedge_3 \mathscr{B})) \Big)}
\end{equation}
Here, $\mathscr{B} = \mathscr{B}_i d\gamma^i$ is a one-form on an auxiliary $\R^3$ which corresponds to an interpretation of the two-form $B$ on $M_4$ as a differential form that only has 3 independent components (because of its self-duality),\footnote{Such a correspondence between $\mathscr B$ and $B$ has also been exploited by Haydys in~\cite[eqn.~(37)]{haydys2010fukaya}.}  where $\star_3$ and $\wedge_3$ are the Hodge star operator and exterior product on this auxiliary $\R^3$, respectively, such that the one-form $\star_3 (\mathscr{B} \wedge_3 \mathscr{B})$ on $\R^3$ can be interpreted as another two-form $B'$ on $M_4$ whose components are $B'_{\mu \nu} \sim [B_{\mu\rho}, B^{\rho}_{\nu}]$.

Thus, the summation in `$k$' in \eqref{5d partition floer functional} is over all isolated and non-degenerate critical points of \eqref{morse potential hw} in $\mathfrak A_4$ that are also fixed.\footnote{As we explain next, the aforementioned critical points correspond to Vafa-Witten configurations on $M_4$. For them to be isolated, the actual dimension of their moduli space must be zero. For an appropriate choice of $G$ and $M_4$ such that the actual dimension of the instanton sub-moduli space is zero, this can be true. We shall assume such a choice of $G$ and $M_4$ henceforth; specifically, we choose $G$ and $M_4$ such that $\text{dim}(G) \, (1 + b^+_2) = 4kh$, where $b^+_2$ is the positive second betti number of $M_4$, $k$ is the instanton number on $M_4$, and $h$ is the dual Coxeter number of $G$.  As for their non-degeneracy, a suitable perturbation of $V_4(A, B)$ which can be effected by introducing physically-trivial $\mathcal Q$-exact terms to the action, would ensure this. We would like to thank R.P.~Thomas for discussions on this point. \label{isolated}}

Critical points of $V_4(A,B)$ are configurations in $\mathfrak A_4$ that set the RHS of \eqref{flow on m5 = m4 x R 2/V} to zero; they therefore correspond to configurations on $M_4$ that set the RHS of \eqref{flow on m5 = m4 x R} to zero. i.e., they are Vafa-Witten (VW) configurations on $M_4$. In summary, the partition function \eqref{5d partition floer functional} is an algebraic sum of \emph{fixed} VW configurations on $M_4$ in $\mathfrak{A}_4$.

Notice that \eqref{flow on m5 = m4 x R 2/V} are gradient flow equations, and they govern the classical trajectory of the SQM model from one time-invariant VW configuration on $M_4$ to another in $\mathfrak A_4$. Hence, just as in~\cite{blau1993topological, ong2022vafa}, the equivalent SQM model will physically define a novel gauge-theoretic Floer homology theory.

Specifically, the \emph{time-invariant VW configurations on $M_4$} in $\mathfrak A_4$, i.e., the time-independent solutions to the 4d equations
\begin{equation}
\label{VW configuration}
\boxed{ \begin{aligned}
       F^{+} - \star_3(\mathscr{B} \wedge_3 \mathscr{B}) &= 0\\
       D \star B &= 0
    \end{aligned}    }
\end{equation}
will generate the chains of a Floer complex with \emph{Morse functional $V_4(A, B)$ in \eqref{morse potential hw}}, where HW flow lines, described by time-varying solutions to the \emph{gradient flow equations \eqref{flow on m5 = m4 x R 2/V}}, are the Floer differentials such that the number of outgoing flow lines at each time-invariant VW configuration on $M_4$ in $\mathfrak A_4$ is the degree $d_k$ of the corresponding chain in the Floer complex.

In other words, we can also write \eqref{5d partition floer functional} as
\begin{equation}\label{5d partition function floer}
   \boxed{   \mathcal{Z}_{\text{HW},M_4 \times \R}(G) =  \sum_k  {\cal F}^{G}_{\text{HW}}(\Psi_{M_4}^k)= \sum_k  \text{HF}^{\text{HW}}_{d_k}(M_4, G) = \mathcal{Z}^{\text{Floer}}_{\text{HW},M_4}(G)}
\end{equation}
where  each ${\cal F}^{G}_{\text{HW}}(\Psi_{M_4}^k)$ can be identified with a \emph{novel} class $\text{HF}^{\text{HW}}_{d_k}(M_4, G)$ that we shall henceforth name an HW Floer homology class assigned to $M_4$ defined by \eqref{flow on m5 = m4 x R 2/V}, \eqref{morse potential hw}, \eqref{VW configuration} and the description above.

Note that $\text{HF}^{\text{HW}}_{d_k}(M_4, G)$ was first conjectured to exist by Haydys in~\cite[\S5]{haydys2010fukaya}. We have thus furnished a purely physical proof of his mathematical conjecture.

\section{A 4d-Haydys-Witten Floer Homology of Three-Manifolds}
\label{section: complex flow on m3 x R}

In this section, we specialize to the case where $M_4=M_3 \times S^1$ with $M_3$ being a closed and compact three-manifold, and perform a Kaluza-Klein (KK) dimensional reduction of HW theory by shrinking $S^1$ to be infinitesimally small. This will allow us to physically derive from its topologically-invariant $\mathcal Q$-cohomology, a novel 4d-HW Floer homology of $M_3$.

\subsection{KK Reduction of HW Theory on \texorpdfstring{$S^1$}{S1} and the Corresponding SQM}
\label{section: kk reduction to m3 x R}

Let the $S^1$-direction of $M_4=M_3 \times S^1$ be along $x^0$, and the `time' direction be along $\mathbb{R}$ (recall it is the coordinate labelled as $t$). $A_0$, the component of the gauge field along $S^1$, is then interpreted as a scalar field on $M_3\times \mathbb{R}$ upon KK reduction along $S^1$.
From \eqref{5d susy variations}, the $\mathcal Q$-variation of $A_{\mu}$ (spanning the $x^0$ and $x^i$ directions,  where $i=1,2,3$), and that of $A_t$, are
\begin{equation}
    \begin{aligned}
        \delta C &= \zeta \,, \qquad \delta A_i = \psi'_i\,,\\
        \delta A_4 &= \psi'_4\,,
    \end{aligned}
\end{equation}
where we have, for later convenience, relabeled $(A_0, \psi_0)$ as the scalars $(C, \zeta) \in \Omega^0(M_3 \times \R)$;  $A_t$ as $A_4$, the `fourth' component of the gauge field on $M_3 \times \R$; and $(\psi_i, \eta)$ in \eqref{5d susy variations} as $(\psi'_i, \psi'_4)$, which now define a fermionic one-form $\psi'_{\mu'} \in \Omega^1(M_3 \times \R)$, where $\mu' = 1,2,3,4$.

Notice then from \eqref{5d susy variations} that
\begin{equation}
    \delta^2A_{\mu'} = 2i D_{\mu'} \sigma \, .
\end{equation}
That is, the nilpotency of $\mathcal{Q}$ up to gauge transformations by $\sigma$ is maintained in the resulting 4d theory.

The other relevant 5d bosonic field which has components along $S^1$ is $B_{\mu\nu}$. However, upon shrinking $S^1$, $B_{\mu\nu}\in \Omega^{2, +}(M_3 \times S^1) \otimes \Omega^0(\R)$ can be interpreted as $B_{i}\in \Omega^1(M_3) \otimes \Omega^0(\R)$, as it really only has three independent components along $M_3$ due to its self-duality properties. For later convenience of explanation, let us continue to regard $B_i$ as $B_{\mu' \nu'}\in \Omega^{2,+}(M_3\times \mathbb{R})$. In other words, upon shrinking $S^1$, $B_{\mu \nu}$ becomes a self-dual two-form $B_{\mu' \nu'}$ in the resulting 4d theory.

We are now ready to determine the conditions under which $S_{\text{bo}}$ in \eqref{boson action in s and k} minimizes such that HW theory localizes when we KK reduce along $S^1$. To this end, first note that (i) the non-vanishing components of $s_{\mu\nu}$ in \eqref{boson action in s and k} are  $s_{0i}$ and $s_{ij}$; (ii) in a KK reduction along $S^1$, the derivatives along $x^0$ must be set to zero; (iii) the zero-modes of $C$ should be set to zero (so as to avoid reducible connections in the moduli space that the path integral localizes to). Then, in temporal gauge $A_4=0$,\footnote{We can appeal to this gauge because we saw in $\S$\ref{subsection: 5d theory/5d to sqm} that the gauge field does not contribute along the `time' direction in the end.} we find that the condition  $k_{\mu} =0$ (on zero-modes) which minimizes $S_{\text{bo}}$, would, from \eqref{s and k}, correspond to
\begin{equation}\label{m3 x R flow 2}
    \frac{d{A}^i}{dt} = \epsilon^{ijk}\bigg(\partial_j B_{k} + [A_j, B_{k}]\bigg)\,,
\end{equation}
and
\begin{equation}\label{m3 x R flow 3}
  0 = D_i B^i \equiv \frac{1}{2} \epsilon^{ijk} D_i B_{jk} \, .
\end{equation}
Similarly, we find that the condition  $s_{\mu\nu} = 0$ (on zero-modes)  which minimizes $S_{\text{bo}}$, would, from \eqref{s and k}, correspond to
\begin{equation}\label{m3 x R flow 1}
    \frac{d{B}^i}{dt}  = \frac{1}{2} \epsilon^{ijk} \bigg( F_{jk} - [B_{j}, B_{k}] \bigg)\,.
\end{equation}
Here, we have used the fact that (i) $B_{\mu '\nu '}\in \Omega^{2,+}(M_3 \times \mathbb{R})$ can be interpreted as a one-form on $M_3$ (as mentioned above), where its nonzero independent components are $B_{4i}=B_i$, so $A_i, B_i \in \Omega^1(M_3) \otimes \Omega^0(\R)$; and (ii) from the self-duality condition $B_{\mu '\nu '}=\frac{1}{2}\epsilon_{\mu '\nu '\rho '\lambda '}B^{\rho '\lambda '}$, we can write $B_i = \frac{1}{2} \epsilon_{ijk}B^{jk}$. Notice also that \eqref{m3 x R flow 2}, \eqref{m3 x R flow 3}, and \eqref{m3 x R flow 1} are effectively the KK reductions of \eqref{flow on m5 = m4 x R} (with $A_0 = C$ set to zero).

Just as in \S\ref{subsection: 5d theory/5d to sqm}, this means that the 4d action  can be written as
\begin{equation}\label{vw sqm}
  \begin{aligned}
    S_{\text{HW}_4} =  \frac{1}{2e^2} \int dt \, \Tr \int_{M_3} d^3x \bigg(
    & \bigg| \dot{A}^i - \epsilon^{ijk}\bigg(\partial_j B_{k} + [A_j, B_{k}]\bigg) \bigg|^2
    + \bigg| \dot{B}^i - \frac{1}{2} \epsilon^{ijk} \bigg( F_{jk} - [B_{j}, B_{k}] \bigg) \bigg|^2
    \\
    & + \bigg| \dot{C} - D^i B_i \bigg|^2_{C=0}
      +\dots
      \bigg)
      \,,
  \end{aligned}
\end{equation}
where ``\dots'' contains the fermionic terms and the subscript ``$C=0$'' means that
the zero-modes of $C$ of that term are set to zero. After suitable rescalings,
the equivalent SQM action can be obtained from \eqref{vw sqm} as
\begin{equation}
  \label{vw sqm final}
  S_{\text{SQM,H}} =  \frac{1}{e^2}\int dt
  \left(
    \left|
      \dot{A}^a
      +  g^{ab}_{\mathfrak{A}_3}\frac{\partial V_3}{\partial {A}^b}
    \right|^2_{C=0}
    + \left|
      \dot{B}^a
      +  g^{ab}_{\mathfrak{A}_3}\frac{\partial V_3}{\partial {B}^b}
    \right|^2_{C=0}
    + \left|
      \dot{C}^a
      + g^{ab}_{\mathfrak{A}_3}\frac{\partial V_3}{\partial {C}^b}
    \right|^2_{C=0}
    + \dots
  \right)
  \,,
\end{equation}
where $({A}^a, B^a, C^a)$ and $a,b$ are coordinates and indices on the space $\mathfrak{A}_3$ of irreducible $(A_i, B_i, C)$ fields on $M_3$, respectively; $g^{ab}_{\mathfrak{A}_3}$ is the metric on $\mathfrak{A}_3$; $V_3(A, B, C)$ is the potential function; and the choice of subscript `H' will be clear shortly.

By the squaring argument~\cite{blau1993topological} applied to \eqref{vw sqm}, the configurations that the equivalent SQM localizes onto are those that set the LHS and RHS of \eqref{m3 x R flow 2}--\eqref{m3 x R flow 1} \emph{simultaneously} to zero. In other words, the equivalent SQM localizes onto time-invariant Hitchin configurations on $M_3$, where by Hitchin configurations, we mean solutions to a 3d analog of Hitchin's equations defined as the vanishing of the RHS of \eqref{m3 x R flow 2}--\eqref{m3 x R flow 1}.

\subsection{A 4d-HW Floer Homology of \texorpdfstring{$M_3$}{M3}}
\label{subsection: vw floer homology}

Since the resulting 4d theory on $M_3 \times \mathbb{R}$ can be interpreted as an SQM model in $\mathfrak{A}_3$, its partition function can, like in \eqref{5d partition floer functional}, be written  as
\begin{equation}\label{4d partition floer functional}
    {\mathcal{Z}_{\text{HW},M_3 \times \R}(G) = \langle 1 \rangle_{{F}_3(\Psi_{M_3})}  = \sum_l  {\cal F}^{G}_{\text{HW}_4}(\Psi_{M_3}^l)}\,,
\end{equation}
where ${\cal F}^{G}_{\text{HW}_4}(\Psi_{M_3}^l)$ in the $\mathcal Q$-cohomology is the $l^{\text{th}}$ contribution to the partition function  that depends on the expression of ${{F}_3(\Psi_{M_3})}$ in the bosonic fields on $M_3$ evaluated over the corresponding solutions of \eqref{m3 x R flow 2}--\eqref{m3 x R flow 1} restricted to $M_3$, and the summation in `$l$' is over all isolated and non-degenerate configurations on $M_3$ in $\mathfrak{A}_3$ that the equivalent SQM localizes onto.\footnote{This presumption that the configurations will be isolated and non-degenerate is justified because (the $\mathcal{Q}$-cohomology of) HW theory is topological in all directions and therefore invariant when we shrink the $S^1$, so if $M_3$ (where $M_3 \times S^1 = M_4$) and $G$ are chosen such as to satisfy the conditions spelt out in footnote~\ref{isolated}, $\mathcal{Z}_{\text{HW},M_3 \times \R}$ will be a discrete and non-degenerate  sum of contributions just like $\mathcal{Z}_{\text{HW},M_4 \times \R}$. We shall henceforth assume such a choice of $M_3$ and $G$  whence the presumption would hold. \label{isolated 2}}

Let us now ascertain what the ${\cal F}^{G}_{\text{HW}_4}(\Psi_{M_3}^l)$'s correspond to. Repeating here the analysis in \S \ref{subsection: HW floer on M4} with \eqref{vw sqm final} as the action for the equivalent SQM model, we find that we can also write \eqref{4d partition floer functional} as
\begin{equation}\label{hitchin floer classes}
  \boxed{    \mathcal{Z}_{\text{HW},M_3 \times \R}(G) = \sum_l {\cal F}^{G}_{\text{HW}_4}(\Psi_{M_3}^l)
    =\sum_l \text{HF}_{d_l}^{\text{HW}_4}(M_3, G) = \mathcal{Z}^{\text{Floer}}_{\text{HW}_4,M_3}(G)}
\end{equation}
where each ${\cal F}^{G}_{\text{HW}_4}(\Psi_{M_3}^l)$ can be identified with a \emph{novel} gauge-theoretic 4d-HW Floer homology class $\text{HF}^{\text{HW}_4}_{d_l}(M_3, G)$ of degree $d_l$ assigned to $M_3$.

Specifically, the \emph{time-invariant Hitchin configurations on $M_3$} in $\mathfrak A_3$ that obey the simultaneous vanishing of the LHS and RHS of the \emph{gradient flow equations}
\begin{equation}\label{flow on m4 = m3 x R}
  \boxed{
    \begin{aligned}
      \frac{d{A}^a}{dt}
      &= -g^{ab}_{{\mathfrak A}_3}\frac{\partial V_3(A, B, C)}{\partial A^{b}} \\
      \frac{d{B}^a}{dt}
      &= -g^{ab}_{{\mathfrak A}_3}\frac{\partial V_3(A, B, C)}{\partial B^{b}} \\
      \frac{d{C}^a}{dt}
      &= -g^{ab}_{{\mathfrak A}_3}\frac{\partial V_3(A, B, C)}{\partial C^{b}}
    \end{aligned}
  }
\end{equation}
will generate the chains of the 4d-HW Floer complex with \emph{Morse functional}\begin{equation}\label{v3 functional}
  \boxed{ V_3(A, B, C) = \int_{M_3} \text{Tr}\, \bigg(
    - F \wedge B
    + \frac{1}{3} B \wedge B \wedge B
    - C \wedge D \star B
    \bigg)}
\end{equation}
in $\mathfrak A_3$, where 4d-HW flow lines, described by time-varying solutions to \eqref{flow on m4 = m3 x R}, are the Floer differentials such that the number of outgoing flow lines at each time-invariant Hitchin configuration on $M_3$ in $\mathfrak A_3$ which corresponds to a time-independent solution of the 3d equations
\begin{equation}
\label{H configuration}
\boxed{ \begin{aligned}
  F - B \wedge B &= 0 \\
  DB &= 0 \\
  D \star B &= 0
    \end{aligned}    }
\end{equation}
is the degree $d_l$ of the corresponding chain in the 4d-HW Floer complex.

\section{A 3d-Haydys-Witten Floer Homology of Two-Manifolds}
\label{section: m2 x R}

In this section, we further specialize to the case where $M_3=M_2 \times S^1$ with $M_2$ being a closed and compact two-manifold, and perform a second KK dimensional reduction of HW theory by shrinking this $S^1$ to be infinitesimally small. This will allow us to physically derive from the topologically-invariant $\mathcal Q$-cohomology, a novel 3d-HW Floer homology of $M_2$.

\subsection{KK Reduction of HW Theory on \texorpdfstring{$S^1 \times S^1$}{S1 x S1} and the Corresponding SQM}

Let the $S^1$-direction of $M_3=M_2 \times S^1$ be along $x^1$. Further KK reduction along this $S^1$ amounts to setting $\partial_1 \to 0$, $A_1 = X$, $B_1 = Y$, where $X, Y \in \Omega^0(M_2)$ are scalars field on $M_2$.

In particular, upon further KK reduction of HW theory along this $S^1$ in $M_3$, \eqref{m3 x R flow 2} and \eqref{m3 x R flow 1} become
\begin{equation}\label{m2 x R bps}
  \begin{aligned}
    \frac{d{A^{\alpha}}}{dt}
    &= \epsilon^{\alpha \beta}\Big( D_{\beta} Y - [X, B_{\beta}] \Big)
      \, ,
    \\
    \frac{d B^{\alpha}}{dt}
    &= \epsilon^{\alpha \beta}\Big( D_{\beta} X + [Y, B_{\beta}] \Big)
      \, ,
  \end{aligned}
\end{equation}
and
\begin{equation}\label{m2 x R bps 2}
  \begin{aligned}
    \frac{d{X}}{dt}
    &= \epsilon^{\alpha \beta} D_{\alpha} B_{\beta}
      \, ,
    \\
    \frac{dY}{dt}
    &= \frac{1}{2}\epsilon^{\alpha \beta}\Big( F_{\alpha\beta} - [B_{\alpha}, B_{\beta}] \Big)
      \, ,
  \end{aligned}
\end{equation}
where $\alpha, \beta$ are indices on $M_2$. Defining a connection and scalar field of a complexified gauge group $G_\C$ on $M_2$ via $\mathcal{A} = A + iB \in \Omega^1(M_2, ad(G_\C))$ and $Z = X + iY \in \Omega^0(M_2, ad(G_\C))$, respectively, \eqref{m2 x R bps} and \eqref{m2 x R bps 2} can be written as
\begin{equation}\label{m2 x R bps 3}
    \begin{aligned}
      \dot{\mathcal{A}}^{\alpha}
      &= i \epsilon^{\alpha \beta} \bar{\mathcal{D}}_{\beta} \bar{Z}
        \, ,
      &\qquad
        \dot{\bar{\mathcal{A}}}^{\alpha}
      &= - i \epsilon^{\alpha \beta} \mathcal{D}_{\beta} Z
        \,,
      \\
      \dot{Z}
      &= \frac{i}{2}\epsilon^{\alpha \beta} \bar{\mathcal{F}}_{\alpha\beta}
        \, ,
      &\qquad
        \dot{\bar{Z}}
      &= - \frac{i}{2}\epsilon^{\alpha \beta} \mathcal{F}_{\alpha\beta}
        \, ,
    \end{aligned}
\end{equation}
where $\mathcal{D}_\alpha \equiv \partial_\alpha + [\mathcal{A}_\alpha, \cdot]$ and $\bar{\mathcal{D}}_{\alpha} \equiv \partial_{\alpha} + [\bar{\mathcal{A}}_{\alpha}, \cdot]$.
In other words, 4d-HW theory, upon a second KK reduction along this $S^1$ to a 3d theory on $M_2 \times \R$, localizes onto configurations that satisfy \eqref{m2 x R bps 3}.

Just as in \S\ref{subsection: 5d theory/5d to sqm}, this means that the 3d action  can be written as
\begin{equation}\label{sqm action m2 x R initial}
    \begin{aligned}
        S_{\text{HW}_3} =  \frac{1}{2e^2}\int dt \, \Tr \, \int_{M_2} d^2 x \Bigg(
      & \big| \dot{\mathcal{A}}^{\alpha}
        - i \epsilon^{\alpha \beta}\bar{\mathcal{D}}_{\beta} \bar{Z}
        \big|^2
        + \big|
        \dot{Z} -
        \frac{i}{2}\epsilon^{\alpha \beta}\bar{\mathcal{F}}_{\alpha \beta}
        \big|^2
        \\
      & + \bigg|
        \dot{\bar{\mathcal{A}}}^{\alpha}
        + i \epsilon^{\alpha \beta} \mathcal{D}_{\beta} Z
        \bigg|^2
        + \bigg|
        \dot{\bar{Z}}
        + \frac{i}{2}\epsilon^{\alpha \beta} \mathcal{F}_{\alpha\beta}
        \bigg|^2
        + \dots
        \Bigg)
        \, ,
    \end{aligned}
\end{equation}
where ``$\dots$'' contains the fermionic terms. After suitable rescalings, the equivalent SQM action can be obtained from \eqref{sqm action m2 x R initial} as
\begin{equation}\label{sqm action m2 x R}
  \begin{aligned}
    S_{\text{SQM,BF}} =  \frac{1}{e^2} \int dt \bigg(
    & \bigg| \dot{\mathcal{A}}^a +  g^{ab}_{\mathfrak{A}_2} \left(\frac{\partial V_2}{\partial \mathcal{A}^{b}}\right)^* \bigg|^2
      + \bigg| \dot{Z}^a  +  g^{ab}_{\mathfrak{A}_2} \left(\frac{\partial V_2}{\partial Z^b}\right)^* \bigg|^2
    \\
    & + \bigg| \dot{\bar{\mathcal{A}}}^a +  g^{ab}_{\mathfrak{A}_2} \left(\frac{\partial V_2}{\partial \bar{\mathcal{A}}^{b}}\right)^* \bigg|^2
      + \bigg| \dot{\bar{Z}}^a  +  g^{ab}_{\mathfrak{A}_2} \left(\frac{\partial V_2}{\partial \bar{Z}^b}\right)^* \bigg|^2
      + \dots
      \bigg)
      \, ,
  \end{aligned}
\end{equation}
where $(\mathcal{A}^a, Z^a, \bar{\mathcal{A}}^a, {\bar Z}^a)$ and $a,b$ are coordinates and indices on the space $\mathfrak{A}_2$ of irreducible $(\mathcal A^\alpha, Z, \bar{\mathcal A}^\alpha, {\bar Z})$ fields on $M_2$, respectively; $g^{ab}_{\mathfrak{A}_2}$ is the metric on $\mathfrak{A}_2$; $V_2(\mathcal{A}, Z, \bar{\mathcal{A}}, \bar{Z})$ is the potential function; and the choice of the subscript `BF' will be clear shortly.

By the squaring argument~\cite{blau1993topological} applied to \eqref{sqm action m2 x R}, the configurations that the equivalent SQM localizes onto are those that set the LHS and RHS of \eqref{m2 x R bps 3} \emph{simultaneously} to zero.
In other words, the equivalent SQM localizes onto time-invariant $G_{\mathbb{C}}$-BF configurations on $M_2$.

\subsection{A 3d-HW Floer Homology of \texorpdfstring{$M_2$}{M2}}

Since the resulting 3d theory on $M_2 \times \mathbb{R}$ can be interpreted as an SQM model of fields in $\mathfrak{A}_2$, its partition function can, like in \eqref{5d partition floer functional}, be written  as
\begin{equation}\label{3d partition floer functional}
{\mathcal{Z}_{\text{HW},M_2 \times \R}(G) = \langle 1 \rangle_{{F}_2(\Psi_{M_2})}  = \sum_p  {\cal F}^{G}_{\text{HW}_3}(\Psi_{M_2}^p)}\,,
\end{equation}
where ${\cal F}^{G}_{\text{HW}_3}(\Psi_{M_2}^p)$ in the $\mathcal Q$-cohomology is the $p^{\text{th}}$ contribution to the partition function  that depends on the expression of ${{F}_2(\Psi_{M_2})}$ in the bosonic fields on $M_2$ evaluated over the corresponding solutions of \eqref{m2 x R bps}--\eqref{m2 x R bps 2} restricted to $M_2$, and the summation in `$p$' is over all isolated and non-degenerate configurations on $M_2$ in $\mathfrak{A}_2$ that the equivalent SQM localizes onto.\footnote{This presumption that the configurations will be isolated and non-degenerate is justified because (the $\mathcal{Q}$-cohomology of) HW theory is topological in all directions and therefore invariant when we further shrink this second $S^1$, so if $M_2$ (where $M_2 \times S^1 \times S^1 = M_4$) and $G$ are chosen such as to satisfy the conditions spelt out in footnote~\ref{isolated}, $\mathcal{Z}_{\text{HW},M_2 \times \R}$ will be a discrete and non-degenerate  sum of contributions just like $\mathcal{Z}_{\text{HW},M_4 \times \R}$. We shall henceforth assume such a choice of $M_2$ and $G$ whence the presumption would hold. \label{footnote: justification of isolated and nondegenerate GC-BF configs on M2}}

Let us now ascertain what the ${\cal F}^{G}_{\text{HW}_3}(\Psi_{M_2}^p)$'s correspond to. Repeating here the analysis in \S \ref{subsection: HW floer on M4} with \eqref{sqm action m2 x R} as the action for the equivalent SQM model, we find that we can also write \eqref{3d partition floer functional} as
\begin{equation}\label{BF floer classes}
    \boxed{
        \mathcal{Z}_{\text{HW},M_2 \times \R}(G)
        = \sum_p {\cal F}^{G}_{\text{HW}_3}(\Psi_{M_2}^p)
        = \sum_p \text{HF}_{d_p}^{\text{HW}_3}(M_2, G_\C)
        = \mathcal{Z}^{\text{Floer}}_{\text{HW}_3,M_2}(G_\C)}
\end{equation}
where each ${\cal F}^G_{\text{HW}_3}(\Psi_{M_2}^p)$ can be identified with a \emph{novel} gauge-theoretic 3d-HW Floer homology class $\text{HF}^{\text{HW}_3}_{d_p}(M_2, G_\C)$ of degree $d_p$ assigned to $M_2$.

Specifically, the \emph{time-invariant $G_\C$-BF configurations on $M_2$} in $\mathfrak A_2$ that obey the simultaneous vanishing of the LHS and RHS of the \emph{gradient flow equations}
\begin{equation} \label{m2 x R flow}
 \boxed{   \begin{aligned}
        \frac{d{\mathcal{A}^{a}}}{dt} &= -g^{ab}_{\mathfrak A_2} \left(\frac{\partial V_2(\mathcal{A}, Z, \bar {\mathcal{A}}, \bar Z)}{\partial \mathcal{A}^{b}}\right)^*
        \qquad&
        \frac{d{\bar{\mathcal{A}}^{a}}}{dt} &= -g^{ab}_{\mathfrak A_2} \left(\frac{\partial V_2(\mathcal{A}, Z, \bar {\mathcal{A}}, \bar Z)}{\partial \bar{\mathcal{A}}^{b}}\right)^*
        \\
        \frac{d{Z}^{a}}{dt}&= - g^{ab}_{\mathfrak A_2} \left(\frac{\partial V_2(\mathcal{A}, Z, \bar {\mathcal{A}}, \bar Z)}{\partial Z^{b}}\right)^*
        \qquad&
        \frac{d{\bar{Z}}^{a}}{dt}&= - g^{ab}_{\mathfrak A_2} \left(\frac{\partial V_2(\mathcal{A}, Z, \bar {\mathcal{A}}, \bar Z)}{\partial \bar{Z}^{b}}\right)^*
    \end{aligned} }
\end{equation}
will generate the chains of the 3d-HW Floer complex with \emph{Morse functional}
\begin{equation}\label{potential funcitonal on space of fields on m2}
 \boxed{   V_2(\mathcal{A}, Z, \bar {\mathcal{A}}, \bar Z) = i \int_{M_2 } \,\text{Tr}\, \Big( Z \wedge \mathcal{F} - \bar{Z} \wedge \bar{\mathcal{F}} \Big) }
\end{equation}
in $\mathfrak A_2$, where 3d-HW flow lines, described by time-varying solutions to \eqref{m2 x R flow}, are the Floer differentials such that the number of outgoing flow lines at each time-invariant $G_\C$-BF configuration on $M_2$ in $\mathfrak A_2$ which corresponds to a time-independent solution of the 2d equations
\begin{equation}\label{BF configurations}
 \boxed{ \begin{aligned}
    \mathcal{F} &= 0 = \bar{\mathcal{F}}\\
    \mathcal{D} Z &= 0 = \bar{\mathcal{D}} \bar{Z}
    \end{aligned} }
\end{equation}
is the degree $d_p$ of the corresponding chain in the 3d-HW Floer complex.

\section{Symplectic Floer Homologies, Flat \texorpdfstring{$G_\C$}{G-C} Connections, and \texorpdfstring{$\theta$}{theta}-Hitchin Space}
\label{section: symplectic floer homology}

In this section, we will specialize to the case where $M_5 = \Sigma \times M_2 \times \mathbb{R}$, and perform a topological reduction of HW theory along $M_2$, a Riemann surface  of genus $g\geq 2$, using the Bershadsky-Johansen-Sadov-Vafa reduction method \cite{bershadsky1995topolreduc}, to obtain a 3d ${\mathcal {N}}=4$ topological sigma model on $\Sigma \times \R$, where $\Sigma = S^1 \times S^1$, $I \times S^1$, and $I \times \R$. By recasting this 3d sigma model as an SQM in the appropriate space, we will be able to derive, from the topologically-invariant $\mathcal Q$-cohomology, a novel symplectic Floer homology in the double loop space of flat $G_\C$ connections on $M_2$, and a novel symplectic intersection Floer homology in the loop space of flat $G_\C$ connections on $M_2$ and the path space of $\theta$-Hitchin space of $M_2$.

\subsection{A 3d Sigma Model on \texorpdfstring{$S^1 \times S^1 \times \R$}{S1 x S1 x R}}
\label{subsection: sadov reduction process}

In $M_5 = S^1 \times S^1 \times M_2 \times \mathbb{R}$, denote by $x^M$ with $M \in \{0, 1\}$ the coordinates of $S^1 \times S^1$, $x^m$ with $m \in \{2, 3\}$ the coordinates of $M_2$, and $t$ the coordinate of $\R$. Topological reduction of the action along $M_2$ can be achieved by scaling the metric along $M_2$ to the vanishing limit, as was done in \cite{bershadsky1995topolreduc}.\footnote{%
  This method of topologically reducing along a Riemann surface $M_2$ was also used by Harvey-Moore-Strominger in \cite{harvey-1995-reduc-s} for the specific case of $M_2 = T^2$, just before Bershadsky-Johansen-Sadov-Vafa in \cite{bershadsky1995topolreduc}.
  \label{ft:bjsv method same as hms}
}

By writing in \eqref{5d action terms removed for sqm} the metric of $M_2$ as $g_{M_2} \rightarrow \epsilon g_{M_2}$, we can ignore the terms with positive powers of $\epsilon$ since they will not survive when we take the limit $\epsilon \rightarrow 0$. Only those terms with zero power of $\epsilon$ are of interest since they will survive when we take the limit $\epsilon \rightarrow 0$. These terms can be identified as those with a single contraction in $M_2$ in \eqref{5d action terms removed for sqm}, whence the relevant 5d action is
\begin{equation}
  \label{3d reduced action}
  \begin{aligned}
    S_{\epsilon^0}
    = \frac{1}{e^2} \int d^4 x dt \,
    \text{Tr} \bigg(
    & \frac{1}{2} |F_{tm} + D^M B_{Mm}|^2
      + \left\vert F^+_{Mm}
      - \frac{1}{4} [B_{Mn}, B_m^n]
      - \frac{1}{4} [B_{MN}, B_m^N]
      - \frac{1}{2} D_t B_{Mm}\right\vert^2 \\
    & - 2 D_m \sigma D^m \bar\sigma
      - i \tilde{\eta}D_{m}\psi^{m}
      - i \eta D_{m}\tilde{\psi}^{m}
      -  \tilde{\psi}_{m}D_t\psi^{m}
      - \tilde{\chi}_{mM} D_t \chi^{mM} \\
    & - 2i (\tilde{\psi}_{m} D_{M} - \tilde{\psi}_M D_m) \chi^{mM}
      - 2i (\psi_m D_M - \psi_M D_m )\tilde{\chi}^{mM} \\
    & - i\sigma
      (
      2 \{ \tilde{\chi}_{mM}, \tilde{\chi}^{mM} \}
      - \{ \tilde{\psi}_m, \tilde{\psi}^m \}
      )
      - i\bar\sigma
      (
      2 \{ \chi_{mM}, \chi^{mM} \}
      - \{ \psi_m, \psi^m \}
      )
      \bigg)\,.
  \end{aligned}
\end{equation}
That said, there are also terms with negative powers of $\epsilon$ which will blow up when we take the limit $\epsilon \rightarrow 0$, and they are those with more than one contraction along $M_2$. As such, these terms need to be set to $0$ to ensure the finiteness of the reduction process, whence we will have the conditions
\begin{subequations}
  \label{3d unique finiteness equations}
  \begin{align}
    F^+_{mn}
    - \frac{1}{4} [B_{mM}, B_n^M]
    - \frac{1}{2} D_t B_{mn}
    = D^m B_{mM}
    &= 0 \, ,
      \label{3d unique finiteness equations for target space} \\
    D^n B_{nm}
    &= 0 \,.
      \label{3d unique finiteness equations for B_ab}
  \end{align}
\end{subequations}

Re-expressing in \eqref{3d reduced action}, the coordinates of $M_2$ as complex coordinates defined by $z = x^2 + ix^3$, the final expression for the relevant 5d action that will survive when we take the limit $\epsilon \rightarrow 0$ is\footnote{To arrive at the following expression, we have used the fact that $B_{\mu\nu}$ is a self-dual 2-form in $\Omega^{2, +}(M_4, ad(G))$, and that the terms $B_{MN} = \pm B_{01} = \pm B_{23} = B_{mn}$ in \eqref{3d reduced action}, vanish. This fact can be understood as follows.
Firstly, one can exploit the self-duality of $B_{\mu\nu}$ to define a section $\phi = B_{0z} dx^0 \wedge dz - i B_{1z} dx^1 \wedge dz + B_{0\zb} dx^0 \wedge d\zb + i B_{1\zb} dx^1 \wedge d\zb$ of $\Omega^1(S^1 \times S^1) \otimes \Omega^1(M_2)$, where the remaining independent component $B_{01}$ can then be interpreted as a scalar 0-form w.r.t rotations on $M_5$.
Secondly, one can understand from \eqref{3d unique finiteness equations for B_ab} that $B_{01}$ generates infinitesimal gauge transformations whilst keeping $A_z$ and $A_{\zb}$ constant. Thus, to ensure that gauge connections are irreducible whence the moduli space of \eqref{3d unique finiteness equations} (which will play a relevant role in our resulting 3d theory) is well-behaved, we shall consider the case that $B_{01} = 0$. \label{phi definition}}
\begin{equation}\label{3d reduced action in complex coordinates}
    \begin{aligned}
      S_{\epsilon^0}
      = \frac{1}{e^2} \int d^2 x |dz|^2 dt \, \Tr \Big(
        & \, D_t A_z D_t A^z
        + D^{M} B_{Mz} D_{N} B^{Nz}
        \\
        & \, + D^M A_z D_M A^z
        + \hlf D_t B_{Mz} D_t B^{Mz}
        + \dots \Big) \, ,
    \end{aligned}
\end{equation}
where ``$\dots$'' contains the fermionic terms and terms with the scalars $\sigma$ and $\bar\sigma$ in \eqref{3d reduced action}. Note that the terms with covariant derivatives in $M_2$ ($D_z$ or $D_\zb$) have been omitted as they will not contribute when we take the limit $\epsilon \rightarrow 0$.\footnote{When $M_2$ is shrunken away in the vanishing limit of $\epsilon \to 0$, it is clear that the partial derivatives $\partial_{z, \zb}$ will not contribute. The remaining terms of the covariant derivatives will be paired to auxiliary fields, which will be integrated out in a later step. Therefore, we have chosen not to present them in the current expression.}

The finiteness conditions in \eqref{3d unique finiteness equations} can now be expressed as
\begin{equation}\label{3d finiteness conditions in complex coordinates}
    \begin{aligned}
        F_{z\zb} - \frac{1}{2} [B_{Mz}, B^M_\zb] &= 0 \, , \\
        D^z B_{Mz} &= 0 \, .
    \end{aligned}
\end{equation}
Let us denote by $\mathbf{V} \coloneq \partial_0 + \partial_1$ a unit vector along $S^1 \times S^1$, $F_{z\zb} = F$ the field strength two-form on $M_2$, and $\varphi \coloneq \iota_\mathbf{V} \phi = (B_{0z} - i B_{1z}) dz + (B_{0\zb} + i B_{1\zb}) d\zb$ a section of $\Omega^0(S^1 \times S^1 \times \mathbb{R}) \otimes \Omega^1(M_2)$ obtained by performing an interior product of the section $\phi$ with the unit vector $\mathbf{V}$ (see definition of $\phi$ in footnote~\ref{phi definition}). Then, with an appropriate scaling of $\varphi$, we can also express \eqref{3d finiteness conditions in complex coordinates} as
\begin{equation}
\label{Hitchin on M2}
    \boxed{
    \begin{aligned}
        F - \varphi \wedge \varphi &= 0 \\
        D \varphi = D^* \varphi &= 0
    \end{aligned}
    }
\end{equation}
This is just Hitchin's equations for $G$ on $M_2$!

Recall that in \eqref{Hitchin on M2}, $F_{z\bar{z}} = \partial_z A_\zb - \partial_{\bar{z}} A_z + [A_z, A_{\bar{z}}]$, $\varphi = \varphi_z dz + \varphi_\zb d\zb$, where
$\varphi_z = B_{0z} - i B_{1z}$ and $\varphi_\zb = B_{0\zb} + i B_{1\zb}$. These $A$ and $B$ components are exactly those that appear in $S_{\epsilon^0}$ in \eqref{3d reduced action in complex coordinates}. What this means is that when we take the limit $\epsilon \to 0$ and shrink $M_2$ away, $S_{\epsilon^0}$ in \eqref{3d reduced action in complex coordinates} becomes an action of a 3d theory on $S^1 \times S^1 \times \mathbb{R}$ with the bosonic $A$'s and $B$'s as scalars and one-forms, respectively, which can be interpreted as coordinates of the space of solutions $(A_{M_2}, \varphi)$ to \eqref{Hitchin on M2} (modulo gauge equivalence). In other words, we will have a  3d $\mathcal{N} = 4$ topological sigma model with symplectic target $\mathcal{M}^G_{\text{H}}(M_2)$, the Hitchin moduli space of $G$ on $M_2$. For $M_2$ a Riemann surface of genus $g \ge 2$, $\mathcal{M}^G_{\text{H}}(M_2)$ is well-behaved (see also footnote~\ref{phi definition}). We shall henceforth assume this to be the case.

Let us now determine the sigma model action. Via the cotangent bases $(\alpha_{iz}, \alpha_{\ib \zb})$ and $(\beta_{iz}, \beta_{\ib\zb})$ that can be defined for the base and fiber of the hyperk\"{a}hler target space $\mathcal{M}^G_{\text{H}}(M_2)$, we can write the bosons in \eqref{3d reduced action in complex coordinates} as
\begin{equation}\label{3d boson field correspondence}
    \begin{aligned}
        A_{z} &= X^i \alpha_{iz} \, ,
        \qquad & \qquad
        A_\zb &= X^{\ib} \alpha_{\ib \zb} \, ,
        \\
        B_{Mz} &= Y^i_M \beta_{iz} \, ,
        \qquad & \qquad
        B_{M \zb} &= Y^\ib_M \beta_{\ib\zb} \, ,
    \end{aligned}
\end{equation}
where $(X, Y)$ are coordinates of the base and fiber of $\mathcal{M}^G_{\text{H}}(M_2)$, respectively.

What about the fermions in \eqref{3d reduced action in complex coordinates}? Well, they can be interpreted as tangent vectors of $\mathcal{M}^G_{\text{H}}(M_2)$, and can thus be written as
\begin{equation}\label{3d fermion field correspondence}
    \begin{aligned}
        \psi_{z} &= \lambda^i_1 \alpha_{iz} \, ,
        \qquad & \qquad
        \psi_\zb &= \lambda^{\ib}_1 \alpha_{\ib \zb} \, ,
        \\
        \td\psi_z &= \lambda^i_2\alpha_{iz} \, ,
        \qquad & \qquad
        \td\psi_\zb &= \lambda^{\ib}_2 \alpha_{\ib \zb} \, ,
        \\
        \chi_{Mz} &= \rho^i_{1M}\beta_{iz} \, ,
        \qquad & \qquad
        \chi_{M\zb} &= \rho^{\ib}_{1M} \beta_{\ib \zb} \, ,
        \\
        \td\chi_{Mz} &= \rho^i_{2M} \beta_{iz} \, ,
        \qquad & \qquad
        \td\chi_{M\zb} &= \rho^{\ib}_{2M} \beta_{\ib \zb} \, ,
    \end{aligned}
\end{equation}
where $(\lambda, \rho)$ are tangent vectors of the base and fiber of $\mathcal{M}^G_{\text{H}}(M_2)$, respectively.

Finally, let us topologically reduce \eqref{3d reduced action in complex coordinates} along $M_2$ by taking the limit $\epsilon \rightarrow 0$. Noting that the fields $\sigma$, $\bar\sigma$, $A_t$, $A_M$, $\eta$, $\td\eta$, $\psi_M$, and $\td\psi_M$ can be integrated out to get the curvature and Christoffel connection terms for the 3d fermions (as was done in the context of an SQM in equations \eqref{5d scalar fields} and \eqref{A_t eom}), the resultant action is\footnotemark{}
\begin{equation}\label{3d sigma model action}
    S_{\text{3d} }
    = \frac{1}{e^2} \int d^2x \, dt \, \Big( L_1 + L_2 \Big) \, ,
\end{equation}
\begin{equation*}
  \label{3d sigma action-l1 s1 times s1-scaled}
  \begin{aligned}
    L_1 = \gij \Big(
    & \partial_t X^i \partial_t X^\jb
      + \partial_t Y^i \partial_t Y^\jb
      + \partial_0 X^i \partial_0 X^\jb
      + \partial_1 X^i \partial_1 X^\jb
      + \partial_0 Y^i \partial_0 Y^\jb
      + \partial_1 Y^i \partial_1 Y^\jb
    \\
    & - \lambda^i_2 \nabla_t \lambda^\jb_1
    - \rho^i_2 \nabla_t \rho^\jb_1
    + 2i \lambda^i_2 \nabla_0 \rho^\jb_1
    + 2i \lambda^i_2 \nabla_1 \rho^\jb_1
    + 2i \lambda^i_1 \nabla_0 \rho^\jb_2
    + 2i \lambda^i_1 \nabla_1 \rho^\jb_2
    \Big) \, ,
  \end{aligned}
\end{equation*}
\begin{equation*}
  \label{3d sigma action-l2 s1 times s1-scaled}
  \begin{aligned}
    L_2 = \Omega_{i \jb k \lb} \Big(
    \lambda_2^i \lambda_2^{\bar{\jmath}}
    - \rho_2^i \rho_2^{\bar{\jmath}}
    \Big)
    \Big(
    \lambda_1^i \lambda_1^{\bar{\jmath}}
    - \rho_1^i \rho_1^{\bar{\jmath}}
    \Big) \, ,
  \end{aligned}
\end{equation*}
where (i) $\Omega_{i\jb k\lb}$ is the Riemann curvature tensor of the target space $\mathcal{M}^G_{\text{H}}(M_2)$, and (ii) $\nabla_{\kappa} \Psi^i = \partial_{\kappa} \Psi^i + \Gamma^i_{jk} \Psi^j \partial_{\kappa}(X^k + Y^k)$ and $\nabla_{\kappa} \Psi^{\bar{\imath}} = \partial_{\kappa} \Psi^{\bar{\imath}} + \Gamma^{\bar{\imath}}_{\bar{\jmath}\bar{k}} \Psi^{\bar{\jmath}} \partial_{\kappa}(X^{\bar{k}} + Y^{\bar{k}})$ are covariant derivatives of the fermions on the worldvolume with $\kappa \in \{t, 0, 1\}$.
\footnotetext{%
To arrive at the following expression, we have done three things. Firstly, the indices of the worldvolume were lowered via the fact that the metric of $S^1 \times S^1$ is flat, i.e. $\eta^{01} = \eta^{10} = 0$ and $\eta^{00} = \eta^{11} = 1$.
Secondly, the $Y$-, $\rho_1$-, and $\rho_2$-fields were written only in their $M=0$ index -- this is because from the self-duality of $B$, $\chi$, $\td\chi$, we have, in the field correspondences \eqref{3d boson field correspondence} and \eqref{3d fermion field correspondence}, a duality between different $M$ indices of the $Y$-, $\rho_1$-, and $\rho_2$-fields, respectively, i.e., $Y^i_0 = -i Y^i_1$, $Y^\ib_0 = i Y^\ib_1$, $\rho^i_{\alpha 0} = -i\rho^i_{\alpha 1}$, and $\rho^\ib_{\alpha 0} = i\rho^\ib_{\alpha 1}$, for $\alpha \in \{1, 2\}$.
Thirdly, the single $M=0$ index is omitted for the aforementioned fields as it is effectively the only index, i.e., its specification is redundant.
}

The supersymmetric variations descend from \eqref{5d susy variations} for the fields that survive the topological reduction from $M_5$ to $S^1 \times S^1 \times \mathbb{R}$, and they are given by
\begin{equation}
  \label{3d sigma model susy variations}
  \begin{aligned}
    \delta X^i
    &= \lambda^i_1
      \, ,
    & \qquad
      \delta \lambda^i_1
    &= 0
      \, ,
    & \qquad
      \delta \lambda^i_2
    &= - \partial_t X^i
      + \partial_0 Y^i
      + i \partial_1 Y^i
      + i \Gamma^i_{jk} \lambda_2^j \left( \lambda_1^k + \rho_1^k \right)
      \, ,
    \\
    \delta X^\ib
    &= \lambda^\ib_1
      \, ,
    & \qquad
     \delta \lambda^\ib_1
    &= 0
      \, ,
    & \qquad
     \delta \lambda^\ib_2
    &= - \partial_t X^\ib
      + \partial_0 Y^\ib
      - i \partial_1 Y^\ib
      + i \Gamma^{\bar{\imath}}_{\bar{\jmath}\bar{k}} \lambda_2^{\bar{\jmath}} \left( \lambda_1^{\bar{k}} + \rho_1^{\bar{k}} \right)
      \, ,
    \\
    \delta Y^i
    &= \rho^i_1
      \, ,
    & \qquad
      \delta \rho^i_1
    &= 0
      \, ,
    & \qquad
      \delta \rho^i_2
    &= \partial_t Y^i
      + \partial_0 X^i
      - i \partial_1 X^i
      + i \Gamma^i_{jk} \rho_2^j \left( \lambda_1^k + \rho_1^k \right)
      \, ,
    \\
     \delta Y^\ib
    &= \rho^\ib_1
      \, ,
    & \qquad
     \delta \rho^{\ib}_1
    &= 0
      \, ,
    & \qquad
      \delta \rho^\ib_2
    &= \partial_t Y^\ib
      + \partial_0 X^\ib
      + i \partial_1 X^\ib
      + i \Gamma^{\bar{\imath}}_{\bar{\jmath}\bar{k}} \rho_2^{\bar{\jmath}} \left( \lambda_1^{\bar{k}} + \rho_1^{\bar{k}} \right)
      \, .
  \end{aligned}
\end{equation}

\subsection{An SQM in the Double Loop Space of Flat \texorpdfstring{$G_C$}{G-C} Connections}

Relabelling the coordinates $(x^0, x^1)$ as $(r, s)$ for convenience, we get the BPS equations of our 3d $\mathcal{N} = 4$ topological sigma model with target $\mathcal{M}^G_{\text{H}}(M_2)$ as
\begin{equation}
    \label{3d sigma model BPS equations}
    \begin{aligned}
        \partial_t X^i &= \partial_r Y^i + i \partial_s Y^i \, ,
        & \quad
        \partial_t X^\ib &= \partial_r Y^\ib - i \partial_s Y^\ib \, ,
        \\
        \partial_t Y^\ib &= - \partial_r X^\ib - i \partial_s X^\ib \, ,
        & \quad
        \partial_t Y^i &= - \partial_r X^i + i \partial_s X^i \, .
    \end{aligned}
\end{equation}
In fact, it can be determined from the \eqref{3d sigma model BPS equations} that $\mathcal{M}^G_{\text{H}}(M_2) \equiv \mathcal{M}^{G_\C}_{\text{flat}}(M_2)$, the moduli space of flat $G_\C$ connections on $M_2$.\footnote{%
  By performing KK reductions of \eqref{3d sigma model BPS equations} along either of the $S^1$'s, the BPS equations of the resulting 2d sigma model will tell us that the holomorphic coordinates of $\mathcal{M}^G_{\text{H}}(M_2)$ are (complex) linear combinations of $X$ and $Y$, i.e. the 2d sigma model sees $\mathcal{M}^G_{\text{H}}(M_2)$ in either complex structure $J$ or $K$, depending on the choice of $S^1$ that is being reduced. As the KK reduction is a topological process on the 3d worldvolume that will not alter the target space, it can be inferred that the 3d sigma model will also see $\mathcal{M}^G_{\text{H}}(M_2)$ in either complex structure $J$ or $K$, whence $\mathcal{M}^G_{\text{H}}(M_2) = \mathcal{M}^{G_\C}_{\text{flat}}(M_2)$.
  \label{footnote: determining 3d sigma model complex structure}
}

Therefore, it is more appropriate to re-express the BPS equations in \eqref{3d sigma model BPS equations} as
\begin{equation}
    \label{3d sigma model BPS equations complexified}
    \begin{aligned}
        \partial_t Z^i &= - i \partial_r Z^i - \partial_s \bar{Z}^i \, ,
        &\quad
        \partial_t Z^\ib &= - i \partial_r Z^\ib + \partial_s \bar{Z}^\ib \, ,
        \\
        \partial_t \bar{Z}^\ib &= i \partial_r \bar{Z}^\ib - \partial_s {Z}^\ib \, ,
        &\quad
        \partial_t \bar{Z}^i &= i \partial_r \bar{Z}^i + \partial_s Z^i \, ,
    \end{aligned}
\end{equation}
where $Z^i \coloneq X^i + iY^i$, $Z^{\ib} \coloneq {X}^{\ib} + i {Y}^{\ib}$,  $\bar{Z}^i \coloneq X^i - i Y^i$, and $\bar{Z}^{\ib} \coloneq {X}^{\ib} - i {Y}^{\ib}$, are the coordinates of $\mathcal{M}^{G_\C}_{\text{flat}}(M_2)$.

Just as in \S\ref{subsection: 5d theory/5d to sqm}, we can now rewrite the 3d action in \eqref{3d sigma model action} (involving only the bosons) as
\begin{equation}
    \label{sqm action 3d sigma initial}
    \begin{aligned}
    S_{\text{3d}} = \frac{1}{e^2} \int dt \int_{S^1 \times S^1} dr ds \Big(
    & | \dot{Z}^i + i \partial_r Z^i + \partial_s \bar{Z}^i |^2
      + | \dot{Z}^\ib + i \partial_r Z^\ib - \partial_s \bar{Z}^\ib |^2
    \\
    & + | \dot{\bar{Z}}^\ib - i \partial_r \bar{Z}^\ib + \partial_s Z^\ib |^2
      + | \dot{\bar{Z}}^i - i \partial_r \bar{Z}^i - \partial_s Z^i |^2
      + \dots
      \Big)
      \, ,
    \end{aligned}
\end{equation}
where ``\dots'' refer to fermionic terms in \eqref{3d sigma model action}. After suitable rescalings, the equivalent SQM action can be obtained from \eqref{sqm action 3d sigma initial}
as
\begin{equation}
    \label{sqm action 3d sigma}
    \begin{aligned}
        S_{\text{SQM}, \Sigma} = \frac{1}{e^2} \int dt \, \bigg(
        & \left| \dot{Z}^a + g^{a\bb}_{L^2\mathcal{M}^{G_\C}_{\text{flat}}(M_2)} \left(\frac{\partial V_\Sigma}{\partial Z^b}\right)^* \right|^2
        + \left| \dot{Z}^\ab + g^{\ab b}_{L^2\mathcal{M}^{G_\C}_{\text{flat}}(M_2)} \left(\frac{\partial V_\Sigma}{\partial Z^\bb}\right)^* \right|^2
        \\
        & + \left| \dot{\bar{Z}}^\ab + g^{\ab b}_{L^2\mathcal{M}^{G_\C}_{\text{flat}}(M_2)} \left(\frac{\partial V_\Sigma}{\partial \bar{Z}^\bb}\right)^* \right|^2
        + \left| \dot{\bar{Z}}^a + g^{a \bb}_{L^2\mathcal{M}^{G_\C}_{\text{flat}}(M_2)} \left(\frac{\partial V_\Sigma}{\partial \bar{Z}^b}\right)^* \right|^2
          + \dots
          \bigg)
          \, ,
    \end{aligned}
\end{equation}
where $(Z^a, Z^\ab, \bar{Z}^a, \bar{Z}^\ab)$ and $g^{a\bb}_{L^2\mathcal{M}^{G_\C}_{\text{flat}}(M_2)}$ are the coordinates and the metric on the double loop space $L^2\mathcal{M}^{G_\C}_{\text{flat}}(M_2)$ of maps from $S^1 \times S^1$ to the symplectic manifold $\mathcal{M}^{G_\C}_{\text{flat}}(M_2)$, respectively; and $V_\Sigma(Z,\bar{Z})$ is the potential function. In other words, we equivalently have an SQM in $L^2\mathcal{M}^{G_\C}_{\text{flat}}(M_2)$.

\subsection{A Symplectic Floer Homology in the Double Loop Space of Flat \texorpdfstring{$G_\C$}{G-C} Connections}

Since the resulting 3d theory on $S^1 \times S^1 \times \mathbb{R}$ can be interpreted as an SQM model in $L^2\mathcal{M}^{G_\C}_{\text{flat}}(M_2)$, its partition function can be written  as
\begin{equation}
  \label{3d partition symplectic floer functional}
  \mathcal{Z}_{\text{HW},S^1 \times S^1 \times \mathbb{R}}(G) = \sum_q  {\cal F}^{q}_{L^2\mathcal{M}^{G_\C}_{\text{flat}}(M_2)} \, ,
\end{equation}
where the $q^{\text{th}}$ contribution ${\cal F}^{q}_{L^2\mathcal{M}^{G_\C}_{\text{flat}}(M_2)}$ to the partition function, is in the $\mathcal Q$-cohomology, and the summation in `$q$' is over all isolated and non-degenerate configurations on $S^1 \times S^1$ in $L^2\mathcal{M}^{G_\C}_{\text{flat}}(M_2)$ that the equivalent SQM localizes onto.\footnote{That the configurations will be isolated and non-degenerate will be clear shortly.}

Via a similar analysis to that in \S \ref{subsection: HW floer on M4} with \eqref{sqm action 3d sigma} as the action for the equivalent SQM model, we find that we can also write \eqref{3d partition symplectic floer functional} as
\begin{equation}\label{L^2M floer classes}
  \boxed{
    \mathcal{Z}_{\text{HW},S^1 \times S^1 \times \mathbb{R}}(G)
    = \sum_q  {\cal F}^{q}_{L^2\mathcal{M}^{G_\C}_{\text{flat}}(M_2)}
    = \sum_q \text{HSF}^{\text{hol}}_{d_q}({L^2\mathcal{M}^{G_\C}_{\text{flat}}(M_2)})
    = \mathcal{Z}^{\text{SympFloer}}_{L^2\mathcal{M}^{G_\C}_{\text{flat}}(M_2)}
  }
\end{equation}
where each ${\cal F}^{q}_{L^2\mathcal{M}^{G_\C}_{\text{flat}}(M_2)}$ can be identified with a \emph{novel} symplectic Floer homology class  $\text{HSF}^{\text{hol}}_{d_q}({L^2\mathcal{M}^{G_\C}_{\text{flat}}(M_2)})$ of degree $d_q$ in  $L^2\mathcal{M}^{G_\C}_{\text{flat}}(M_2)$.

Specifically, the \textit{time-invariant maps in $L^2\mathcal{M}^{G_\C}_{\text{flat}}(M_2)$} that obey the simultaneous vanishing of the LHS and RHS of the \textit{gradient flow equations}
\begin{equation}
    \label{3d sigma flow}
    \boxed{
    \begin{aligned}
        \frac{dZ^a}{dt}
        &= - g^{a \bb}_{L^2{\mathcal{M}^{G_\C}_{\text{flat}}}} \left(\frac{\partial V_\Sigma(Z, \bar{Z})}{\partial Z^b}\right)^*
        \qquad&
        \frac{d\bar{Z}^a}{dt}
        &= - g^{a \bb}_{L^2\mathcal{M}^{G_\C}_{\text{flat}}} \left(\frac{\partial V_\Sigma(Z, \bar{Z})}{\partial \bar{Z}^b}\right)^*
        \\
        \frac{dZ^\ab}{dt}
        &= - g^{\ab b}_{L^2{\mathcal{M}^{G_\C}_{\text{flat}}}} \left(\frac{\partial V_\Sigma(Z, \bar{Z})}{\partial Z^\bb}\right)^*
        \qquad &
        \frac{d\bar{Z}^\ab}{dt}
        &= - g^{\ab b}_{L^2{\mathcal{M}^{G_\C}_{\text{flat}}}} \left(\frac{\partial V_\Sigma(Z, \bar{Z})}{\partial \bar{Z}^\bb}\right)^*
    \end{aligned}
    }
\end{equation}
will generate the chains of the symplectic Floer complex with \emph{Morse functional}
\begin{equation}
    \label{potential functions for 3d sigma}
    \boxed{
    V_\Sigma(Z, \bar{Z})
    = \int_{S^1 \times S^1} dr ds \, g_{i\jb} \Big( Z^i \partial_s Z^\jb - i Z^i \partial_r \bar{Z}^\jb - \bar{Z}^i \partial_s \bar{Z}^\jb + i \bar{Z}^i \partial_r Z^\jb \Big)
    }
\end{equation}
such that flow lines, described by time-varying solutions to \eqref{3d sigma flow}, are the Floer differentials whereby the number of outgoing flow lines at each time-invariant map in $L^2\mathcal{M}^{G_\C}_{\text{flat}}(M_2)$ which corresponds to a time-independent solution of the 2d equations
\begin{equation}
    \label{critical points of 3d sigma potential}
 \boxed{   \begin{aligned}
        i \partial_r Z^i + \partial_s \bar{Z}^i &= 0 = i \partial_r \bar{Z}^i + \partial_s Z^i
        \\
        i \partial_r \bar{Z}^\ib - \partial_s {Z}^\ib &= 0 = i \partial_r Z^\ib - \partial_s \bar{Z}^\ib
    \end{aligned} }
\end{equation}
 on $S^1 \times S^1$, is the degree $d_q$ of the corresponding chain in the symplectic Floer complex.

Notice that we can equivalently combine the terms in \eqref{critical points of 3d sigma potential} to get $ \partial_\wb Y^i = 0 = \partial_w \bar{Y}^\ib$ and $ \partial_\wb \bar{X}^\ib = 0 = \partial_w X^i$,
which describes simultaneous holomorphic and antiholomorphic maps from $S^1 \times S^1$ to the fiber and base of $\mathcal{M}^{G}_{\text{H}}(M_2)$, respectively, so the time-invariant maps in $L^2\mathcal{M}^{G_\C}_{\text{flat}}(M_2)$ that they correspond to (thus the ``hol'' label in the superscript of $\text{HSF}^{\text{hol}}_{d_q}$), are indeed isolated and non-degenerate.\footnote{The dimension $d$ of the moduli space of (anti)holomorphic maps from a genus $g$ Riemann surface to a target space $\mathcal{M}$ is given by $d = c_1(\mathcal{M}) \cdot \beta + \text{dim}\mathcal{M} (1-g)$, where $c_1$ is the first Chern class and $\beta$ is the degree of the map. In our case, $g=1$ and $c_1(\mathcal{M}) = 0$, so $d=0$. Thus, the critical point set of $V_\Sigma$ in $L^2\mathcal{M}^{G_\C}_{\text{flat}}(M_2)$ is, in our case, zero-dimensional, and is therefore made up of isolated points which, one can, up to the addition of physically-inconsequential $\mathcal{Q}$-exact terms in the action to deform $V_\Sigma$, assume to be non-degenerate.}

\subsection{An SQM in the Interval Space of the Loop Space of Flat \texorpdfstring{$G_\C$}{G-C} Connections}
\label{subsection: sqm of sigma models on interval space of loop flat gc}

Replacing the $S^1$ along the $s$-direction with an interval $I$, the action \eqref{3d sigma model action} can be expressed as
\begin{equation}
    \label{3d sigma model action replace s1}
    \begin{aligned}
      S_{\text{3d}} = \frac{1}{e^2} \int dt \int_{I \times S^1} dr ds \, \Big(
      & | \dot{Z}^i + i \partial_r Z^i + \partial_s \bar{Z}^i |^2
        + | \dot{Z}^\ib + i \partial_r Z^\ib - \partial_s \bar{Z}^\ib |^2
      \\
      & + | \dot{\bar{Z}}^\ib - i \partial_r \bar{Z}^\ib + \partial_s Z^\ib |^2
        + | \dot{\bar{Z}}^i - i \partial_r \bar{Z}^i - \partial_s Z^i |^2
      + \dots
      \Big) \, ,
    \end{aligned}
\end{equation}
where ``$\dots$'' are boundary and fermionic terms.

\bigskip\noindent\textit{A 2d Sigma Model on $\R \times I$ with Target the Loop Space of Flat $G_\C$ Connections}
\vspace*{0.5em}

\enlargethispage{5pt}
After suitable rescalings, it can be recast as\footnote{%
  Here, we have made use of Stokes' theorem and the fact that $S^1$ has no boundary to note that $\partial_s X$, $\partial_s Y$ should vanish in their integration over $S^1$.}
\begin{equation}
  \label{2d sigma model in loop space action}
  S_{\text{2d}, I \rightarrow S^1} = \frac{1}{e^2} \int dt \int_I dr \, \Big(
  | \dot{Z}^A + i \partial_r Z^A |^2
  + | \dot{Z}^{\bar{A}} + i \partial_r Z^{\bar{A}} |^2
  + | \dot{\bar{Z}}^{\bar{A}} - i \partial_r \bar{Z}^{\bar{A}} |^2
  + | \dot{\bar{Z}}^A - i \partial_r \bar{Z}^A |^2
  + \dots
  \Big)
  \, ,
\end{equation}
where $(Z^A, Z^{\bar{A}}, \bar{Z}^A, \bar{Z}^{\bar{A}})$ are the coordinates on the loop space $L\mathcal{M}^{G_\C}_{\text{flat}}(M_2)$ of maps from $S^1$ to the symplectic manifold $\mathcal{M}^{G_\C}_{\text{flat}}(M_2)$. In other words, the 3d sigma model on $\R \times I \times S^1$ with target $\mathcal{M}^{G_\C}_{\text{flat}}(M_2)$ is equivalent to a 2d sigma model on $\R \times I$ with target $L\mathcal{M}^{G_\C}_{\text{flat}}(M_2)$.

\bigskip\noindent\textit{A 2d A-model with Branes $\mathscr{L}_0$ and $\mathscr{L}_1$}
\vspace*{0.5em}

The 2d sigma model describes open strings with worldsheet $\R \times I$ propagating in $L\mathcal{M}^{G_\C}_{\text{flat}}(M_2)$, ending on branes $\mathscr{L}_0$ and $\mathscr{L}_1$. Let $w = \hlf (t + ir)$. Then, its BPS equations can be read off from \eqref{2d sigma model in loop space action} and expressed as
\begin{equation}
    \label{2d sigma R x I - holomorphic maps BPS I}
    \begin{aligned}
        \partial_\wb Z^A &= 0
        \, , &
        \partial_\wb Z^{\bar{A}} &= 0
        \, , \\
        \partial_w \bar{Z}^{\bar{A}} &= 0
        \, , &
        \partial_w \bar{Z}^A &= 0
        \, .
    \end{aligned}
\end{equation}
In other words, the 2d sigma model is characterized by holomorphic maps of $\R \times I$ into $L\mathcal{M}^{G_\C}_{\text{flat}}(M_2)$. As such, our 2d sigma model is an A-model, and $\mathscr{L}_0$ and $\mathscr{L}_1$ are A-branes in $L\mathcal{M}^{G_\C}_{\text{flat}}(M_2)$, where $L\mathcal{M}^{G_\C}_{\text{flat}}(M_2)$ can also be regarded as $L\mathcal{M}^{G}_{\text{H}}(M_2)$ in complex structure $J$.\footnote{To understand this claim, first, note that the subspace of constant maps in $L\mathcal{M}^{G}_{\text{H}}(M_2)$ is actually  $\mathcal{M}^{G}_{\text{H}}(M_2)$, so $L\mathcal{M}^{G}_{\text{H}}(M_2)$ would necessarily inherit the hyperk\"{a}hler structure of $\mathcal{M}^{G}_{\text{H}}(M_2)$, i.e., $L\mathcal{M}^{G}_{\text{H}}(M_2)$ can also be in complex structure $I$, $J$, or $K$. Second, $\mathcal{M}^{G}_{\text{H}}(M_2)$ in complex structure $J$ or $K$ is $\mathcal{M}^{G_\C}_{\text{flat}}(M_2)$, and in our case where the holomorphic coordinates of $\mathcal{M}^{G_\C}_{\text{flat}}(M_2)$ are the $Z$'s, it would mean that $\mathcal{M}^{G_\C}_{\text{flat}}(M_2)$ results from $\mathcal{M}^{G}_{\text{H}}(M_2)$ being in complex structure $J$. Altogether, this means that we can, in our case, regard $L\mathcal{M}^{G_\C}_{\text{flat}}(M_2)$ as $L\mathcal{M}^{G}_{\text{H}}(M_2)$ in complex structure $J$. \label{footnote:LM in complex structure IJK}
\label{footnote: LM-flat as complex J}} $\mathscr{L}_0$ and $\mathscr{L}_1$ are therefore necessarily of type $(*, \text{A}, *)$ in complex structure $(I,J,K)$.

\bigskip\noindent\textit{An SQM in the Interval Space of the Loop Space of Flat $G_\C$ Connections}
\vspace*{0.5em}

Just as in \S\ref{subsection: 5d theory/5d to sqm}, the equivalent SQM action can be obtained from \eqref{2d sigma model in loop space action} as
\begin{equation}
    \label{sqm action on interval space to loop space of higgs}
    \begin{aligned}
        S_{\text{SQM},I \rightarrow S^1} = \frac{1}{e^2} \int dt \bigg(
        & \left| \dot{Z}^\alpha + g^{\alpha\bar{\beta}}_{I \rightarrow S^1} \left( \frac{\partial V_{I \rightarrow S^1}}{\partial Z^\beta} \right)^* \right|^2
        +  \left| \dot{Z}^{\bar{\alpha}} + g^{\bar{\alpha}\beta}_{I \rightarrow S^1} \left( \frac{\partial V_{I \rightarrow S^1}}{\partial Z^{\bar{\beta}}} \right)^* \right|^2 \\
        & +  \left| \dot{\bar{Z}}^{\bar{\alpha}} + g^{\bar{\alpha}\beta}_{I \rightarrow S^1} \left( \frac{\partial V_{I \rightarrow S^1}}{\partial \bar{Z}^{\bar{\beta}}} \right)^* \right|^2
        +  \left| \dot{\bar{Z}}^\alpha + g^{\alpha\bar{\beta}}_{I \rightarrow S^1} \left( \frac{\partial V_{I \rightarrow S^1}}{\partial \bar{Z}^\beta} \right)^* \right|^2
          + \dots
          \bigg)
          \, ,
    \end{aligned}
\end{equation}
where $(Z^\alpha, Z^{\bar{\alpha}}, \bar{Z}^\alpha, \bar{Z}^{\bar{\alpha}})$ and $g^{\alpha\bar{\beta}}_{I \rightarrow S^1}$ are the coordinates and metric on the interval space $\mathcal{M}(\mathscr{L}_0, \mathscr{L}_1)_{L\mathcal{M}^{G_\C, M_2}_{\text{flat}}}$ of smooth trajectories from $\mathscr{L}_0$ to $\mathscr{L}_1$ in $L\mathcal{M}^{G_\C}_{\text{flat}}(M_2)$, respectively; and $V_{I \rightarrow S^1}(Z, \bar{Z})$ is the potential function. In other words, we equivalently have an SQM in $\mathcal{M}(\mathscr{L}_0, \mathscr{L}_1)_{L\mathcal{M}^{G_\C, M_2}_{\text{flat}}}$.

\subsection{A Symplectic Intersection Floer Homology in the Loop Space of Flat \texorpdfstring{$G_\C$}{G-C} Connections}
\label{subsection: HSF in loop space of flat gc}

Since the resulting 2d A-model on $\R \times I$ with target $L\mathcal{M}^{G_\C}_{\text{flat}}(M_2)$ and action \eqref{2d sigma model in loop space action}, can be interpreted as an SQM in $\mathcal{M}(\mathscr{L}_0, \mathscr{L}_1)_{L\mathcal{M}^{G_\C, M_2}_{\text{flat}}}$ with action \eqref{sqm action on interval space to loop space of higgs}, its partition function can be expressed as
\begin{equation}
    \label{2d partition symplectic floer functional to loop space}
    \mathcal{Z}_{\text{HW},\R \times I \rightarrow S^1}(G) = \sum_u  {\cal F}^{u}_{L\mathcal{M}^{G_\C, M_2}_{\text{flat}}} \, ,
\end{equation}
where the $u^{\text{th}}$ contribution ${\cal F}^{u}_{L\mathcal{M}^{G_\C, M_2}_{\text{flat}}}$ to the partition function, is in the $\mathcal{Q}$-cohomology, and the summation in `$u$' is over all isolated and non-degenerate configurations on $I$ in $\mathcal{M}(\mathscr{L}_0, \mathscr{L}_1)_{L\mathcal{M}^{G_\C, M_2}_{\text{flat}}}$ that the equivalent SQM localizes onto.\footnote{That the configurations on $I$ will be isolated and non-degenerate can be justified as follows. As we will show, the aforementioned configurations on $I$ are constant paths which correspond to the intersection points of ${\mathscr L}_0$ and ${\mathscr L}_1$. These points can always be made isolated and non-degenerate by adding physically-inconsequential $\mathcal{Q}$-exact terms to the SQM action which will (1) correspond to a deformation of the 2d sigma model worldsheet such that the ${\mathscr L}_0$ and ${\mathscr L}_1$ branes can be moved to intersect only at isolated points, and (2) deform the SQM potential accordingly such that its critical points will be non-degenerate.\label{footnote: isolation and non-degen of intersection points in loop space}} Let us now determine what ${\cal F}^{u}_{L\mathcal{M}^{G_\C, M_2}_{\text{flat}}}$ is.

To this end, first note that the BPS equations of the 2d A-model in \eqref{2d sigma R x I - holomorphic maps BPS I} can also be written as
\begin{equation}
    \label{2d sigma R x I to loops flow equation}
    \begin{aligned}
        \frac{\partial Z^A}{\partial t} + i \frac{\partial Z^A}{\partial r} &= 0
        \, , \qquad &
        \frac{\partial Z^{\bar{A}}}{\partial t} + i \frac{\partial Z^{\bar{A}}}{\partial r} &= 0
        \, , \\
        \frac{\partial \bar{Z}^{\bar{A}}}{\partial t} - i \frac{\partial \bar{Z}^{\bar{A}}}{\partial r} &= 0
        \, , \qquad &
        \frac{\partial \bar{Z}^A}{\partial t} - i \frac{\partial \bar{Z}^A}{\partial r} &= 0
        \, .
    \end{aligned}
\end{equation}
By the squaring argument \cite{blau1993topological} applied to \eqref{sqm action on interval space to loop space of higgs}, we find that its equivalent SQM  will localize onto configurations that set the LHS and RHS of the expression within the squared terms therein \emph{simultaneously} to zero. In other words, its equivalent SQM will localize onto time-invariant critical points of $V_{I \rightarrow S^1}$ that correspond to 2d A-model configurations that obey
\begin{equation}
    \begin{aligned}
        \frac{\partial Z^A}{\partial r} &= 0
        \, , \qquad &
        \frac{\partial Z^{\bar{A}}}{\partial r} &= 0
        \, , \\
        \frac{\partial \bar{Z}^{\bar{A}}}{\partial r} &= 0
        \, , \qquad &
        \frac{\partial \bar{Z}^A}{\partial r} &= 0
        \, .
    \end{aligned}
\end{equation}
These are constant intervals, which correspond to stationary trajectories in $\mathcal{M}(\mathscr{L}_0, \mathscr{L}_1)_{L\mathcal{M}^{G_\C, M_2}_{\text{flat}}}$, i.e., intersection points of $\mathscr{L}_0$ and $\mathscr{L}_1$.

Second, via a similar analysis to that in \S\ref{subsection: HW floer on M4}, with \eqref{sqm action on interval space to loop space of higgs} as the action for the equivalent SQM model, we thus find that we can also express \eqref{2d partition symplectic floer functional to loop space} as
\begin{equation}\label{interval to loop space floer classes}
  \boxed{
  \mathcal{Z}_{\text{HW},\R \times I \rightarrow S^1}(G)
    = \sum_u  {\cal F}^{u}_{L\mathcal{M}^{G_\C, M_2}_{\text{flat}}}
    = \sum_u \text{HSF}^{\text{Int}}_{d_u}({L\mathcal{M}^{G_\C, M_2}_{\text{flat}}}, \mathscr{L}_0, \mathscr{L}_1)
    = \mathcal{Z}^{\text{IntSympFloer}}_{L\mathcal{M}^{G_\C, M_2}_{\text{flat}}}}
\end{equation}
where each ${\cal F}^{u}_{L\mathcal{M}^{G_\C,M_2}_{\text{flat}}}$ can be identified with $\text{HSF}^{\text{Int}}_{d_u}({L\mathcal{M}^{G_\C,M_2}_{\text{flat}}}, \mathscr{L}_0, \mathscr{L}_1)$, a symplectic intersection Floer homology class, that is generated by the intersection points of $\mathscr{L}_0$ and $\mathscr{L}_1$ in $L\mathcal{M}^{G_\C}_{\text{flat}}(M_2)$, where its degree $d_u$ is counted by the Floer differential realized as the flow lines of the SQM whose gradient flow equations are the expression within the squared terms in \eqref{sqm action on interval space to loop space of higgs} set to zero.

\subsection{An SQM in the Interval Space of the Path Space of \texorpdfstring{$\theta$}{theta}-Hitchin Space}
\label{section: sqm of sigma models on interval space of path higgs}

\bigskip\noindent\textit{$\theta$-deformed 3d Sigma Model BPS Equations}
\vspace*{0.5em}

Let us further relabel the coordinate $r$ as $\tau$. By replacing the $S^1$ along the $\tau$-direction with $\R$, and the remaining $S^1$ along the $s$-direction with $I$, we will have a 3d sigma model on $\R \times I \times \R$, i.e., $I \times \R^2$. For later convenience, we shall take $\tau$ to be the temporal direction (instead of $t$).

Notice that we can make use of the rotational symmetry of $\R^2$ to rotate the 3d sigma model about $I$ by an angle $\theta$. The BPS equations \eqref{3d sigma model BPS equations} (which continue to apply here even though we have replaced $S^1 \times S^1$ with $\R \times I$) then become
\begin{equation}
    \label{3d sigma model BPS equations theta deformed}
    \begin{aligned}
        \partial_\tau X^i &= - \partial_t Y^i + i  \partial_s\big( X^i \cos(\theta) +  Y^i \sin(\theta) \big) \, , \\
        \partial_\tau Y^i &= \partial_t X^i - i  \partial_s\big(Y^i \cos(\theta) - X^i \sin(\theta) \big) \, , \\
        \partial_\tau {X}^\ib &= - \partial_t {Y}^\ib - i \partial_s\big( {X}^\ib \cos(\theta) +  {Y}^\ib \sin(\theta) \big) \, , \\
        \partial_\tau {Y}^\ib &= \partial_t {X}^\ib + i \partial_s\big( \bar{Y}^\ib \cos(\theta) - {X}^\ib \sin(\theta) \big) \, ,
    \end{aligned}
\end{equation}
which can be viewed as $\theta$-deformed 3d sigma model BPS equations.

\bigskip\noindent\textit{$\theta$, and Flat $G_\C$ Connections or Higgs Bundles on $M_2$}
\vspace*{0.5em}

To determine what role $\theta$ plays, let us take a few specific values of $\theta$ and discuss the consequences. When $\theta = \pi/2$, \eqref{3d sigma model BPS equations theta deformed} becomes
\begin{equation}
    \label{3d sigma model BPS equations theta=pi/2}
    \begin{aligned}
        \partial_\tau X^i &= - \partial_t Y^i + i \partial_s Y^i \, ,
        & \quad
        \partial_\tau Y^i &= \partial_t X^i + i \partial_s X^i \, ,
        \\
        \partial_\tau {X}^\ib &= - \partial_t {Y}^\ib - i \partial_s {Y}^\ib \, ,
        & \quad
        \partial_\tau {Y}^\ib &= \partial_t {X}^\ib - i \partial_s {X}^\ib \, ,
    \end{aligned}
\end{equation}
which is similar in form to \eqref{3d sigma model BPS equations} (where the time-derivative of the $(X, Y)$ fields is equal to the spatial derivatives of the $(Y, X)$ fields), whence it can be determined that the target space of the 3d sigma model is once again $\mathcal{M}^{G_\C}_{\text{flat}}(M_2)$.\footnote{%
The same reasoning in footnote \ref{footnote: determining 3d sigma model complex structure} applies here.}

When $\theta = \pi$, \eqref{3d sigma model BPS equations theta deformed} becomes
\begin{equation}
    \label{3d sigma model BPS equations theta=pi}
    \begin{aligned}
        \partial_\tau X^i &= - \partial_t Y^i - i \partial_s X^i \, ,
        & \quad
        \partial_\tau Y^i &= \partial_t X^i + i \partial_s Y^i \, ,
        \\
        \partial_\tau {X}^\ib &= - \partial_t {Y}^\ib + i \partial_s {X}^\ib \, ,
        & \quad
        \partial_\tau {Y}^\ib &= \partial_t X^i - i \partial_s {Y}^\ib \, ,
    \end{aligned}
\end{equation}
for which it can be determined that the symplectic target space of the 3d sigma model is $ \mathcal{M}^G_{\text{Higgs}}(M_2)$, the moduli space of Higgs bundles of $G$ on $M_2$.\footnote{%
To see this, we can adapt the reasoning employed in footnote \ref{footnote: determining 3d sigma model complex structure}, where the BPS equations of the resulting 2d sigma model will tell us that the holomorphic coordinates of $\mathcal{M}^G_{\text{H}}(M_2)$ are independently $X$ and $Y$, i.e., the 2d sigma model now sees $\mathcal{M}^G_{\text{H}}(M_2)$ in complex structure $I$. Thus, it can be inferred that the 3d sigma model will also see $\mathcal{M}^G_{\text{H}}(M_2)$ in complex structure $I$, whence $\mathcal{M}^G_{\text{H}}(M_2) = \mathcal{M}^G_{\text{Higgs}}(M_2)$.}

This tells us that the complex structure of the target space $\mathcal{M}^G_{\text{H}}(M_2)$
varies as we vary $\theta$. Therefore, let us henceforth denote the target space in complex structure $\mathcal{J}(\theta)$, where $\mathcal{J}(\theta)$ is a linear combination of the complex structures $I$, $J$, $K$ that depends on $\theta$, as $\mathcal{M}^G_{\text{H}, \theta}(M_2)$, the $\theta$-Hitchin moduli space of $G$ on $M_2$.

\bigskip\noindent\textit{The $\theta$-rotated 3d Sigma Model Action}
\vspace*{0.5em}

The action of the $\theta$-rotated 3d sigma model is
\begin{equation}
    \label{3d sigma model theta deformed}
    \begin{aligned}
      S_{\text{3d}}(\theta) = \frac{1}{e^2} \int_{I \times \R} d\tau ds \int_{\R} dt \, \Big(
      & | \partial_\tau{X}^i + p^i(\theta) |^2
        + | \partial_\tau{Y}^i - q^i(\theta) |^2 \\
      & + | \partial_\tau{{X}}^\ib + \pb^\ib(\theta) |^2
        + | \partial_\tau{{Y}}^\ib - \qb^\ib(\theta) |^2
        + \dots
        \Big) \, ,
    \end{aligned}
\end{equation}
where
\begin{equation}
    \begin{aligned}
        p^i(\theta) &= \partial_t Y^i - i  \partial_s\big( X^i \cos(\theta) +  Y^i \sin(\theta) \big) \, , \\
        q^i(\theta) &= \partial_t X^i - i  \partial_s\big(Y^i \cos(\theta) - X^i \sin(\theta) \big) \, , \\
        \pb^\ib(\theta) &= \partial_t {Y}^\ib + i \partial_s\big( {X}^\ib \cos(\theta) + {Y}^\ib \sin(\theta) \big) \, , \\
        \qb^\ib(\theta) &= \partial_t {X}^\ib + i \partial_s\big( {Y}^\ib \cos(\theta) - {X}^\ib \sin(\theta) \big) \, ,
    \end{aligned}
\end{equation}
and  ``$\dots$'' refer to boundary and fermionic terms.

\bigskip\noindent\textit{A 2d Sigma Model on $\R \times I$ with Target the Path Space of $\theta$-Hitchin Space}
\vspace*{0.5em}

As before, we can recast the $\theta$-rotated 3d sigma model on $\R \times I \times \R$ as a 2d sigma model on $\R \times I$ with target the path space $\mathcal{M}\big(\R, \mathcal{M}^G_{\text{H}, \theta}(M_2)\big)$ of maps from $\R$ to $\mathcal{M}^G_{\text{H}, \theta}(M_2)$, whereby the action is
\begin{equation}
  \label{2d sigma model in path space action}
  \begin{aligned}
    S_{\text{2d},I \rightarrow \R}(\theta) = \frac{1}{e^2} \int d\tau \int_I ds \, \Big(
    & | \partial_\tau{X}^C + p^C(\theta) |^2
      + | \partial_\tau{Y}^C - q^C(\theta) |^2 \\
    & + | \partial_\tau{{X}}^{\bar{C}} + \pb^{\bar{C}}(\theta) |^2
      + | \partial_\tau{{Y}}^{\bar{C}} - \qb^{\bar{C}}(\theta) |^2
      + \dots
      \Big)
      \, ,
  \end{aligned}
\end{equation}
where $(X^C, Y^C, {X}^{\bar{C}}, {Y}^{\bar{C}})$ are the coordinates on $\mathcal{M}\big(\R, \mathcal{M}^G_{\text{H}, \theta}(M_2)\big)$, and
\begin{equation}
    \begin{aligned}
        p^C &= - i  \partial_s\big( X^C \cos(\theta) +  Y^C \sin(\theta) \big) \, , \\
        q^C &= - i  \partial_s\big(Y^C \cos(\theta) - X^C \sin(\theta) \big) \, , \\
        \pb^{\bar{C}} &= i \partial_s\big( {X}^{\bar{C}} \cos(\theta) + {Y}^{\bar{C}} \sin(\theta) \big) \, , \\
        \qb^{\bar{C}} &= i \partial_s\big( {Y}^{\bar{C}} \cos(\theta) - {X}^{\bar{C}} \sin(\theta) \big) \, .
    \end{aligned}
\end{equation}

\bigskip\noindent\textit{A 2d $\text{A}_\theta$-model with Branes $\mathscr{P}_0(\theta)$ and $\mathscr{P}_1(\theta)$}
\vspace*{0.5em}

We thus have a 2d sigma model that describes open strings with worldsheet $\R \times I$ propagating in $\mathcal{M}\big(\R, \mathcal{M}^G_{\text{H}, \theta}(M_2)\big)$, ending on branes $\mathscr{P}_0(\theta)$ and $\mathscr{P}_1(\theta)$. Let $w(\theta)$ and ${\bar w}(\theta)$ be generalized coordinates on the worldsheet such that $\partial_{w(\theta)} = \partial_\tau - i e^{i\theta} \partial_s$, and $\partial_{\wb(\theta)} = \partial_\tau + i e^{-i\theta} \partial_s$. Then, its BPS equations can be read off from \eqref{2d sigma model in path space action} and expressed as
\begin{equation}
\label{2d sigma R x I - holomorphic maps BPS II}
    \begin{aligned}
    \partial_{\wb(-\theta)} \bar{Z}^{\bar{C}} + \partial_{\wb(\theta)} Z^{\bar{C}} &= 0 \, ,
    &\quad
    \partial_{\wb(\theta)} Z^C - \partial_{\wb(-\theta)} \bar{Z}^C &= 0 \, ,
    \\
    \partial_{w(-\theta)} Z^C + \partial_{w(\theta)} \bar{Z}^C &= 0 \, ,
    &\quad
    \partial_{w(\theta)} \bar{Z}^{\bar{C}} - \partial_{w(-\theta)} Z^{\bar{C}} &= 0 \, ,
    \end{aligned}
\end{equation}
where $Z^C \coloneq X^C + iY^C$, $Z^{\bar{C}} \coloneq X^{\bar{C}} + i Y^{\bar{C}}$, $\bar{Z}^C \coloneq X^C - i Y^C$, and ${\bar Z}^{\bar{C}} \coloneq X^{\bar{C}} - iY^{\bar{C}}$.

When $\theta = 0$, \eqref{2d sigma R x I - holomorphic maps BPS II} can be written as
\begin{equation}
    \label{2d sigma R x I - holomorphic maps BPS II theta=0}
    \begin{aligned}
    \partial_{\wb} {X}^{\bar{C}} &= 0 \, ,
    &\quad
   \partial_{{\bar w}} Y^C &= 0 \, ,
    \\
\partial_{w} X^C &= 0 \, ,
    &\quad
    \partial_w {Y}^{\bar{C}} &= 0 \,  ,
       \end{aligned}
\end{equation}
where $w = w(0) = \hlf(\tau + is)$ and $\wb = \wb(0) = \hlf(\tau - is)$. In other words, the 2d sigma model is characterised by holomorphic/antiholomorphic maps of $\R \times I$ into $\mathcal{M}\big(\R, \mathcal{M}^G_{\text{Higgs}}(M_2)\big)$. That is, it is a standard A-model.

When $\theta = \pi$, \eqref{2d sigma R x I - holomorphic maps BPS II} can be written as
\begin{equation}
    \label{2d sigma R x I - holomorphic maps BPS II theta=pi}
    \begin{aligned}
    \partial_{\wb} X^C &= 0 \, ,
    &\quad
    \partial_{\wb} Y^{\bar{C}} &= 0 \, ,
    \\
    \partial_{w} X^{\bar{C}} &= 0 \, ,
    &\quad
    \partial_{w} Y^C &= 0 \, ,
    \end{aligned}
\end{equation}
where $w = \wb(\pi) = \hlf(\tau + is)$ and $\wb = w(\pi) = \hlf(\tau - is)$. In other words, the 2d sigma model is characterised by holomorphic/antiholomorphic maps of $\R \times I$ into  $\mathcal{M}\big(\R, \mathcal{M}^G_{\text{Higgs}}(M_2)\big)$. That is, it is a standard A-model.

When $\theta = \pi/2$, \eqref{2d sigma R x I - holomorphic maps BPS II} can be written as
\begin{equation}
    \label{2d sigma R x I -holomorphic maps BPS II theta=pi/2}
    \begin{aligned}
    \partial_{\wb}(X^{\bar{C}} + Y^{\bar{C}}) &= 0 \, ,
    &\quad
    \partial_{\wb}(X^C - Y^C) &= 0 \, ,
    \\
    \partial_{w}(X^C + Y^C) &= 0 \, ,
    &\quad
    \partial_{w}(X^{\bar{C}} - Y^{\bar{C}}) &= 0 \, .
    \end{aligned}
\end{equation}
In other words, the 2d sigma model is characterised by holomorphic/antiholomorphic maps of $\R \times I$ into $\mathcal{M}\big(\R, \mathcal{M}^{G_\C}_{\text{flat}}(M_2)\big)$. That is, it is a standard A-model.

As such, our 2d sigma model is a $\theta$-generalized A-model with $\theta$-generalized branes. We will henceforth call it an $\text{A}_\theta$-model with $\text{A}_\theta$-branes
$\mathscr{P}_0(\theta)$ and $\mathscr{P}_1(\theta)$ in $\mathcal{M}\big(\R, \mathcal{M}^G_{\text{H}, \theta}(M_2)\big)$, where if $\theta = 0$, $\pi$ and $\pi/2$, it is just a standard A-model with standard A-branes. From \eqref{2d sigma R x I - holomorphic maps BPS II theta=0}--\eqref{2d sigma R x I -holomorphic maps BPS II theta=pi/2}, one can see that the standard A-branes at $\theta = 0$, $\pi$ and $\pi/2$ are necessarily of type $(\text{A}, *, *)$, $(\text{A}, *, *)$ and $(*, *, \text{A})$ in complex structure $(I,J,K)$, respectively.

\bigskip\noindent\textit{An SQM in the Interval Space of the Path Space of $\theta$-Hitchin Space}
\vspace*{0.5em}

Once again, we can obtain the equivalent SQM action as
\begin{equation}
    \label{sqm action on interval space to path space of higgs}
    \begin{aligned}
        S_{\text{SQM},I \rightarrow \R} = \frac{1}{e^2} \int d\tau \bigg(
        & \left| \partial_\tau{X}^\gamma + g^{\gamma \bar{\delta}}_{I \rightarrow \R} \left( \frac{\partial V_{I \rightarrow \R}}{\partial X^\delta} \right)^* \right|^2
        +  \left| \partial_\tau{\bar{X}}^{\bar{\gamma}} + g^{\bar{\gamma} \delta}_{I \rightarrow \R} \left( \frac{\partial V_{I \rightarrow \R}}{\partial \bar{X}^{\bar{\delta}}} \right)^* \right|^2 \\
        & +  \left| \partial_\tau{Y}^\gamma + g^{\gamma \bar{\delta}}_{I \rightarrow \R} \left( \frac{\partial V_{I \rightarrow \R}}{\partial Y^\delta} \right)^* \right|^2
        +  \left| \partial_\tau{\bar{Y}}^{\bar{\gamma}} + g^{\bar{\gamma} \delta}_{I \rightarrow \R} \left( \frac{\partial V_{I \rightarrow \R}}{\partial \bar{Y}^{\bar{\delta}}} \right)^* \right|^2
        + \dots
        \bigg)
        \, ,
    \end{aligned}
\end{equation}
where $(X^\gamma, Y^\gamma, \bar{X}^{\bar{\gamma}}, \bar{Y}^{\bar{\gamma}})$  and $g^{\gamma \bar{\delta}}_{I \rightarrow \R}$ are, respectively, the coordinates and metric on the interval space $\mathcal{M}(\mathscr{P}_0, \mathscr{P}_1)_{\mathcal{M}(\R, \mathcal{M}^{G, M_2}_{\text{H}, \theta})}$ of smooth trajectories from $\mathscr{P}_0(\theta)$ to $\mathscr{P}_1(\theta)$ in $\mathcal{M}\big(\R, \mathcal{M}^G_{\text{H}, \theta}(M_2)\big)$; and $V_{I \rightarrow \R}(X^\gamma, Y^\gamma, \bar{X}^{\bar{\gamma}}, \bar{Y}^{\bar{\gamma}}, \theta)$ is the potential function.

In other words, we equivalently have an SQM in $\mathcal{M}(\mathscr{P}_0, \mathscr{P}_1)_{\mathcal{M}(\R, \mathcal{M}^{G, M_2}_{\text{H}, \theta})}$.

\subsection{A Symplectic Intersection Floer Homology in the Path Space of \texorpdfstring{$\theta$}{theta}-Hitchin Space}
\label{subsection: HSF in path space of theta-Hitchin}

Likewise, since the 2d A$_\theta$-model on $\R \times I$ with target $\mathcal{M}(\R, \mathcal{M}^G_{\text{H}, \theta}(M_2))$ can be interpreted as an SQM in $\mathcal{M}(\mathscr{P}_0, \mathscr{P}_1)_{\mathcal{M}(\R, \mathcal{M}^{G, M_2}_{\text{H}, \theta})}$ with action \eqref{sqm action on interval space to path space of higgs}, its partition function can be expressed as
\begin{equation}
    \label{2d partition symplectic floer functional to path space}
    \mathcal{Z}_{\text{HW}^\theta,\R \times I \rightarrow \R}(G) = \sum_v {\cal F}^{v}_{\mathcal{M}(\R, \mathcal{M}^{G, M_2}_{\text{H}, \theta})} \, ,
\end{equation}
where the $v^{\text{th}}$ contribution ${\cal F}^{v}_{\mathcal{M}(\R, \mathcal{M}^{G,M_2}_{\text{H}, \theta})}$ to the partition function, is in the $\mathcal{Q}$-cohomology, and the summation in `$v$' is over all isolated and non-degenerate configurations on $I$ in $\mathcal{M}(\mathscr{P}_0, \mathscr{P}_1)_{\mathcal{M}(\R, \mathcal{M}^{G, M_2}_{\text{H}, \theta})}$ that the equivalent SQM localizes onto.\footnote{The same reasoning as footnote~\ref{footnote: isolation and non-degen of intersection points in loop space} applies here.} Let us determine what ${\cal F}^{v}_{\mathcal{M}(\R, \mathcal{M}^{G,M_2}_{\text{H}, \theta})}$ is.

To this end, first note that the BPS equations of the 2d A$_\theta$-model in \eqref{2d sigma R x I - holomorphic maps BPS II} can also be written as
\begin{equation}
    \label{2d sigma R x I to loops flow equation theta deformed}
    \begin{aligned}
        \frac{\partial X^C}{\partial \tau} - i \frac{\partial}{\partial s} \big( X^C \cos(\theta) +  Y^C \sin(\theta) \big) &= 0
        \, , \quad &
        \frac{\partial Y^C}{\partial \tau} + i \frac{\partial}{\partial s}\big(Y^C \cos(\theta) - X^C \sin(\theta) \big) &= 0
        \, , \\
        \frac{\partial \bar{X}^{\bar{C}}}{\partial \tau} + i \frac{\partial}{\partial s} \big( \bar{X}^{\bar{C}} \cos(\theta) + \bar{Y}^{\bar{C}} \sin(\theta) \big) &= 0
        \, , \quad &
        \frac{\partial \bar{Y}^{\bar{C}}}{\partial \tau} - i \frac{\partial}{\partial s} \big( \bar{Y}^{\bar{C}} \cos(\theta) - \bar{X}^{\bar{C}} \sin(\theta) \big) &= 0
        \, .
    \end{aligned}
\end{equation}
By the squaring argument \cite{blau1993topological} applied to \eqref{sqm action on interval space to path space of higgs}, we find that its equivalent SQM  will localize onto configurations that set the LHS and RHS of the expression within the squared terms therein \emph{simultaneously} to zero. In other words, its equivalent SQM will localize onto time-invariant critical points of $V_{I \rightarrow \R}$ that correspond to 2d A$_\theta$-model configurations that obey
\begin{equation}
    \begin{aligned}
        \frac{\partial}{\partial s} \big( X^C \cos(\theta) +  Y^C \sin(\theta) \big) &= 0
        \, , \quad &
        \frac{\partial}{\partial s}\big(Y^C \cos(\theta) - X^C \sin(\theta) \big) &= 0
        \, , \\
        \frac{\partial}{\partial s} \big( \bar{X}^{\bar{C}} \cos(\theta) + \bar{Y}^{\bar{C}} \sin(\theta) \big) &= 0
        \, , \quad &
        \frac{\partial}{\partial s} \big( \bar{Y}^{\bar{C}} \cos(\theta) - \bar{X}^{\bar{C}} \sin(\theta) \big) &= 0
        \, .
    \end{aligned}
\end{equation}
These are constant intervals, which correspond to stationary trajectories in $\mathcal{M}(\mathscr{P}_0, \mathscr{P}_1)_{\mathcal{M}(\R, \mathcal{M}^{G, M_2}_{\text{H}, \theta})}$, i.e., intersection points of $\mathscr{P}_0(\theta)$ and $\mathscr{P}_1(\theta)$.

Second, via a similar analysis to that in \S\ref{subsection: HW floer on M4}, with \eqref{sqm action on interval space to path space of higgs} as the action for the equivalent SQM model, we thus find that we can also express \eqref{2d partition symplectic floer functional to path space} as
\begin{equation}\label{interval to path space floer classes}
  \boxed{
  \mathcal{Z}_{\text{HW}^\theta,\R \times I \rightarrow \R}(G)
    = \sum_v  {\cal F}^{v}_{\mathcal{M}(\R, \mathcal{M}^{G,M_2}_{\text{H}, \theta})}
    = \sum_v \text{HSF}^{\text{Int}}_{d_v}(\mathcal{M}\big(\R, \mathcal{M}^{G,M_2}_{\text{H}, \theta}\big) , \mathscr{P}_0, \mathscr{P}_1)
    = \mathcal{Z}^{\text{IntSympFloer}}_{\mathcal{M}\big(\R, \mathcal{M}^{G,M_2}_{\text{H}, \theta}\big)}}
\end{equation}
where each ${\cal F}^{v}_{\mathcal{M}(\R, \mathcal{M}^{G,M_2}_{\text{H}, \theta})}$ can be identified with a symplectic intersection Floer homology class $\text{HSF}^{\text{Int}}_{d_v}(\mathcal{M}\big(\R, \mathcal{M}^{G,M_2}_{\text{H}, \theta}\big) , \mathscr{P}_0, \mathscr{P}_1)$ that is generated by the intersection points of $\mathscr{P}_0(\theta)$ and $\mathscr{P}_1(\theta)$ in $\mathcal{M}(\R, \mathcal{M}^G_{\text{H}, \theta}(M_2))$, where its degree $d_v$ is counted by the Floer differential realized as the flow lines of the SQM whose gradient flow equations are the expression within the squared terms in \eqref{sqm action on interval space to path space of higgs} set to zero.

\section{Haydys-Witten Atiyah-Floer Correspondences}
\label{section: hw atiya-floer correspondence}

In this section, we consider HW theory on $M_5 = M_3 \times S^1 \times \R$. We decompose $M_5$ into two pieces by performing a Heegaard split of $M_3$ along a Riemann surface $\Sigma$. Then, by the topological invariance of HW theory under a shrinking of $\Sigma$, and the results of the previous section, we will be able to derive an HW Atiyah-Floer correspondence between HW Floer homology of $M_3 \times S^1$ and symplectic intersection Floer homology of the loop space of Higgs bundles, which, in turn, implies its 4d version between 4d-HW Floer homology of $M_3$ and symplectic intersection Floer homology of the space of Higgs bundles. We will also derive a duality between our 4d-HW Atiyah-Floer correspondence and the VW Atiyah-Floer correspondence in~\cite{ong2022vafa}.

\subsection{Splitting HW Theory on \texorpdfstring{$M_5 = M_3 \times M_1 \times \R$}{M5 = M3 x M1 x R}}
\label{Heegaard Split of M5}

\bigskip\noindent\textit{Splitting HW Theory on $M_5 = M_3 \times M_1 \times \R$ via Heegaard Splitting $M_3$}
\vspace*{0.5em}

Let us perform a Heegaard split of $M_3 = M_3' \bigcup_\Sigma M_3''$ along $\Sigma$, as shown in Fig.~\ref{fig:heegaard split m3}, whence we can view $M_3'$ and $M_3''$ as nontrivial fibrations of $\Sigma$ over intervals $I'$ and $I''$, where $\Sigma$ goes to zero size at one end of each of the intervals. The metric on $M_3'$ and $M_3''$ can then be written as
\begin{equation}
    ds^2_{\text{3d}} = (dx^B)^2 + f(x^B)(g_\Sigma)_{ab}dx^a dx^b \, ,
\end{equation}
where $x^{a,b}$ are coordinates on the Riemann surface $\Sigma$; $x^B$ is a coordinate on $I'$ and $I''$; and $f(x^B)$ is a scalar function along $I'$ and $I''$. Therefore,  $M_5 = M_5' \bigcup_\mathscr B M_5''$, where $M_5' = M_3' \times M_1 \times \R$, $M_5'' = M_3'' \times M_1 \times \R$, and $\mathscr B$ is their common boundary. This is illustrated in Fig.~\ref{fig:heegaard split m5}.

\usetikzlibrary{patterns.meta}
\begin{figure}
    \centering
    \begin{subfigure}{0.45\textwidth}
        \centering
        \begin{tikzpicture}[auto]
            \draw[blue, pattern={Lines[angle=45,distance=4pt]},pattern color=blue, thick] (0,0) ellipse (0.4 and 1);
            \draw (0,1) arc(90:270:2 and 1);
            \draw[fill=white,draw=none] (0,0) circle[radius=0.3] node [scale=1.5] (sigma-left) {$\Sigma$};
            \node [left of=sigma-left,scale=1.5] {$M_3'$};
            \draw[blue, thick] (1,1) arc(90:270:0.4 and 1);
            \draw (1,-1) arc(270:450:2 and 1);
            \node [right=0.5cm of sigma-left,scale=1.5] {$M_3''$};
            \node [below=1.3cm of sigma-left] {};
        \end{tikzpicture}
        \caption{$M_3$ as a connected sum of three-manifolds $M_3'$ and $M_3''$ along the Riemann surface $\Sigma$.}
        \label{fig:heegaard split m3}
    \end{subfigure}
    \hfill
    \begin{subfigure}{0.45\textwidth}
        \centering
        \begin{tikzpicture}[auto]
            \def \eliA {1.5} 
            \def \eliB {0.4} 
            \def \eliAngle {15} 
            \def \faceLength {1.2} 
            \def \sepLength {0.6} 
            \def \nodeScale {1.5} 
            \draw ({\eliA*sin(\eliAngle) - \sepLength}, {\eliB*cos(\eliAngle) + \faceLength})
            arc({90 - \eliAngle}:{270 - \eliAngle}:{\eliA} and {\eliB});
            \draw[blue, pattern={Lines[angle=45,distance=4pt]},pattern color=blue, thick]
                ({\eliA*sin(\eliAngle) - \sepLength}, {\eliB*cos(\eliAngle) + \faceLength})
                -- ({-\eliA*sin(\eliAngle) - \sepLength}, {-\eliB*cos(\eliAngle) + \faceLength})
                -- ({-\eliA*sin(\eliAngle) - \sepLength}, {-\eliB*cos(\eliAngle) - \faceLength})
                -- ({\eliA*sin(\eliAngle) - \sepLength}, {\eliB*cos(\eliAngle) - \faceLength})
                -- ({\eliA*sin(\eliAngle) - \sepLength}, {\eliB*cos(\eliAngle) + \faceLength});
            \draw[fill=white,draw=none] ({-\sepLength},0) circle[radius=0.3] node [scale=\nodeScale] (sigma-left) {$\mathscr{B}$};
            \draw ({-\eliA - \sepLength}, {\faceLength})
            -- ({-\eliA - \sepLength}, {-\faceLength});
            \draw ({-\eliA - \sepLength}, {-\faceLength})
            arc(180:{270 - \eliAngle}:{\eliA} and {\eliB});
            \node [scale=\nodeScale] at ({-\eliA*(1 + sin(\eliAngle))/2 - \sepLength}, 0) (M5-prime) {$M_5'$};
            \draw ({\eliA*sin(\eliAngle) + \sepLength}, {\eliB*cos(\eliAngle) + \faceLength})
            arc({360 + 90 - \eliAngle}:{270 - \eliAngle}:{\eliA} and {\eliB});
            \draw[blue, thick] ({\eliA*sin(\eliAngle) + \sepLength}, {\eliB*cos(\eliAngle) + \faceLength})
            -- ({-\eliA*sin(\eliAngle) + \sepLength}, {-\eliB*cos(\eliAngle) + \faceLength})
            -- ({-\eliA*sin(\eliAngle) + \sepLength}, {-\eliB*cos(\eliAngle) - \faceLength});
            \draw ({\eliA + \sepLength}, {\faceLength})
            -- ({\eliA + \sepLength}, {-\faceLength});
            \draw ({-\eliA*sin(\eliAngle) + \sepLength}, {-\eliB*cos(\eliAngle) - \faceLength}) arc({270 - \eliAngle}:360:{\eliA} and {\eliB});
            \node [scale=\nodeScale] at ({\eliA*(1 - sin(\eliAngle))/2 + \sepLength}, 0) (M5-primeprime) {$M_5''$};
            \node [below=1.5cm of M5-primeprime] {};
        \end{tikzpicture}
        \caption{$M_5$ splits into five-manifolds $M_5'$ and $M_5''$ along their common boundary $\mathscr{B}$.}
        \label{fig:heegaard split m5}
    \end{subfigure}
    \caption[Caption for Heegaard split figure]{Heegaard splits of $M_3$ and $M_5$.\footnotemark}
    \label{fig:Heegaard split figure}
\end{figure}
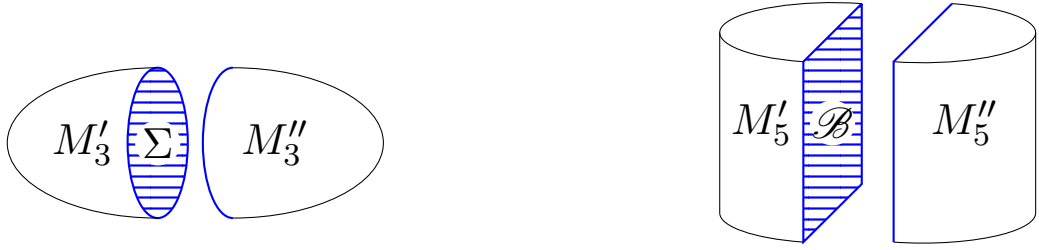

Because of the topological invariance of HW theory, we are free to perform a Weyl rescaling of the corresponding Heegaard split metrics on $M_5'$ and $M_5''$ to
\begin{equation}
    ds^2_{\text{5d}} = \frac{1}{f(x^B)} \left[ (dx^A)^2 + (ds)^2 + (dx^B)^2 \right] + (g_\Sigma)_{ab}dx^a dx^b \, ,
\end{equation}
where $x^A$ and $s$ are the coordinates on $\R$ and $M_1$, respectively. The prefactor of $f(x^B)^{-1}$ is effectively a scaling factor on $\R \times M_1 \times I'$ and $\R \times M_1 \times I''$, whence their topologies are left unchanged. We can thus regard $M_5 = M_5' \bigcup_\mathscr B M_5'' = (\R \times I' \times M_1 \times \Sigma) \bigcup_{\mathscr B} (\R \times I'' \times M_1 \times \Sigma)$, where $\mathscr B =  \Sigma \times M_1 \times \R$.
\footnotetext{This diagram is adapted from \cite[Fig.~2]{ong2022vafa}.}

Hence, HW theory on $M_5$ can be regarded as a union of an HW theory on $\R \times I' \times M_1 \times \Sigma$ and an HW theory on $\R \times I'' \times M_1 \times \Sigma$ along their common boundary $\mathscr B  =  \Sigma \times M_1 \times \R$.

\subsection{An HW Atiyah-Floer Correspondence of \texorpdfstring{$M_3 \times S^1$}{M3 x S1}}

\bigskip\noindent\textit{HW Theory on $M_5'$($M_5''$)  as a 2d A-model on $\R \times I'$($I''$) with Branes $\mathscr L_0, \mathscr L_1'$ ($\mathscr L_0'', \mathscr L_1$) in $L\mathcal{M}^{G_\C}_{\text{flat}}(\Sigma)$}
\vspace*{0.5em}

Let $M_1 = S^1$. Notice that topological HW theory on $M_5' = \R \times I' \times S^1 \times \Sigma$ and $M_5'' = \R \times I'' \times S^1 \times \Sigma$, when $\Sigma$ is shrunken, can be interpreted as a 3d sigma model on $\R \times I' \times S^1$ and $\R \times I'' \times S^1$, respectively, whose action is \eqref{3d sigma model action replace s1}. In turn, these 3d sigma models can be interpreted as 2d sigma models with action \eqref{2d sigma model in loop space action}.

That is, HW theory on $M_5' = \R \times I' \times S^1 \times \Sigma$ and $M_5'' = \R \times I'' \times S^1 \times \Sigma$ can also be interpreted as a 2d A-model on $\R \times I'$ and $\R \times I''$ with branes  $(\mathscr L_0, \mathscr L_1')$ and ($\mathscr L_0'', \mathscr L_1$) in $L\mathcal{M}^{G_\C}_{\text{flat}}(\Sigma)$, respectively, whose action is \eqref{2d sigma model in loop space action}.

\bigskip\noindent\textit{An HW Atiyah-Floer Correspondence of $M_3 \times S^1$}
\vspace*{0.5em}

Therefore, the union of an HW theory on $M_5'$ and an HW theory on $M_5''$ along their common boundary $\mathscr B$ to get an HW theory on $M_5$, can be interpreted as the union of a 2d A-model on $\R \times I'$  with branes $(\mathscr L_0, \mathscr L_1')$ and a 2d A-model on $\R \times I''$  with branes $(\mathscr L_0'', \mathscr L_1)$ along their common boundary $\mathscr R = \mathscr L_1' = \mathscr L_0''$ to get a 2d A-model on $\R \times I$ with branes $(\mathscr L_0, \mathscr L_1)$. This is illustrated in Fig.~\ref{fig:gluing 2d sigma models}.

\tikzset{
    box/.style={rectangle, text centered, minimum height=10em,text width=8mm,draw,fill=gray!40},
    shortwave/.style={*-*,thick,decorate,decoration={snake,amplitude=5pt,segment length=1.3cm}},
    longwave/.style={*-*,thick,decorate,decoration={snake,amplitude=5pt,segment length=1.7cm}}
}
\usetikzlibrary{arrows,arrows.meta,decorations.pathmorphing}
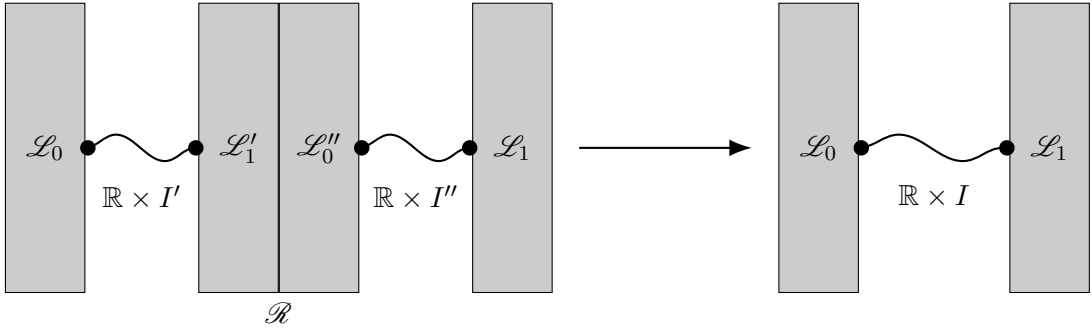
\begin{figure}
    \centering
    \begin{tikzpicture}[auto,shorten >=-2pt,shorten <=-2pt]
        \node [box] (ori-l0) {$\mathscr{L}_0$};
        \node [box, right=1.5cm of ori-l0] (ori-l1-prime) {$\mathscr{L}_1'$};
        \node [box, right=0cm of ori-l1-prime] (ori-l0-prime) {$\mathscr{L}_0''$};
        \node [box, right=1.5cm of ori-l0-prime] (ori-l1) {$\mathscr{L}_1$};
        \path (ori-l1-prime) -- (ori-l0-prime) coordinate[midway] (R-aux);
        \node [below=5em of R-aux] (R) {$\mathscr{R}$};
        \draw [shortwave] (ori-l0) -- node[below=10pt] {$\R \times I'$} (ori-l1-prime);
        \draw [shortwave] (ori-l0-prime) -- node[below=10pt] {$\R \times I''$} (ori-l1);
        \node [box, right=3cm of ori-l1] (fin-l0) {$\mathscr{L}_0$};
        \node [box, right=2cm of fin-l0] (fin-l1) {$\mathscr{L}_1$};
        \draw [longwave] (fin-l0) -- node[below=10pt] {$\R \times I$} (fin-l1);
        \draw [-{Latex[length=3mm]}, thick, shorten >=10pt, shorten <=10pt] (ori-l1) -- (fin-l0);
    \end{tikzpicture}
    \caption{Union of 2d A-models along their common boundary $\mathscr{R}$.}
    \label{fig:gluing 2d sigma models}
\end{figure}

\enlargethispage{5pt}
In other words, $G$-type HW theory on $M_3 \times S^1 \times \R$ can be interpreted as a 2d A-model on $\R \times I$ with branes $(\mathscr L_0, \mathscr L_1)$ in $L\mathcal{M}^{G_\C}_{\text{flat}}(\Sigma)$ whose action is \eqref{2d sigma model in loop space action}. This means that we can equate their partition functions in \eqref{5d partition function floer} and \eqref{interval to loop space floer classes}, respectively, to get
\begin{equation}
\label{Atiyah-Floer partition equality S1}
 {   \sum_k  \text{HF}^{\text{HW}}_{d_k}(M_3 \times S^1, G) = \sum_u \text{HSF}^{\text{Int}}_{d_u}({L\mathcal{M}^{G_\C,\Sigma}_{\text{flat}}}, \mathscr{L}_0, \mathscr{L}_1)}\,,
\end{equation}
where $\mathscr{L}_0, \mathscr{L}_1$ are necessarily isotropic-coisotropic of $(\text{A}, \text{A}, \text{B})$ type in complex structure $(I, J, K)$.\footnote{%
In $\S$\ref{subsection: sqm of sigma models on interval space of loop flat gc}, we saw that the branes $\mathscr{L}_{0,1}$ must be of type  $(*, \text{A}, *)$ in complex structure $(I,J,K)$; this means they can either be $(\text{A}, \text{A}, \text{B})$ or $(\text{B}, \text{A}, \text{A})$ branes (as $(\text{B}, \text{A}, \text{B})$ and $(\text{A}, \text{A}, \text{A})$ branes are not possible~\cite[footnote 24]{kapustin2006electmagnet}). Since in this case, we require that $\mathscr{L}_{0,1}$, when restricted to $\mathcal{M}^{G_\C, \Sigma}_{\text{flat}}$, ought to correspond to branes that wrap $\mathcal{M}^{G_\C, \Sigma}_{\text{flat}}$ where flat $G_\C$ connections on $\Sigma$ are extendable to $M_3$, i.e., the branes must be Lagrangian~\cite{kapustin2008notequant} and therefore isotropic-coisotropic~\cite{mcduff2017introdsympltopol}, it would mean that $\mathscr{L}_{0,1}$ must also be isotropic-coisotropic. Although $(\text{A}, \text{A}, \text{B})$  and $(\text{B}, \text{A}, \text{A})$ branes are both isotropic-coisotropic,  we choose $\mathscr{L}_{0,1}$ to be $(\text{A}, \text{A}, \text{B})$ branes for later convenience.}

We would now like to ascertain if there is a one-to-one correspondence between $(k, d_k)$ and $(u, d_u)$, which is to ask if there is a degree-to-degree isomorphism between the HW Floer homology and the symplectic intersection Floer homology.

To ascertain if there is a one-to-one correspondence between `$k$' and `$u$', first, note that each `$k$' refers to a time-invariant critical point of $V_4$ in $\mathfrak{A}_4$ which corresponds to a time-invariant solution of the 5d BPS equations \eqref{5d bps hw} on $M_3 \times S^1 \times \R$. Second, note that each `$u$' refers to a time-invariant critical point of $V_{I \to S^1}$ in $\mathcal{M}(\mathscr{L}_0, \mathscr{L}_1)_{{L\mathcal{M}^{G_\C,\Sigma}_{\text{flat}}}}$ which corresponds to a time-invariant solution of the 3d BPS equations given by the expression within the squared terms in \eqref{3d sigma model action replace s1} set to zero. Third, note that the 3d BPS equations are a direct topological reduction of the 5d BPS equations along $\Sigma \subset M_3$ (which we can regard as $I \times \Sigma$ in our context), whence there is a one-to-one correspondence between time-invariant solutions of the latter and former. Altogether, this means that there is a one-to-one correspondence between `$k$' and `$u$'.

To ascertain if there is a one-to-one correspondence between `$d_k$' and `$d_u$', first, note that the flow lines between time-invariant critical points of $V_4$ in $\mathfrak{A}_4$ that realize the Floer differential of $\text{HF}^{\text{HW}}_{*}$ which counts `$d_k$', correspond to solutions of the gradient flow equations \eqref{flow on m5 = m4 x R 2/V} which, in turn, correspond to solutions of the 5d BPS equations \eqref{5d bps hw} on $M_3 \times S^1 \times \R$. Second, note that the flow lines between time-invariant critical points of $V_{I \to S^1}$ in $\mathcal{M}(\mathscr{L}_0, \mathscr{L}_1)_{{L\mathcal{M}^{G_\C,\Sigma}_{\text{flat}}}}$ that realize the Floer differential of $\text{HSF}^{\text{Int}}_{*}$ which counts `$d_u$', correspond to solutions of the gradient flow equations given by the expression within the squared terms in \eqref{sqm action on interval space to loop space of higgs} set to zero which, in turn, correspond to solutions of the 3d BPS equations given by the expression within the squared terms in \eqref{3d sigma model action replace s1} set to zero. Third, note again that the 3d BPS equations are a direct topological reduction of the 5d BPS equations along $\Sigma \subset M_3$ (which we can regard as $I \times \Sigma$ in our context), whence there is a one-to-one correspondence between solutions of the latter and former. Altogether, this means that there is a one-to-one correspondence between `$d_k$' and `$d_u$'.

Therefore, we indeed have a one-to-one correspondence between $(k, d_k)$ and $(u, d_u)$ in \eqref{Atiyah-Floer partition equality S1}, whence we would have the following degree-to-degree isomorphism between the HW Floer and symplectic intersection Floer homology classes
\begin{equation}
    \label{Atiyah-Floer HW - S1}
    \boxed{
    \text{HF}^{\text{HW}}_*(M_3 \times S^1, G) \cong \text{HSF}^{\text{Int}}_*({L\mathcal{M}^{G,\Sigma}_{\text{Higgs}}}, \mathscr{L}_0, \mathscr{L}_1)
    }
\end{equation}
where we have exploited the fact that the isotropic-coisotropic branes $\mathscr{L}_0, \mathscr{L}_1$ are of $(\text{A}, \text{A}, \text{B})$ type in complex structure $(I, J, K)$, whence we have a choice to also interpret them as A-branes in complex structure $I$ that live in ${L\mathcal{M}^{G,\Sigma}_{\text{Higgs}}}$.

That is, we have an  HW Atiyah-Floer correspondence of $M_3 \times S^1$!

\subsection{A 4d-HW Atiyah-Floer Correspondence of \texorpdfstring{$M_3$}{M3}}
\label{subsection: 7d-hw atiyah-floer correspondence}

If we let $S^1 \to 0$, the LHS of \eqref{Atiyah-Floer partition equality S1} will, according to $\S$\ref{section: complex flow on m3 x R}, become \eqref{hitchin floer classes}, i.e.,
\begin{equation}
 {   \sum_l  \text{HF}^{\text{HW}_4}_{d_l}(M_3, G) }\,,
\end{equation}
while the RHS of \eqref{Atiyah-Floer partition equality S1} will, according to $\S$\ref{subsection: sqm of sigma models on interval space of loop flat gc}-\ref{subsection: HSF in loop space of flat gc}, become
\begin{equation}
\label{Atiyah-Floer partition equality HSF^int Mflat Gc}
 {  \sum_w \text{HSF}^{\text{Int}}_{d_w}({\mathcal{M}^{G_\C,\Sigma}_{\text{flat}}}, {L}_0, {L}_1)}\,,
\end{equation}
where the  branes $L_0$ and $L_1$ are Lagrangian of $(\text{A}, \text{A}, \text{B})$ type in complex structure $(I, J, K)$.\footnote{$L_0, L_1$ are the restrictions of the isotropic-coisotropic branes $\mathscr{L}_0, \mathscr{L}_1$ to the subspace ${\mathcal{M}^{G,\Sigma}_{{H}}} \subset L{\mathcal{M}^{G,\Sigma}_{{H}}}$, where on a finite-dimensional symplectic manifold such as ${\mathcal{M}^{G,\Sigma}_{{H}}}$, branes that are isotropic-coisotropic are Lagrangian.
\label{restrictions of L-branes}}

In other words, we have
\begin{equation}
\label{Atiyah-Floer partition equality HW_4}
 {   \sum_l  \text{HF}^{\text{HW}_4}_{d_l}(M_3, G) = \sum_w \text{HSF}^{\text{Int}}_{d_w}({\mathcal{M}^{G_\C,\Sigma}_{\text{flat}}}, {L}_0, {L}_1)}\,.
\end{equation}

By applying the same arguments that led us to \eqref{Atiyah-Floer HW - S1}, we find that \eqref{Atiyah-Floer partition equality HW_4} will mean that
\begin{equation}
    \label{Atiyah-Floer HW_4}
    \boxed{
    \text{HF}^{\text{HW}_4}_*(M_3, G) \cong \text{HSF}^{\text{Int}}_*({\mathcal{M}^{G,\Sigma}_{\text{Higgs}}}, {L}_0, {L}_1)
    }
\end{equation}
That is, we have a 4d-HW Atiyah-Floer correspondence of $M_3$!

\subsection{A 5d ``Rotation'' and a 4d-HW/VW Atiyah-Floer Correspondence Duality}

It was shown in \cite[eqn.(5.12)]{ong2022vafa} that 4d VW theory on $M_3 \times \R$  gives, instead of \eqref{Atiyah-Floer HW_4},
\begin{equation}
\label{VW partition equality HW_4}
\boxed {    \text{HF}^{\text{inst}}_{*}(M_3, G_\C) \cong  \text{HSF}^{\text{Int}}_{*}({\mathcal{M}^{G_\C,\Sigma}_{\text{flat}}}, {L}_0', {L}_1')}
\end{equation}
where $\text{HF}^{\text{inst}}_{*}(M_3, G_\C)$ is $G_\C$-instanton Floer homology, and $L'_{0,1}$ are Lagrangian branes of $(\text{A}, \text{B}, \text{A})$ type in complex structure $(I, J, K)$.

Clearly, this indicates that 4d-HW theory does not coincide with VW theory. Indeed, the BPS equations of 4d-HW theory on $M_3 \times \R$, given by \eqref{m3 x R flow 2}--\eqref{m3 x R flow 1}, \textit{differ} from the BPS equations of VW theory on $M_3 \times \R$ given by \cite[eqn.~(4.11)]{ong2022vafa} -- the expressions for the time-derivatives of the $A$ and $B$ fields are swopped in the former and latter. In turn, this means that HW theory on $M_3 \times S^1 \times \R$, upon KK reduction along $S^1$, does not become VW theory.

\bigskip\noindent\textit{A 5d ``Rotation'' and Dual 4d Reductions of HW Theory}
\vspace*{0.5em}

Nevertheless, it is observed that if one were to swop the directions of $S^1$ and $\R$, i.e., consider HW theory on $M_3 \times \R \times S^1$ instead, upon KK reduction along the $S^1$,  the 4d-HW BPS equations that replace \eqref{m3 x R flow 2}--\eqref{m3 x R flow 1} will coincide with \cite[eqn.~(4.11)]{ong2022vafa}. In other words, HW theory on $M_3 \times S^1 \times \R$, with the $S^1 \times \R$ cylinder ``rotated'' by $\pi/2$ about its midpoint, will, upon KK reduction along the $S^1$, become VW theory.
Since HW theory should be invariant under this topologically-trivial ``rotation'' of the cylinder, it would mean that 4d-HW and VW theory on $M_3 \times \R$ are \textit{dual} 4d reductions of HW theory on $S^1 \times M_3 \times \R$.

\bigskip\noindent\textit{A 4d-HW/VW Atiyah-Floer Correspondence Duality}
\vspace*{0.5em}

Thus, we have the following  4d-HW/VW Atiyah-Floer correspondence duality
\begin{equation}
    \label{4d-HW/VW AF Duality}
    \boxed{
    \text{HF}^{\text{HW}_4}_*(M_3, G) \cong \text{HSF}^{\text{Int}}_*({\mathcal{M}^{G,\Sigma}_{\text{Higgs}}}, {L}_0, {L}_1) \xleftrightarrow[]{\text{5d ``rotation}"}
   \text{HF}^{\text{inst}}_{*}(M_3, G_\C) \cong  \text{HSF}^{\text{Int}}_{*}({\mathcal{M}^{G_\C,\Sigma}_{\text{flat}}}, {L}_0', {L}_1') }
\end{equation}
This duality, which relates (a) Hitchin configurations/Higgs pairs on $M_3$ and $\Sigma$ to (b) flat $G_\C$ connections on $M_3$ and $\Sigma$, that is due to a topologically-trivial ``rotation'' of the space orthogonal to $M_3$ in HW theory, will manifest itself again in $\S$\ref{section: Fukaya-Seidel} where we study HW theory on $M_3 \times \R^2$.

\bigskip\noindent\textit{A Continuous Family of Dual 4d Reductions of HW Theory}
\vspace*{0.5em}

Notice that the product space $M_3 \times S^1 \times \R$ can also be regarded as a trivial fibration of the $S^1 \times \R$ cylinder over the $M_3$ base. Thus, we can ``rotate'' the $S^1 \times \R$ fiber by an arbitrary angle, and the space would be unchanged topologically. This means that HW theory on $M_3 \times S^1 \times \R$ would be invariant under an arbitrary ``rotation'' of the $S^1 \times \R$ cylinder.

In turn, this suggests that we actually have a continuous family of dual 4d reductions of HW theory on $M_3 \times \R$, where if the angle of ``rotation'' is 0 or $\pi/2$, they are the 4d-HW or VW theory, respectively.

\section{Topological Invariance,  5d \texorpdfstring{``S-duality''}{"S-duality"}, and a Web of Relations}
\label{section: s-duality and web of relations}

In this section, we will first exploit the topological invariance of HW theory to relate the Floer homologies we derived in $\S$\ref{section: hw floer homology}-$\S$\ref{section: hw atiya-floer correspondence} to one another. We will then proceed to exploit a 5d ``S-duality'' of HW theory to obtain Langlands dual relations between loop/toroidal group generalizations of 4d/3d-HW Floer homologies, respectively. Finally, we summarize these findings in a web of relations amongst these Floer homologies.

\subsection{Topological Invariance of HW Theory and the Floer Homologies}

\bigskip\noindent\textit{Relating the Floer Homologies from Sections 3 to 6}
\vspace*{0.5em}

Notice that the topological invariance of (the $\mathcal{Q}$-cohomology of) HW theory in all directions means that we can relate the partition functions \eqref{5d partition function floer}, \eqref{hitchin floer classes}, \eqref{BF floer classes} and \eqref{L^2M floer classes} as
\begin{equation}\label{HW = HW_4 = HW_3}
  \boxed{   \sum_k  \text{HF}^{\text{HW}}_{d_k}(M_4, G)
    \xleftrightarrow[]{M_4 = M_3 \times {\hat S}^1} \sum_l \text{HF}_{d_l}^{\text{HW}_4}(M_3, G)
     \xleftrightarrow[]{M_3 = M_2 \times {\hat S}^1} \sum_p \text{HF}_{d_p}^{\text{HW}_3}(M_2, G_\C) }
\end{equation}
and
\begin{equation}\label{HW = HSF}
  \boxed{   \sum_k  \text{HF}^{\text{HW}}_{d_k}(M_4, G)
    \xleftrightarrow[]{M_4 = {\hat M}_2 \times S^1 \times S^1} \sum_q \text{HSF}^{\text{hol}}_{d_q}({L^2\mathcal{M}^{G_\C}_{\text{flat}}(M_2)})
  }
\end{equation}
where $S^1$ is a circle of fixed radius; ${\hat S}^1$ is a circle of variable radius; and $\hat{M}_2$ is a Riemann surface of genus $g \geq 2$ of variable size.

Note that \eqref{HW = HW_4 = HW_3} and \eqref{HW = HSF} are consistent relations in that there is a one-to-one correspondence in the summations over `$k$', `$l$', `$p$' and `$k$', `$q$'. Specifically, each `$k$', `$l$', `$p$' corresponds to a solution of (the simultaneous vanishing of the LHS and RHS of) \eqref{flow on m5 = m4 x R}, \eqref{m3 x R flow 2}--\eqref{m3 x R flow 1}, \eqref{m2 x R bps}--\eqref{m2 x R bps 2}, respectively, where \eqref{m2 x R bps}--\eqref{m2 x R bps 2} is obtained via a KK reduction of \eqref{m3 x R flow 2}--\eqref{m3 x R flow 1} which is in turn obtained via a KK reduction of \eqref{flow on m5 = m4 x R}. Similarly, each `$k$', `$q$' corresponds to a solution of (the simultaneous vanishing of the LHS and RHS of) \eqref{flow on m5 = m4 x R}, \eqref{3d sigma model BPS equations complexified}, respectively, where  \eqref{3d sigma model BPS equations complexified} is obtained via a topological reduction of \eqref{flow on m5 = m4 x R}.

In short, we have a \emph{novel} equivalence of gauge-theoretic and symplectic Floer homologies of these four, three, two-manifolds, respectively!

\bigskip\noindent\textit{The Floer Homologies from Section 7}
\vspace*{0.5em}

Furthermore, from \eqref{Atiyah-Floer HW - S1}, we have
\begin{equation}
\label{Atiyah-Floer partition equality S1 - web}
\boxed {   \sum_k  \text{HF}^{\text{HW}}_{d_k}(M_4, G) \xleftrightarrow[]{M_4 = M_3 \times S^1} \sum_u \text{HSF}^{\text{Int}}_{d_u}({L\mathcal{M}^{G,\Sigma}_{\text{Higgs}}}, \mathscr{L}_0, \mathscr{L}_1)}
\end{equation}
where $\Sigma$ is the Heegaard surface of the Heegaard split of $M_3$. From \eqref{Atiyah-Floer HW_4}, we have
\begin{equation}
\label{Atiyah-Floer partition equality HW_4 - web}
\boxed {   \sum_l  \text{HF}^{\text{HW}_4}_{d_l}(M_3, G) = \sum_w \text{HSF}^{\text{Int}}_{d_w}({\mathcal{M}^{G,\Sigma}_{\text{Higgs}}}, {L}_0, {L}_1)}
\end{equation}
From \eqref{4d-HW/VW AF Duality}, we have
\begin{equation}
    \label{4d-HW/VW AF Duality Partition Functions}
    \boxed{
    \begin{aligned}
    \sum_l \text{HF}^{\text{HW}_4}_{d_l}(M_3, G)
    &= \sum_w \text{HSF}^{\text{Int}}_{d_w}({\mathcal{M}^{G,\Sigma}_{\text{Higgs}}}, {L}_0, {L}_1) \\
    & \displaystyle\left\updownarrow\vphantom{\int}\right. \text{5d ``rotation}" \\
    \sum_m \text{HF}^{\text{inst}}_{d_m}(M_3, G_\C)
    &= \sum_n \text{HSF}^{\text{Int}}_{d_n}({\mathcal{M}^{G_\C,\Sigma}_{\text{flat}}}, {L}_0', {L}_1')
    \end{aligned}
   }
\end{equation}


\bigskip\noindent\textit{An Equivalent to the Floer Homology of Section 5}
\vspace*{0.5em}

Last but not least, another relation can be identified by performing a topological reduction along $M_2$ of the 3d gauge theory with action \eqref{sqm action m2 x R initial} (via the same procedure as that in $\S$\ref{subsection: sadov reduction process}). Specifically, from the finiteness conditions of the action under the process of topological reduction, we find that we will get a 1d sigma model with a target space $\mathcal{M}^{G_\C}_{\text{BF}}(M_2)$ of $G_\C$-BF configurations on $M_2$, where the BPS equations are  $\dot{\mathcal{A}}^a = 0 = \dot{\bar{\mathcal{A}}}^a$ and $\dot{Z}^a = 0 = \dot{\bar{Z}}^a$, which define time-invariant configurations.

As we have a standard 1d sigma model, its $\mathcal{Q}$-cohomology can be identified as the de Rham cohomology of $\mathcal{M}^{G_\C}_{\text{BF}}(M_2)$ of dimension zero,\footnote{One can understand why this is true from footnote~\ref{footnote: justification of isolated and nondegenerate GC-BF configs on M2}, where the isolated and non-degenerate configurations therein refer to $G_\C$-BF configurations on $M_2$.} whence we only have de Rham classes of zero-forms.
Thus, the partition function can be written as
\begin{equation}
    \label{de-rham BF}
    \mathcal{Z}_{\text{HW},\R}(G) = \sum_p {\text{H}}^0_{\text{dR}}\big(\mathcal{M}^{G_\C}_{\text{BF}}(M_2)\big) \, ,
\end{equation}
where $p$ is a sum over all time-invariant $G_\C$-BF configurations on $M_2$, and the de Rham classes ${\text H}^0_{\text{dR}}$ are constants.

Via the topological invariance of HW theory, this ought to be equal to the partition function in \eqref{BF floer classes} which also sums over all time-invariant $G_\C$-BF configurations on $M_2$, i.e.,
\begin{equation}
    \label{de-rham BF correspondence}
    \boxed{
    \sum_p \text{HF}_{d_p}^{\text{HW}_3}(M_2, G_\C) = \sum_p {\text H}^0_{\text{dR}}\big(\mathcal{M}^{G_\C}_{\text{BF}}(M_2)\big)
    }
\end{equation}

\bigskip\noindent\textit{Summarizing the Relations Amongst the Gauge-theoretic and Symplectic Floer Homologies}
\vspace*{0.5em}

We can summarize the various relations amongst the gauge-theoretic and symplectic Floer homologies we have obtained via the topological invariance of HW theory in Fig. \ref{fig:web of floer relations}, where the sizes of $\hat S^1$, $\hat M_2$, and $\Sigma$ can be varied.

\usetikzlibrary{arrows,automata,positioning,calc}
\begin{figure}
    \centering
    \begin{tikzpicture}[%
        auto,%
        block/.style={draw, rectangle},%
        every edge/.style={draw, <->},%
        relation/.style={scale=0.8, sloped, anchor=center, align=center},%
        vertRelation/.style={scale=0.8, anchor=center, align=center},%
        shorten >=4pt,%
        shorten <=4pt,%
        ]
        \def \verRel {2} 
        \def \horRel {1.8} 
        \node[block, ultra thick] (HW5) {$ \sum_k \text{HF}^{\text{HW}}_{d_k}(M_4, G)$};
        \node[block, below={\verRel} of HW5] (HW4) {$\sum_l \text{HF}_{d_l}^{\text{HW}_4}(M_3, G)$};
        -
        \node[block, left={\horRel} of HW4] (HSF-LM) {$\sum_u \text{HSF}^{\text{Int}}_{d_u}({L\mathcal{M}^{G,\Sigma}_{\text{Higgs}}}, \mathscr{L}_0, \mathscr{L}_1)$};
        \node[block, below={\verRel} of HW4] (HW3) {$\sum_p \text{HF}_{d_p}^{\text{HW}_3}(M_2, G_\C)$};
        -
        \node[block, right={\horRel} of HW3] (HSF-L2M) {$\sum_q \text{HSF}^{\text{hol}}_{d_q}({L^2\mathcal{M}^{G_\C}_{\text{flat}}(M_2)})$};
        -
        \node[block, left={\horRel} of HW3] (HSF-M) {$\sum_w \text{HSF}^{\text{Int}}_{d_w}({\mathcal{M}^{G,\Sigma}_{\text{Higgs}}}, {L}_0, {L}_1)$};
        \node[block, below={\verRel} of HW3] (VW4) {$\sum_m \text{HF}^{\text{inst}}_{d_m}(M_3, G_\C)$};
        -
        \node[block, below left={sqrt(\verRel*\verRel + \horRel*\horRel)} of VW4] (VW-HSF) {$\sum_n \text{HSF}^{\text{Int}}_{d_n}({\mathcal{M}^{G_\C,\Sigma}_{\text{flat}}}, {L}_0', {L}_1')$};
        \node[block, below={\verRel} of HSF-L2M] (HW1) {$\sum_p {\text H}^0_{\text{dR}}\big(\mathcal{M}^{G_\C}_{\text{BF}}(M_2)\big)$};
        \draw
        (HW5) edge[dashed] node[vertRelation, right] {$M_4 = M_3 \times \hat{S}^1$} (HW4)
        (HW4) edge[dashed] node[vertRelation, right] {$M_3 = M_2 \times \hat{S}^1$} (HW3)
        (HW5.south east) edge[dashed] node[relation, above] {$M_4 = \hat{M}_2 \times S^1 \times S^1$} (HSF-L2M)
        (HW5.south west) edge node[relation, above] {$M_3 = M_3' \bigcup_\Sigma M_3''$} node[relation, below] {$M_4 = M_3 \times S^1$} (HSF-LM)
        (HW4.south west) edge node [relation, name=HW4-HSF, above] {$M_3 = M_3' \bigcup_\Sigma M_3''$} (HSF-M)
        (VW4.south west) edge node [relation, name=VW4-VW-HSF, below] {$M_3 = M_3' \bigcup_\Sigma M_3''$} (VW-HSF)
        ($(HW4-HSF) + (0.5em, -0.5em)$) edge node[vertRelation, right] {5d ``rotation''} (VW4-VW-HSF)
        (HW3.south east) edge[dashed] node[relation, above] {$M_2 = \hat{M}_2$} (HW1);
    \end{tikzpicture}
    \caption{Relations amongst gauge-theoretic and symplectic Floer homologies from the topological invariance of HW theory.}
    \label{fig:web of floer relations}
\end{figure}
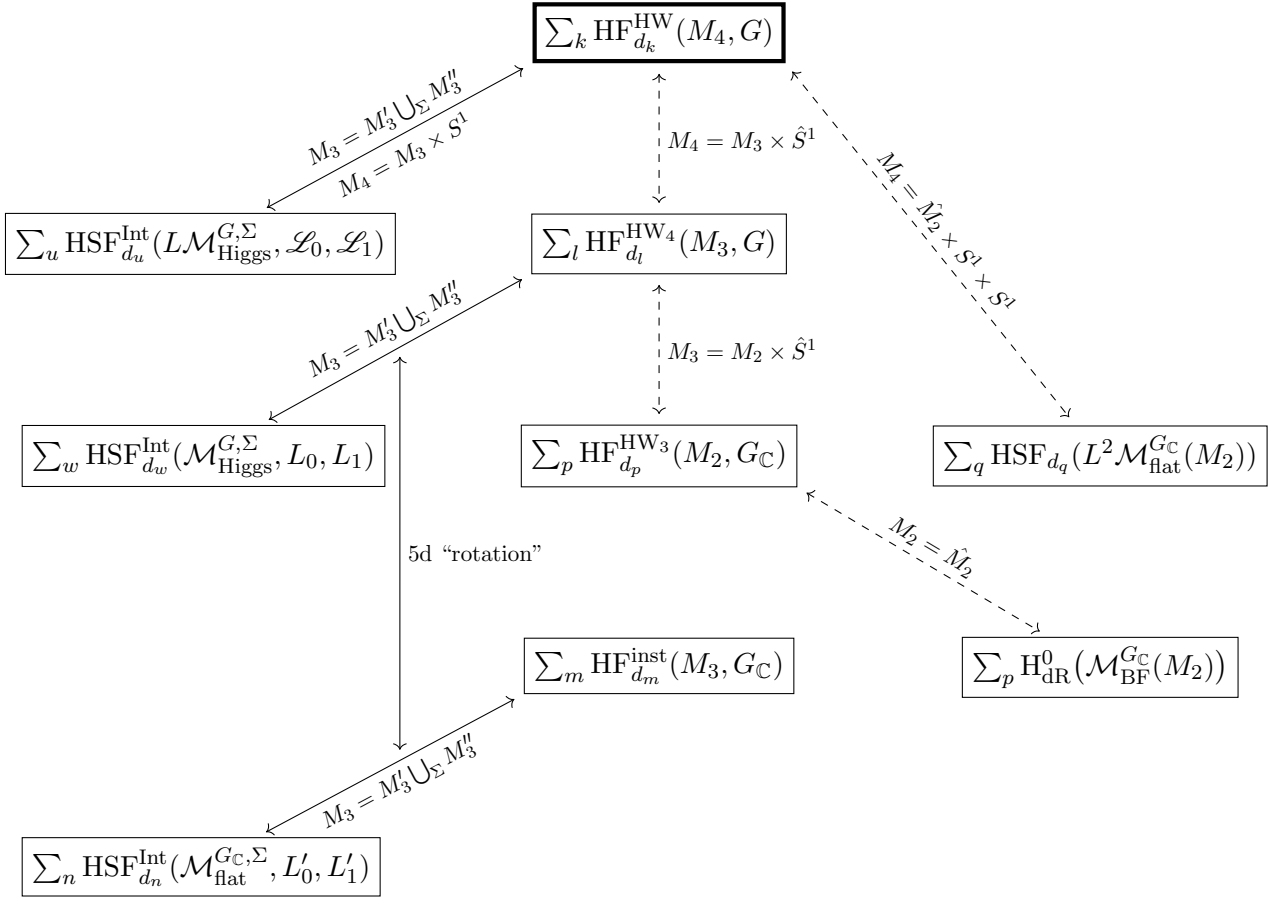

\subsection{5d \texorpdfstring{``S-duality''}{"S-duality"} of HW Theory as a Langlands Duality of Loop/Toroidal Group 4d/3d-HW Floer Homology}

\bigskip\noindent\textit{5d ``S-duality'' of HW Theory}
\vspace*{0.5em}

5d $\mathcal{N}=2$ SYM on $X_4 \times S^1$ is known to enjoy an ``S-duality'' in the following sense~\cite{tachikawa2011sdualit}. The theory with a  nonsimply-laced gauge group of coupling $e$ and circle radius $R_5$, is equivalent to one with a simply-laced gauge group of coupling $1/e$ and circle radius $1/R_5$ that is twisted around the circle.

Specifically, HW theory on $M_3 \times \R \times S^1$ with gauge group   $G = SO(N+1)$, $USp(2N -2)$, $F_4$ or $G_2$ of coupling $e$ and $S^1$-radius $R_5$,  is equivalent to HW theory on $M_3 \times \R \times {\tilde S}^1_\sigma$  with gauge group $\tilde{G} = SU(N)$, $SO(2N)$, $E_6$ or $SO(8)$ of coupling $1/e$ and ${\tilde S}^1_\sigma$-radius $1/R_5$ that undergoes a $\sigma =\Z_2$, $\Z_2$, $\Z_2$ or $\Z_3$ outer automorphism as we go around the ${\tilde S}^1_\sigma$, respectively. That is,
\begin{equation}\label{S-dual PF 1}
    \mathcal{Z}_{\text{HW},M_3 \times S^1 \times \R}(G)  \xleftrightarrow[]{\text{5d ``S-duality''}}     \mathcal{Z}_{\text{HW},M_3 \times \tilde S^1_\sigma \times \R}(\tilde G)\,.
\end{equation}
From \eqref{5d partition function floer}, this means that
\begin{equation}\label{S-dual Floer homology}
{  \sum_k \text{HF}_{d_k}^{\text{HW}}(M_3 \times S^1, G)    \xleftrightarrow[]{\text{5d ``S-duality''}}   \sum_{\tilde k} \text{HF}_{d_{\tilde k}}^{\text{HW}}(M_3 \times \tilde{S}^1_\sigma, \tilde G) }\,.
\end{equation}

The aforementioned outer automorphism in the ``S-dual'' simply-laced 5d theory and HW Floer homology can be better understood as follows. To this end, first note that a 5d theory on $X_4 \times {\tilde S}^1$ with gauge group $\tilde G$ can be regarded as a 4d theory on $X_4$ with gauge group $L\tilde G$, the loop group of $\tilde G$ generated by maps $\gamma: {\tilde S}^1 \to \tilde G$. Then, 5d-HW theory on $M_3 \times \R \times {\tilde S}^1_\sigma$ with gauge group $\tilde G$ where there is an outer automorphism $\sigma$ as we go around ${\tilde S}^1_\sigma$, can be regarded as a 4d-HW theory on $M_3 \times \R$ with gauge group $L_\sigma \tilde G$, where $L_\sigma \tilde G$ is a $\sigma$-twisted loop group generated by maps $\gamma_\sigma: {\tilde S}^1_\sigma \to \tilde G$ such that $\gamma_\sigma(2\pi) = \sigma(\gamma_\sigma(0))$. In other words, we can also express \eqref{S-dual PF 1} as
\begin{equation}\label{S-dual 4d HW}
      \mathcal{Z}^{\text{4d}}_{\text{HW},M_3 \times \R}(LG) \xleftrightarrow[]{\text{5d ``S-duality''}}    \mathcal{Z}^{\text{4d}}_{\text{HW},M_3 \times \R}(L_\sigma \tilde G)\,.
\end{equation}
Consequently, $\text{HF}_{d_k}^{\text{HW}}(M_3 \times S^1, G)$ can be understood as an $LG$ Floer homology class that corresponds to a critical point in the SQM formulation of the 4d-HW theory on $M_3 \times \R$ with gauge group $LG$, while $\text{HF}_{d_{\tilde k}}^{\text{HW}}(M_3 \times \tilde{S}^1_\sigma, \tilde G)$ can be understood as an $L_\sigma\tilde{G}$ Floer homology class that corresponds to a critical point in the SQM formulation of the 4d-HW theory on $M_3 \times \R$ with gauge group $L_\sigma \tilde{G}$.

\bigskip\noindent\textit{More about the 4d-HW Theory with Loop Gauge Group}
\vspace*{0.5em}

In order to describe the $LG$ and $L_\sigma \tilde{G}$ Floer homology classes, we will first need to determine the above 4d-HW theory on $M_3 \times \R$ with gauge group $L_\omega\mathcal{G}$ and underlying fifth-circle $\mathcal{S}^1_\omega$, where $\omega = \Z_n$ for $n=1$, 2 or 3. In this case, the loop algebra $L_\omega \mathfrak g$ of $L_\omega \mathcal{G}$ would obey the twisted periodicity condition
\begin{equation}
\label{loop algebra}
    {\mathfrak g} \otimes {\mathcal{P}}(2\pi) = \omega ( {\mathfrak g}) \otimes e^{\frac{2 \pi i \cdot j}{n}} {\mathcal{P}}(0)\,, \qquad j = 0, \dots, n-1\,,
\end{equation}
where ${\mathcal{P}}$ is a smooth function on $\mathcal{S}^1_\omega$, and $\omega ( {\mathfrak g})$ is a Lie algebra whose Dynkin diagram is an outer $\Z_n$-automorphism of the Dynkin diagram of the Lie algebra ${\mathfrak g}$ of $\mathcal{G}$.

Next, note that the 4d-HW theory which interprets 5d-HW theory in terms of a loop gauge group can be obtained via the same methods we used in the previous sections to obtain the SQM's which interpret the gauge theories and sigma models. In particular, we can  ``absorb'' the integration over $\mathcal{S}^1_\omega$ in the 5d action to cast it as a 4d action where the fields are now valued not in $\mathfrak g$ but $L_\omega \mathfrak g$. Specifically, the 4d action would be given by
\begin{equation}\label{4d HW action}
    S_{\text{4d-HW}} =  \frac{1}{e^2}\int dt \, \text{Tr}^{L_\omega} \int_{M_3} d^3x \, \bigg(
    \bigg| \dot{A}^i_\omega - \epsilon^{ijk}\bigg(\partial_j B^\omega_{k} + [A^\omega_j, B^\omega_{k}]\bigg) \bigg|^2
    + \bigg| \dot{B}^i_\omega - \frac{1}{2}\epsilon^{ijk}\bigg( F^\omega_{jk} - [B^\omega_{j}, B^\omega_{k}]\bigg) \bigg|^2
    + \dots
    \bigg) \, ,
\end{equation}
where $\text{Tr}^{L_\omega}$ is the trace over $L_\omega \mathfrak g$; the $\omega$ sub(super)script means that the 4d field is circle-valued obeying the same twisted periodicity condition as $\mathcal{P}$ in \eqref{loop algebra}; and  ``\dots'' contains the fermionic terms.

\bigskip\noindent\textit{Loop Group 4d-HW Floer Homology}
\vspace*{0.5em}

Notice that \eqref{4d HW action} has the same form as \eqref{vw sqm}. This means that the SQM formulation of \eqref{4d HW action} would also have the same form as \eqref{vw sqm final}. In turn, this means that we can write the 4d-HW partition function as
\begin{equation}
\label{4d HW LG Floer}
    \mathcal{Z}^{\text{4d}}_{\text{HW},M_3 \times \R}(L_\omega \mathcal{G}) =  \sum_l \text{HF}_{{d_l}}^{\text{HW}_4}(M_3, L_\omega \mathcal{G})\,,
\end{equation}
where $\text{HF}^{\text{HW}_4}_{d_l}(M_3, L_\omega \mathcal{G})$ is a loop group generalization of the 4d-HW Floer homology class in $\S$\ref{section: complex flow on m3 x R},  defined by time-invariant $L_\omega \mathcal{G}$ Hitchin configurations on $M_3$ and $L_\omega \mathcal{G}$ generalizations of \eqref{flow on m4 = m3 x R}--\eqref{H configuration}.

\bigskip\noindent\textit{5d ``S-duality'' as a Langlands Duality of Loop Group 4d-HW Floer Homologies}
\vspace*{0.5em}

Hence, \eqref{4d HW LG Floer} means that we can also express \eqref{S-dual 4d HW} as
\begin{equation}\label{S-dual 4d HW LG Floer}
   {   \sum_l \text{HF}_{{d_l}}^{\text{HW}_4}(M_3, L{G}) \xleftrightarrow[]{\text{5d ``S-duality''}}    \sum_{\tilde l} \text{HF}_{{d_{\tilde l}}}^{\text{HW}_4}(M_3, L_\sigma \tilde{G})}\,.
\end{equation}
In turn, this can be further expressed as
\begin{equation}\label{Langlands-dual 4d HW LG Floer}
  \boxed {   \sum_l \text{HF}_{{d_l}}^{\text{HW}_4}(M_3, L{G}) \xleftrightarrow[]{\text{5d ``S-duality''}}    \sum_{\tilde l} \text{HF}_{{d_{\tilde l}}}^{\text{HW}_4}(M_3, (LG)^\vee)}
\end{equation}
where $(LG)^\vee$ is the Langlands dual of $LG$ in the sense that their loop algebras are Langlands dual of each other.\footnote{To understand this, first note that ${\tilde{\mathfrak g}}_{\text{aff}}^\sigma$, the $\sigma$-twisted affine Lie algebra of $\tilde G$, is the Langlands dual ${{\mathfrak g}}^\vee_{\text{aff}}$ of ${{\mathfrak g}}_{\text{aff}}$, the untwisted affine Lie algebra of $G$ \cite{kac1983infindimen}, where in particular, we have the Lie algebra embedding $\mathfrak g \hookrightarrow {\mathfrak g}_{\text{aff}}$ and that of its Langlands dual $\mathfrak g^\vee \hookrightarrow {\mathfrak g}^\vee_{\text{aff}}$.  Second, note that (twisted) loop algebras are just (twisted) affine algebras specialized to zero central extension. Altogether, this means that $L_\sigma \tilde{\mathfrak g}$ is the Langlands dual $(L\mathfrak g)^\vee$ of $L \mathfrak g$. \label{langlands dual loop algebras}}

\bigskip\noindent\textit{A 3d-HW Theory with Toroidal Gauge Group}
\vspace*{0.5em}

If we further let $M_3 = M_2 \times S^1$, we can recast 4d-HW theory as a 3d-HW theory by repeating the calculation of ``absorbing'' the integration over $S^1$ in \eqref{4d HW action}. Then, we can also express \eqref{S-dual 4d HW} as\footnote{We will get a 3d-HW theory with no extra scalars $X$ and $Y$ because we are not KK reducing along the $S^1$'s.}
\begin{equation}\label{S-dual 3d HW}
      \mathcal{Z}^{\text{3d}}_{\text{HW},M_2 \times \R}(LLG)_{\scriptscriptstyle {X, Y = 0}} \xleftrightarrow[]{\text{5d ``S-duality''}}    \mathcal{Z}^{\text{3d}}_{\text{HW},M_2 \times \R}(LL_\sigma \tilde G)_{\scriptscriptstyle {X, Y = 0}} \, .
\end{equation}

\bigskip\noindent\textit{Toroidal Group 3d-HW Floer Homology}
\vspace*{0.5em}

Via the same arguments that led us to \eqref{4d HW LG Floer}, we can also write the 3d-HW partition function as
\begin{equation}
\label{3d HW LG Floer}
    \mathcal{Z}^{\text{3d}}_{\text{HW},M_2 \times \R}(LL_\omega \mathcal{G})_{\scriptscriptstyle {X, Y = 0}} =  \sum_q \text{HF}_{{d_q}}^{\text{HW}_3}(M_2, LL_\omega \mathcal{G}_\C)_{\scriptscriptstyle {Z, \bar{Z} = 0}}\,,
\end{equation}
where $\text{HF}^{\text{HW}_3}_{d_q}(M_2, LL_\omega \mathcal{G}_\C)_{\scriptscriptstyle {Z, \bar{Z} = 0}}$ is a toroidal group generalization of the 3d-HW Floer homology class in $\S$\ref{section: m2 x R},  defined by time-invariant  flat $LL_\omega \mathcal{G}_\C$ connections on $M_2$ corresponding to critical points of an $LL_\omega \mathcal{G}_\C$ generalization of the Chern-Simons functional, with gradient flow equations an $LL_\omega \mathcal{G}_\C$ generalization of \eqref{m2 x R flow} (with ${Z, \bar{Z} = 0}$).

\bigskip\noindent\textit{5d ``S-duality'' as a Langlands Duality of Toroidal Group 3d-HW Floer Homologies}
\vspace*{0.5em}

Hence, \eqref{3d HW LG Floer} means that we can also express \eqref{S-dual 3d HW} as
\begin{equation}\label{S-dual 3d HW LG Floer}
   \sum_q \text{HF}_{{d_q}}^{\text{HW}_3}(M_2, LL{G}_\C)_{\scriptscriptstyle {Z, \bar{Z} = 0}} \xleftrightarrow[]{\text{5d ``S-duality''}}    \sum_{\tilde q} \text{HF}_{{d_{\tilde q}}}^{\text{HW}_3}(M_2, LL_\sigma \tilde{G}_\C)_{\scriptscriptstyle {Z, \bar{Z} = 0}}\,.
\end{equation}
In turn, via footnote~\ref{langlands dual loop algebras}, this can be further expressed as
\begin{equation}\label{Langlands-dual 3d HW LG Floer}
   \boxed{   \sum_q \text{HF}_{{d_q}}^{\text{HW}_3}(M_2, LL{G}_\C)_{\scriptscriptstyle {Z, \bar{Z} = 0}} \xleftrightarrow[]{\text{5d ``S-duality''}}    \sum_{\tilde q} \text{HF}_{{d_{\tilde q}}}^{\text{HW}_3}(M_2, L(LG_\C)^\vee)_{\scriptscriptstyle {Z, \bar{Z} = 0}}}
\end{equation}
where $L(LG_\C)^\vee$ can be regarded as the Langlands dual of $LLG_\C$ in the sense that the loop algebras of $(LG_\C)^\vee$ and $LG_\C$ are Langlands dual.

\bigskip\noindent\textit{An Equivalent to the LHS of \eqref{Langlands-dual 3d HW LG Floer}}
\vspace*{0.5em}

From the same reasoning we used to arrive at \eqref{de-rham BF}, we can perform a topological reduction along $M_2$ of the 3d theory with gauge group $LLG$, and write the resulting partition function as
\begin{equation}
    \label{de-rham flat GC}
    \mathcal{Z}_{\text{HW},\R}(LLG)_{\scriptscriptstyle X, Y = 0} = \sum_q {\text{H}}^0_{\text{dR}}\big(\mathcal{M}^{LLG_\C}_{\text{flat}}(M_2)\big) \, ,
\end{equation}
where $q$ is a sum over all time-invariant flat $LLG_\C$ connections on $M_2$, and the de Rham classes $H^0_{\text{dR}}$ are constants.

Via the topological invariance of HW theory, this ought to be equal to the partition function on the LHS of \eqref{Langlands-dual 3d HW LG Floer} which also sums over all time-invariant flat $LLG_\C$ connections on $M_2$, i.e.
\begin{equation}
    \label{de-rham flat GC correspondence}
    \boxed{
    \sum_q \text{HF}_{d_q}^{\text{HW}_3}(M_2, LLG_\C)_{\scriptscriptstyle Z, \bar{Z} = 0} = \sum_q {\text{H}}^0_{\text{dR}}\big(\mathcal{M}^{LLG_\C}_{\text{flat}}(M_2)\big)
    }
\end{equation}

\bigskip\noindent\textit{Summarizing the Relations Between the Loop/Torodial Group Floer Homologies}
\vspace*{0.5em}

We can summarize the relations between the loop and toroidal group Floer homologies we have obtained via the 5d ``S-duality'' of HW theory  in Fig. \ref{fig:web of torodial/loop floer relations}, where the size of $\hat M_2$ can be varied.

\usetikzlibrary{arrows,automata,positioning}
\begin{figure}
    \centering
    \begin{tikzpicture}[%
        auto,%
        block/.style={draw, rectangle},%
        every edge/.style={draw, <->},%
        relation/.style={scale=0.8, sloped, anchor=center, align=center},%
        vertRelation/.style={scale=0.8, anchor=center, align=center},%
        shorten >=4pt,%
        shorten <=4pt,%
        ]
        \def \verRel {2} 
        \def \horRel {1.8} 
        \node[block, ultra thick] (HW5) {$ \sum_k \text{HF}^{\text{HW}}_{d_k}(M_4, G)$};
        \node[above={\verRel} of HW5] (aHW5) {};
        \node[below={\verRel} of HW5] (bHW5) {};
        \node[block, left={2*\horRel} of aHW5] (HW4-LG) {$\sum_l \text{HF}_{{d_l}}^{\text{HW}_4}(M_3, L{G})$};
        -
        \node[block, above={\verRel} of HW4-LG] (HW4-LGv) {$\sum_{\tilde l} \text{HF}_{{d_{\tilde l}}}^{\text{HW}_4}(M_3, (LG)^\vee)$};
        \node[block, right={2*\horRel} of aHW5] (HW3-L2G) {$\sum_q \text{HF}_{{d_q}}^{\text{HW}_3}(M_2, LL{G}_\C)_{\scriptscriptstyle {Z, \bar{Z} = 0}}$};
        -
        \node[block, above={\verRel} of HW3-L2G)] (HW3-L2Gv) {$\sum_{\tilde q} \text{HF}_{{d_{\tilde q}}}^{\text{HW}_3}(M_2, L(LG_\C)^\vee)_{\scriptscriptstyle {Z, \bar{Z} = 0}}$};
        -
        \node[block, right={\horRel} of HW5] (HW1-L2G) {$\sum_q {\text H}^0_{\text{dR}}\big(\mathcal{M}^{LLG_\C}_{\text{flat}}(M_2)\big)$};
        \draw
        (HW5.north west) edge node[relation, below] {$M_4 = M_3 \times {S}^1$}(HW4-LG)
        (HW4-LG) edge node[vertRelation, right] {5d ``S-duality''\\$G$ nonsimply-laced} (HW4-LGv)
        (HW5.north east) edge node[relation, below] {$M_4 = M_2 \times S^1 \times S^1$} (HW3-L2G)
        (HW3-L2G) edge node[vertRelation, right] {5d ``S-duality''\\$G_\C$ nonsimply-laced} (HW3-L2Gv)
        (HW3-L2G) edge[dashed] node[vertRelation, right] {$M_2 = \hat{M}_2$} (HW1-L2G)
        ;
    \end{tikzpicture}
    \caption{Relations between loop/toroidal Floer homologies from the 5d ``S-duality'' of HW theory.}
    \label{fig:web of torodial/loop floer relations}
\end{figure}
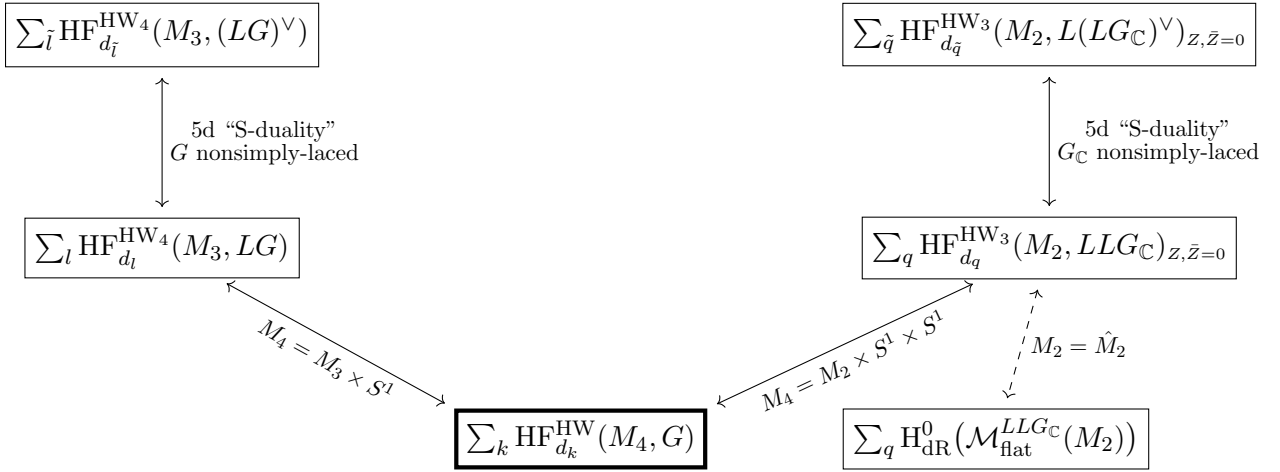

\subsection{A Web of Relations Amongst the Floer Homologies}

Altogether, we have, from the concatenation of Fig.~\ref{fig:web of floer relations} and Fig.~\ref{fig:web of torodial/loop floer relations} at their starting point that is the respective box in bold, a rich web of relations as depicted in Fig.~\ref{fig:web of relations combined}, where the sizes of $\hat S^1$, $\hat M_2$, and $\Sigma$ can be varied.

\usetikzlibrary{arrows,automata,positioning,calc}
\begin{figure}
    \centering
    \begin{adjustbox}{max totalsize={\textwidth}{\textheight},center}
        \begin{tikzpicture}[%
            auto,%
            block/.style={draw, rectangle},%
            every edge/.style={draw, <->, dashed},%
            relation/.style={scale=0.8, sloped, anchor=center, align=center},%
            vertRelation/.style={scale=0.8, anchor=center, align=center},%
            shorten >=4pt,%
            shorten <=4pt,%
            ]
            \def \verRel {2.2} 
            \def \horRel {1.8} 
            \node[block, ultra thick] (HW5) {$ \sum_k \text{HF}^{\text{HW}}_{d_k}(M_4, G)$};
            \node[above={\verRel} of HW5] (aHW5) {};
            \node[below={\verRel} of HW5] (bHW5) {};
            \node[block, below={\verRel} of HW5] (HW4) {$\sum_l \text{HF}_{d_l}^{\text{HW}_4}(M_3, G)$};
            -
            \node[block, left={\horRel} of HW4] (HSF-LM) {$\sum_u \text{HSF}^{\text{Int}}_{d_u}({L\mathcal{M}^{G,\Sigma}_{\text{Higgs}}}, \mathscr{L}_0, \mathscr{L}_1)$};
            \node[block, below={\verRel} of HW4] (HW3) {$\sum_p \text{HF}_{d_p}^{\text{HW}_3}(M_2, G_\C)$};
            -
            \node[block, right={\horRel} of HW3] (HSF-L2M) {$\sum_q \text{HSF}^{\text{hol}}_{d_q}({L^2\mathcal{M}^{G_\C}_{\text{flat}}(M_2)})$};
            -
            \node[block, left={\horRel} of HW3] (HSF-M) {$\sum_w \text{HSF}^{\text{Int}}_{d_w}({\mathcal{M}^{G,\Sigma}_{\text{Higgs}}}, {L}_0, {L}_1)$};
            \node[block, below={\verRel} of HW3] (VW4) {$\sum_m \text{HF}^{\text{inst}}_{d_m}(M_3, G_\C)$};
            -
            \node[block, below left={sqrt(\verRel*\verRel + \horRel*\horRel)} of VW4] (VW-HSF) {$\sum_n \text{HSF}^{\text{Int}}_{d_n}({\mathcal{M}^{G_\C,\Sigma}_{\text{flat}}}, {L}_0', {L}_1')$};
            \node[block, below={\verRel} of HSF-L2M] (HW1) {$\sum_p {\text H}^0_{\text{dR}}\big(\mathcal{M}^{G_\C}_{\text{BF}}(M_2)\big)$};
            \node[block, left={\horRel} of aHW5] (HW4-LG) {$\sum_l \text{HF}_{{d_l}}^{\text{HW}_4}(M_3, L{G})$};
            -
            \node[block, above={\verRel} of HW4-LG] (HW4-LGv) {$\sum_{\tilde l} \text{HF}_{{d_{\tilde l}}}^{\text{HW}_4}(M_3, (LG)^\vee)$};
            \node[block, right={\horRel} of aHW5] (HW3-L2G) {$\sum_q \text{HF}_{{d_q}}^{\text{HW}_3}(M_2, LL{G}_\C)_{\scriptscriptstyle {Z, \bar{Z} = 0}}$};
            -
            \node[block, above={\verRel} of HW3-L2G)] (HW3-L2Gv) {$\sum_{\tilde q} \text{HF}_{{d_{\tilde q}}}^{\text{HW}_3}(M_2, L(LG_\C)^\vee)_{\scriptscriptstyle {Z, \bar{Z} = 0}}$};
            -
            \node[block, right={1.5*\horRel} of HW5] (HW1-L2G) {$\sum_q {\text H}^0_{\text{dR}}\big(\mathcal{M}^{LLG_\C}_{\text{flat}}(M_2)\big)$};
            -
            \node[block, above={3.5*\verRel} of HW1-L2G] (HW1-L2Gv) {$\sum_{\qt} {\text H}^0_{\text{dR}}\big(\mathcal{M}^{L{(LG_\C)^\vee}}_{\text{flat}}(M_2)\big)$};
            \draw
            (HW5) edge node[vertRelation, right] {$M_4 = M_3 \times \hat{S}^1$} (HW4)
            (HW4) edge node[vertRelation, right] {$M_3 = M_2 \times \hat{S}^1$} (HW3)
            (HW5.south east) edge node[relation, above] {$M_4 = \hat{M}_2 \times S^1 \times S^1$} (HSF-L2M)
            (HW5.south west) edge[solid] node[relation, above] {$M_3 = M_3' \bigcup_\Sigma M_3''$} node[relation, below] {$M_4 = M_3 \times S^1$} (HSF-LM)
            (HW4.south west) edge[solid] node [relation, name=HW4-HSF, above] {$M_3 = M_3' \bigcup_\Sigma M_3''$} (HSF-M)
            (VW4.south west) edge[solid] node [relation, name=VW4-VW-HSF, below] {$M_3 = M_3' \bigcup_\Sigma M_3''$} (VW-HSF)
            ($(HW4-HSF) + (0.5em, -0.5em)$) edge[solid] node[vertRelation, right] {5d ``rotation''} (VW4-VW-HSF)
            (HW3.south east) edge node[relation, above] {$M_2 = \hat{M}_2$} (HW1)
            (HW5.north west) edge[solid] node[relation, below] {$M_4 = M_3 \times {S}^1$} (HW4-LG)
            (HW4-LG) edge[solid] node[vertRelation, right] {5d ``S-duality''\\$G$ nonsimply-laced} (HW4-LGv)
            (HW5.north east) edge[solid] node[relation, below] {$M_4 = M_2 \times S^1 \times S^1$} (HW3-L2G)
            (HW3-L2G) edge[solid] node[vertRelation, right] {5d ``S-duality''\\$G_\C$ nonsimply-laced} (HW3-L2Gv)
            (HW4-LG) edge[solid] node[vertRelation, left] {$M_3 = M_3' \bigcup_{\Sigma} M_3''$} (HSF-LM)
            (HW1-L2G) edge node[relation, above] {$M_2 = \hat{M_2}$} (HW3-L2G)
            (HW1-L2G) edge[solid] (HSF-L2M)
            (HW4-LGv.south west) edge[solid, bend right] node[relation, above] {$G$ nonsimply-laced} node[relation, below] {$M_3 = M_3' \bigcup_{\Sigma} M_3''$} (HSF-LM.north west)
            (HW3-L2Gv) edge node[relation, above] {$M_2 = \hat{M}_2$} (HW1-L2Gv)
            (HW1-L2Gv) edge[solid, bend left] node[relation, above] {$G_\C$ nonsimply-laced} (HW1-L2G.north east)
            ;
        \end{tikzpicture}
    \end{adjustbox}
    \caption{A web of relations amongst the Floer homologies.}
    \label{fig:web of relations combined}
\end{figure}

\bigskip\noindent\textit{Langlands Duality in a 4d-HW Atiyah-Floer Correspondence for Loop Gauge Groups}
\vspace*{0.5em}

In particular, from the web, we have, amongst other additional relations,
\begin{equation}\label{Langlands-dual Atiyah-Floer LG}
  \boxed {  \sum_{\tilde l} \text{HF}_{{d_{\tilde l}}}^{\text{HW}_4}(M_3, (LG)^\vee) \xleftrightarrow[\text{$M_3 = M_3' \bigcup_\Sigma M_3''$}]{ \text{$G$ nonsimply-laced}}  \sum_u \text{HSF}^{\text{Int}}_{d_u}({L\mathcal{M}^{G,\Sigma}_{\text{Higgs}}}, \mathscr{L}_0, \mathscr{L}_1)  }
\end{equation}
which is a Langlands duality in a 4d-HW Atiyah-Floer correspondence for loop gauge groups!

\section{A Fukaya-Seidel Type \texorpdfstring{$A_\infty$}{A-infinity}-category of Three-Manifolds}
\label{section: Fukaya-Seidel}

In this section, we will consider the case where $M_5 = M_3 \times \R^2$ with $M_3$ being a closed and compact three-manifold, and recast HW theory as either a 2d model on $\R^2$ or an SQM in path space. In doing so, we would be able to physically realize, via (1) a gauged Landau-Ginzburg (LG) interpretation of the 2d model and its equivalent SQM which define a soliton string theory, and (2) the 5d-HW partition function, a Fukaya-Seidel type (FS) $A_\infty$-category with objects being $G$ Hitchin configurations on $M_3$.

\subsection{HW Theory on \texorpdfstring{$M_3 \times \R^2$}{M3 x R2} as a 2d Model on \texorpdfstring{$\R^2$}{R2} or SQM in Path Space}
\label{subsection: hw theory as 2d model}

For $M_4 = M_3 \times \R$, where we shall now relabel the coordinate $x_0$ of $\R$ as $\tau$ for later convenience, \eqref{5d action terms removed for sqm} will be given by\footnote{%
Similar to the case in $\S$\ref{section: kk reduction to m3 x R}, upon taking a direction of $M_4$ to be $\R$, $B_{\mu\nu}, \chi_{\mu\nu} \in \Omega^{2, +}(M_3 \times \R)$ can be interpreted as $B_i, \chi_i \in \Omega^1(M_3)$ due to its self-duality properties. Thus in the following expression, we will be expressing the aforementioned fields as $B_i$ and $\chi_i$.}
\begin{equation}
  \label{5d action on m3 x r2}
  \begin{aligned}
    S_{\text{HW}, M_3 \times \R^2}
    = \frac{1}{e^2}
    & \int_{M_3 \times \R^2} dt d\tau d^3 x \, \Tr \bigg(
      \frac{1}{2} |F_{t\tau} - D^i B_{i}|^2
      + \frac{1}{2} |F_{ti} + D_\tau B_{i} - \epsilon_{ijk} D^j B^k|^2
    \\
    & + \frac{1}{4} \left|
        D_t B_{i}
        - F_{\tau i}
        - \frac{1}{2} \epsilon_{ijk} \left(F^{jk} - [B^j, B^k] \right)
      \right|^2
      - i \td\eta D_\tau \psi^\tau
      - i \eta D_\tau \td\psi^\tau
      - i \td\eta D_i \psi^i
      - i \eta D_i \td\psi^i
    \\
    & - 2i (\td\psi_{\tau} D_i - \td\psi_i D_\tau) \chi^i
      - 2i (\td\psi_{\tau} D_i - \psi_i D_\tau) \td\chi^i
      - 4i \epsilon^{ijk} (\td\psi_i D_j \chi_k + \psi_i D_j \td\chi_k)
    \\
    & - \td\psi_\tau D_t \psi^\tau
      - \td \psi_i D_t \psi^i
      - 2 \td\chi_i D_t \chi^i
      + 2 D_\tau \sigma D^\tau \bar\sigma
      + 2 D_i \sigma D^i \bar\sigma
    \\
    & + i \sigma \left(
        \{ \td\psi_i, \td\psi^i \}
        - 2 \{ \td\chi_i, \td\chi^i \}
      \right)
      + i \bar\sigma \left(
        \{ \psi_i, \psi^i \}
        - 2 \{ \chi_i, \chi^i \}
      \right)
      \bigg) \, .
  \end{aligned}
\end{equation}


\bigskip\noindent\textit{HW Theory as a 2d Model}
\vspace*{0.5em}

Let us recast HW theory as a 2d model on $\R^2$. To this end, we first expand out the bosonic part of \eqref{5d action on m3 x r2} to get
\begin{equation}
  \label{5d action on m3 x r2 decomposed}
  \begin{aligned}
    S_{\text{HW}, M_3 \times \R^2}
    = \frac{1}{e^2} \int_{\R^2} dt d\tau \, \Tr \int_{M_3} d^3 x \, \bigg(
    & \left|
        \dot{A_\tau} - A^\prime_t + [A_t, A_\tau] - D^i B_i
      \right|^2
      + 2 \sigma^\prime \bar\sigma^\prime
      + 2 D_i \sigma D^i \bar\sigma
    \\
    & + \frac{1}{4} \left|
      \dot{B}_i
      - A^\prime_i
      + [A_t, B_i]
      + D_i A_\tau
      - \frac{1}{2} \epsilon_{ijk} \left(F^{jk} - [B^j, B^k] \right)
      \right|^2
    \\
    & + \left|
      \dot{A}_i
      + B^\prime_i
      - D_i A_t
      + [A_\tau, B_i]
      - \epsilon_{ijk} D^j B^k
      \right|^2
      + \dots
      \bigg)
      \, ,
  \end{aligned}
\end{equation}
where $\dot\Phi \equiv \partial_t \Phi$ and $\Phi^\prime \equiv \partial_\tau \Phi$ for any field $\Phi$, and ``$\dots$'' contains the fermionic terms from \eqref{5d action on m3 x r2}. From this, we can identify the conditions (on the zero-modes) that minimize the action $S_{\text{HW}, M_3 \times \R^2}$ as
\begin{equation}
  \label{m3 x r2 bps}
  \begin{aligned}
    \dot{A}_i
    + B^\prime_i
    - D_i A_t
    + [A_\tau, B_i]
    &= \epsilon_{ijk} D^j B^k
      \, ,
    \\
    \dot{B}_i
    - A^\prime_i
    + [A_t, B_i]
    + D_i A_\tau
    &= \frac{1}{2} \epsilon_{ijk} \left(F^{jk} - [B^j, B^k] \right)
      \, ,
    \\
    \dot{A_\tau} - A^\prime_t + [A_t, A_\tau]
    &= D^i B_i
      \, .
  \end{aligned}
\end{equation}

Next, noting that we are physically free to rotate $\R^2$ about the origin, \eqref{m3 x r2 bps} becomes
\begin{equation}
  \label{m3 x r2 bps 2}
  \begin{aligned}
    \dot{A}_i
    + B^\prime_i
    - D_i A_t
    + [A_\tau, B_i]
    &= \epsilon_{ijk} \left(
      D^j B^k \cos \theta
      - \frac{1}{2} \epsilon_{ijk} \left(F^{jk} - [B^j, B^k] \right) \sin \theta
      \right)
      \, ,
    \\
    \dot{B}_i
    - A^\prime_i
    + [A_t, B_i]
    + D_i A_\tau
    &= \epsilon_{ijk} \left(
      D^j B^k \sin \theta
      + \frac{1}{2} \left(F^{jk} - [B^j, B^k] \right) \cos \theta
      \right)
      \, ,
    \\
    \dot{A_\tau} - A^\prime_t + [A_t, A_\tau]
    &= D^i B_i
      \, ,
  \end{aligned}
\end{equation}
where $\theta$ is the angle of rotation. The motivation for this rotation will be clear in the next subsection.

Just as in \S\ref{subsection: 5d theory/5d to sqm}, this means that \eqref{5d action on m3 x r2 decomposed} can be written as\footnote{%
The terms with $\sigma$ and $\bar\sigma$ are omitted, as we will see that they will be integrated off in the next step.}
\begin{equation}
  \label{5d action on m3 x r2 with complex bps}
  \begin{aligned}
    S_{\text{HW}, M_3 \times \R^2}
    = \frac{1}{e^2} \int_{\R^2} dt d\tau \, \Tr \int_{M_3} d^3 x \, \bigg(
    & | D_\tau B_i + D_t A_i + p_i |^2
      + | D_\tau A_i - D_t B_i + q_i |^2
      + |  F_{\tau t} + r |^2
      + \dots
      \bigg)
      \, ,
  \end{aligned}
\end{equation}
where
\begin{equation}
  \label{5d action on m3 x r2 boosted components}
  \begin{aligned}
    p_i
    &= - \partial_i A_t
      - \epsilon_{ijk} \left(
      D^j B^k \cos\theta
      - \frac{1}{2} \left( F^{jk} - [B^j, B^k] \right) \sin\theta
      \right)
      \, ,
    \\
    q_i
    &= - \partial_i A_\tau
      + \epsilon_{ijk} \left(
      D^j B^k \sin\theta
      + \frac{1}{2} \left( F^{jk} - [B^j, B^k] \right) \cos\theta
      \right)
      \, ,
    \\
    r
    &= D^i B_i
      \, .
  \end{aligned}
\end{equation}

Finally, noting again that the curvature terms for the fermions can be obtained by integrating out the fields $\sigma, \bar\sigma$ (as was done in the context of an SQM in equations \eqref{5d scalar fields}, and a 3d sigma model in \eqref{3d sigma model action}), we can, after suitable rescalings, recast \eqref{5d action on m3 x r2 with complex bps} as an equivalent 2d model action\footnote{To arrive at this expression for the action, we have further integrated out an auxiliary scalar field $H =  D^a B_a$ whose contribution to the action was $|H|^2$. We have also made use of Stokes' theorem and the fact that $M_3$ has no boundary to note that $\partial_i A_t$ and $\partial_i A_\tau$ should vanish in their integration over $M_3$.
\label{footnote:reason for no D*B in the 2d action}}

\begin{equation}
  \label{2d action on r2}
  S_{\text{2d}}
  = \frac{1}{e^2} \int d\tau \int dt  \, \bigg(
  \left| D_\tau B^a + D_t A^a + p^a \right|^2
  + \left| D_\tau A^a - D_t B^a + q^a \right|^2
  + \left|F_{\tau t}\right|^2
  + \dots
  \bigg)
  \, ,
\end{equation}
where $(A^a, B^a)$ and $a, b$ are the coordinates and indices on the space $\mathfrak{A}_{3}$ of irreducible $(A_i, B_i)$ fields on $M_3$, respectively, $F_{\tau t}$ is a field strength on $\R^2$, and the expressions for $p^a, q^a$ are
\begin{equation}
  \label{5d action r2 boosted components}
  \begin{aligned}
    p^a
    &= - \epsilon^{abc} \left(
      D_b B_c \cos\theta
      - \frac{1}{2} \left(F_{bc} - [B_b, B_c]\right) \sin\theta
      \right)
      \, ,
    \\
    q^a
    &= \epsilon^{abc} \left(
      D_b B_c \sin\theta
      + \frac{1}{2} (F_{bc} - [B_b, B_c]) \cos\theta
      \right)
      \, .
  \end{aligned}
\end{equation}

In short, HW theory on $M_3 \times \R^2$ can be regarded as a  2d \textit{gauged} sigma model along the $(t, \tau)$-directions with target $\mathfrak A_{3}$ whose action is \eqref{2d action on r2}.

\bigskip\noindent\textit{The 2d Model as a 1d SQM}
\vspace*{0.5em}

Now that we have recast HW theory as a 2d gauged sigma model on $\R^2$ with target $\mathfrak A_{3}$, we want to single out the direction in $\tau$ as a direction in ``time'' and further recast it as a 1d SQM. After suitable rescalings, the equivalent SQM action can be obtained from \eqref{2d action on r2} as\footnote{%
In 1d, $A_\tau$ has no field strength and is thus non-dynamical. Therefore, it can be integrated out to furnish the Christoffel connection for the fermion kinetic terms as was done in the context of an SQM in \eqref{A_t eom}, leaving us with an SQM without $A_\tau$.
We have also omitted a term $|\partial_{\tau} A_t|^2$ in the following expression as it will just lead to the trivial condition $\partial_{\tau} A_t = 0$.
\label{footnote:2d model integrate off gauge fields}
}
\begin{equation}
    \label{sqm action m3 x R2}
    \begin{aligned}
        S_{\text{SQM,2d}}
        = \frac{1}{e^2} \int d\tau \, \bigg(
            & \left\vert \partial_\tau B^{\alpha} + g^{\alpha \beta}_{\mathcal{M}(\R, \mathfrak{A}_3)} \frac{\partial h}{\partial B^\beta} \right\vert^2
            + \left\vert \partial_\tau A^{\alpha} + g^{\alpha \beta}_{\mathcal{M}(\R, \mathfrak{A}_3)} \frac{\partial h}{\partial A^\beta} \right\vert^2
            + \dots
            \bigg)
            \, ,
    \end{aligned}
\end{equation}
where $(A^\alpha, B^\alpha)$ and $\alpha, \beta$ are coordinates and indices on the path space $\mathcal{M}(\R, \mathfrak{A}_{3})$ of maps from $\R$ to $\mathfrak{A}_{3}$, respectively; $g^{\alpha \beta}_{\mathcal{M}(\R, \mathfrak{A}_3)}$ is the metric on $\mathcal{M}(\R, \mathfrak{A}_{3})$; and $h(A, B)$ is the potential function.

In short, HW theory on $M_3 \times \R^2$ can also be regarded as a 1d SQM along the $\tau$-direction in $\mathcal{M}(\R, \mathfrak{A}_{3})$ whose action is \eqref{sqm action m3 x R2}.

\subsection{Non-constant Paths, Solitons, and HW Configurations}
\label{subsection: solitons of FS cat}

\bigskip\noindent\textit{$\theta$-deformed Non-constant Paths in the SQM}
\vspace*{0.5em}

By the squaring argument~\cite{blau1993topological} applied to \eqref{sqm action m3 x R2}, we find that the equivalent SQM will localize onto configurations that set the LHS and RHS of the expression within the squared terms therein \emph{simultaneously} to zero.  In other words, the equivalent SQM localizes onto $\tau$-invariant  critical points of $h(A, B)$ that obey
\begin{equation}
  \label{soliton equations}
  \begin{aligned}
    \dot{A}^\alpha
    &= \epsilon^{\alpha\beta\gamma} \left(
      D_\beta B_\gamma \cos\theta
      - \hlf \left(F_{\beta\gamma} - [B_\beta, B_\gamma]\right) \sin\theta
      \right)
      \, ,
    \\
    \dot{B}^\alpha
    &= \epsilon^{\alpha\beta\gamma} \left(
      D_\beta B_\gamma \sin\theta
      + \hlf \left(F_{\beta\gamma} - [B_\beta, B_\gamma]\right) \cos\theta
      \right)
      \, .
  \end{aligned}
\end{equation}
These are $\theta$-deformed non-constant paths in $\mathcal{M}(\R, \mathfrak{A}_{3})$.

\bigskip\noindent\textit{$\theta$-deformed Solitons in the 2d Model}
\vspace*{0.5em}

By comparing \eqref{sqm action m3 x R2} with \eqref{2d action on r2}, we find that they correspond, in the equivalent 2d gauged sigma model on $\R^2$, to $\theta$-deformed solitons along the $t$-direction defined by $p^a - D_t A^a + [A_\tau, B^a] = 0$ and $q^a + D_t B^a + [A_\tau, A^a] = 0$, i.e.,
\begin{equation}
  \label{Soliton 2d}
  \begin{aligned}
    \dot{A^a}
    & = - [A_t, A^a]
      - [A_\tau, B^a]
      + \epsilon^{abc} \left(
      D_b B_c \cos\theta
      + \hlf \left(F_{bc} - [B_b, B_c]\right) \sin\theta
      \right)
      \, ,
    \\
    \dot{B^a}
    & = - [A_t, B^a]
      + [A_\tau, A^a]
      + \epsilon^{abc} \left(
      D_b B_c \sin\theta
      + \hlf \left(F_{bc} - [B_b, B_c]\right)
      \cos\theta
      \right)
      \, ,
  \end{aligned}
\end{equation}
with the additional conditions
\begin{equation}
  \label{soliton 2d auxiliary conditions}
  \begin{aligned}
    F_{t\tau} = 0 = D^a B_a \, .
  \end{aligned}
\end{equation}

\bigskip\noindent\textit{$\tau$-independent, $\theta$-deformed HW Configurations in HW Theory}
\vspace*{0.5em}

In turn, by comparing \eqref{2d action on r2} with \eqref{5d action on m3 x r2 with complex bps}, these correspond, in HW theory, to $\tau$-independent, $\theta$-deformed HW configurations on $M_5$ defined by $p_i + D_t A_i + [A_\tau, B_i] = 0$, $q_i - D_t B_i + [A_\tau, A_i] = 0$, and $r - D_t A_\tau =0$, i.e.,
\begin{equation}
  \label{HW 5d}
  \begin{aligned}
    \dot{A}_i
    &= - [A_t, A_i]
      - [A_\tau, B_i]
      + \partial_i A_t
      + \epsilon_{ijk} \left(
      D^j B^k \cos\theta
      - \hlf \left(F^{jk} - [B^j, B^k]\right) \sin\theta
      \right)
      \, ,
    \\
    \dot{B}_i
    &= - [A_t, B_i]
      + [A_\tau, A_i]
      - \partial_i A_\tau
      + \epsilon_{ijk} \left(
      D^j B^k \sin\theta
      + \hlf \left(F^{jk} - [B^j, B^k]\right) \cos\theta
      \right)
      \, ,
    \\
    \dot{A}_\tau
    &= - [A_t, A_\tau]
      \, ,
    \\
    0
    &= D^i B_i
      \, .
  \end{aligned}
\end{equation}

\bigskip\noindent\textit{Soliton Endpoints Corresponding to $\theta$-deformed Hitchin Configurations on $M_3$}
\vspace*{0.5em}

The endpoints of the solitons at $t = \pm \infty$ are fixed. Also, at $t = \pm \infty$, the finite-energy 2d gauge fields $A_\tau, A_t$ ought to decay to zero. In other words, the endpoints of the soliton would be defined by \eqref{Soliton 2d} and \eqref{soliton 2d auxiliary conditions} with $\dot{A}^a = 0 = \dot{B}^a$ and $A_\tau, A_t \to 0$, i.e.,
\begin{equation}
  \label{endpoints soliton equations}
  \begin{aligned}
    \epsilon^{abc} \left(
    D_a B_b \cos\theta
    - \hlf \left(F_{ab} - [B_a, B_b]\right) \sin\theta
    \right)
    &= 0
      \, ,
    \\
    \epsilon^{abc} \left(
    D_a B_b \sin\theta
    + \hlf \left(F_{ab} - [B_a, B_b]\right) \cos\theta
    \right)
    &= 0
      \, ,
    \\
    D^a B_a
    &= 0
      \, ,
  \end{aligned}
\end{equation}
where the condition that the 2d field strength $F_{t\tau} = 0$ is trivially-satisfied.

In turn, they correspond, in HW theory, to the configurations that obey
\begin{equation}
  \label{endpoints HW}
  \begin{aligned}
    \epsilon_{ijk} \left(
    D^j B^k \cos\theta
    - \hlf \left(F^{jk} - [B^j, B^k]\right) \sin\theta
    \right)
    &= 0
      \, ,
    \\
    \epsilon_{ijk} \left(
    D^j B^k \sin\theta
    + \hlf \left(F^{jk} - [B^j, B^k]\right) \cos\theta
    \right)
    &= 0
      \, ,
    \\
    D^i B_i
    &= 0
      \, .
  \end{aligned}
\end{equation}
These are a $\theta$-deformed version of the $G$ Hitchin equations on $M_3$.

\bigskip\noindent\textit{Soliton Endpoints Corresponding to Higgs Pairs or Flat Complexified Connections on $M_3$}
\vspace*{0.5em}

At $\theta = \pi$, \eqref{endpoints HW} will become
\begin{equation} \label{soliton equations endpoints theta=pi}
   \begin{aligned}
     \hlf \epsilon^{ijk} \left(F_{jk} - [B_j, B_k] \right) &= 0 \, , \\
       \epsilon^{ijk} D_j B_k &= 0 \, , \\
        D^i B_i & = 0.
    \end{aligned}
\end{equation}
This tells us that the endpoints of the $\pi$-deformed solitons correspond to $G$ Higgs pairs on $M_3$.

At $\theta = \pi /2$, \eqref{endpoints HW} can also be expressed as
\begin{equation}
    \label{soliton equations endpoints theta=pi/2}
    \begin{aligned}
    \epsilon^{ijk} {\mathcal{F}}_{jk} &= 0 \, .
    \end{aligned}
\end{equation}
where ${\mathcal{F}} = d{{\mathcal{A}}} + { {\mathcal{A}}} \wedge {{\mathcal{A}}}$, and $\mathcal{A}^i = A^i + iB^i$ can be interpreted as a connection of a complexified gauge group $G_\C$ on $M_3$. This tells us that the endpoints of the $\pi/2$-deformed solitons  correspond to flat $G_\C$ connections on $M_3$.

\bigskip\noindent\textit{$\theta$, the Moduli Space of Hitchin Equations on $M_3$, and Isolated Non-degenerate Solitons}
\vspace*{0.5em}

Now recall that the solutions to Hitchin equations on a Riemann surface are Higgs pairs or flat $G_\C$ connections when its hyperk\"{a}hler moduli space is in complex structure $I$ or a linear combination of the complex structures $J$ and $K$, respectively. Thus, when $\theta = \pi$ or $\pi /2$ whence the endpoints of the soliton  correspond to Higgs pairs or flat $G_\C$ connections on $M_3$, respectively, the underlying moduli space of Hitchin equations on $M_3$ is in ``complex structure $I$'' or a linear combination of ``complex structures $J$ and $K$''.

In other words, as we vary $\theta$, the endpoints of the soliton interpolate between being associated with Higgs pairs and flat $G_\C$ connections on $M_3$, while the underlying moduli space of Hitchin equations on $M_3$ interpolate between being in ``complex structure $I$'' and a linear combination of ``complex structures $J$ and $K$''.

For an appropriate choice of $G$ and $M_3$, specifically, if (1) $G$ is compact and $M_3$ is of non-negative Ricci curvature such as a three-sphere or its quotient, or (2) $M_3$ has a finite $G_{\mathbb C}$ representation variety, the endpoints and therefore the solitons themselves would be isolated and non-degenerate.\footnote{For such a choice of $G$ and $M_3$, the moduli space of flat $G_\C$ connections on $M_3$ will be made up of isolated points, and up to physically-inconsequential $\mathcal Q$-exact perturbations of the HW action, these points can also be made non-degenerate. As such, when $\theta = \pi /2$, the endpoints and therefore the solitons themselves would be isolated and non-degenerate. As the physical theory is symmetric under a variation of $\theta$, this observation about the endpoints of the solitons will also hold for any $\theta$. We would like to thank A.~Haydys for discussions on this. \label{footnote: isolation and non-degeneracy of flat GC configs on M3}} We shall henceforth assume such a choice, so that our SQM with action \eqref{sqm action m3 x R2} will localize onto isolated and non-degenerate $\tau$-invariant critical points of its $h(A, B)$ potential, whence its partition function will persist to be a discrete and non-degenerate sum of contributions.

\subsection{The 2d  Model on \texorpdfstring{$\R^2$}{R2} and an Open String Theory in \texorpdfstring{$\mathfrak A_3$}{A3}}
\label{subsection: open string theory of FS cat}

\bigskip\noindent\textit{Flow Lines of the SQM as BPS Worldsheets of the 2d Model}
\vspace*{0.5em}

The classical trajectories or flow lines of the equivalent SQM  governed by the gradient flow equations (defined by setting the expression within the squared terms in \eqref{sqm action m3 x R2} to zero)
\begin{equation}
\begin{aligned}
    \frac{dA^{\alpha}}{d \tau} & = - g^{\alpha \beta}_{\mathcal{M}(\R, \mathfrak{A}_3)} \frac{\partial h}{\partial A^\beta} \, , \\
    \frac{dB^{\alpha}}{d \tau} & = - g^{\alpha \beta}_{\mathcal{M}(\R, \mathfrak{A}_3)} \frac{\partial h}{\partial B^\beta} \, ,
\end{aligned}
\end{equation}
go from one $\tau$-invariant critical point of $h$ to another in $\mathcal{M}(\R, \mathfrak A_{3})$. They therefore correspond, in the equivalent 2d gauged sigma model with target $\mathfrak A_{3}$, to worldsheets that have at $\tau = \pm \infty$, the $\theta$-deformed solitons $\gamma_{\pm}(t, \theta)$ defined by \eqref{Soliton 2d} and \eqref{soliton 2d auxiliary conditions} with the finite-energy 2d gauge fields $A_\tau, A_t \to 0$, i.e.,
\begin{equation}
  \label{soliton equations in 2d model}
  \begin{aligned}
    \frac{d{A}^a}{dt}
    &= \epsilon^{abc} \left(
      D_b B_c \cos\theta
      - \hlf \left(F_{bc} - [B_b, B_c]\right) \sin\theta
      \right)
      \, ,
    \\
    \frac{d{B}^a}{dt}
    &= \epsilon^{abc} \left(
      D_b B_c \sin\theta
      + \hlf \left(F_{bc} - [B_b, B_c]\right) \cos\theta
      \right)
      \, ,
    \\
    0
    &= D^a B_a
      \, ,
  \end{aligned}
\end{equation}
whose endpoints $\gamma(\pm \infty, \theta)$ at $t = \pm \infty$ are defined by
\begin{equation}
  \label{soliton equations endpoints in 2d model}
  \begin{aligned}
    \epsilon^{abc} \left(
    D_b B_c \cos\theta
    - \hlf \left(F_{bc} - [B_b, B_c]\right) \sin\theta
    \right)
    &= 0
      \, ,
    \\
    \epsilon^{abc} \left(
    D_b B_c \sin\theta
    + \hlf \left(F_{bc} - [B_b, B_c]\right) \cos\theta
    \right)
    &= 0
      \, ,
    \\
    D^a B_a
    &= 0
      \, ,
  \end{aligned}
\end{equation}
which is just \eqref{soliton equations in 2d model} with $\dot {A^a} = 0 = \dot {B^a}$.\footnote{Notice that the soliton can translate in the $\tau$-direction due to its ``center of mass'' motion, and because it is $\tau$-invariant, the soliton is effectively degenerate. This reflects the fact that generically, each critical point of $h$ is degenerate and does not correspond to a point but a real line $\R$ in $\mathcal{M} (\R, {\mathfrak A}_3)$. Nonetheless, one can perturb $h$ via the addition of physically-inconsequential $\mathcal{Q}$-exact terms to the SQM action, and collapse the degeneracy such that the critical points really correspond to points in $\mathcal{M} (\R, {\mathfrak A}_3)$. This is tantamount to factoring out the center of mass degree of freedom of the soliton, and fixing it at $\tau = \pm \infty$.} 

Just as the flow lines are governed by the gradient flow equations which really are the BPS equations of the SQM, the corresponding worldsheets will be governed by the BPS equations of the 2d model (defined by setting the expression within the squared terms in \eqref{2d action on r2} to zero)
\begin{equation}
  \label{m3 x r2 bps without gauge fields}
  \begin{aligned}
    \frac{DA^a}{Dt} + \frac{DB^a}{D\tau}
    &= \epsilon^{abc} \left(
      D_b B_c \cos\theta
      - \hlf \left(F_{bc} - [B_b, B_c]\right) \sin\theta
      \right)
      \, ,
    \\
    \frac{DB^a}{Dt} - \frac{DA^a}{D\tau}
    &= \epsilon^{abc} \left(
      D_b B_c \sin\theta
      + \hlf \left(F_{bc} - [B_b, B_c]\right) \cos\theta
      \right)
      \, ,
    \\
    \frac{dA_t}{d\tau} - \frac{dA_\tau}{dt}
    &= [A_t, A_\tau]
      \, ,
  \end{aligned}
\end{equation}
as well as the condition on the auxiliary field in footnote \ref{footnote:reason for no D*B in the 2d action}, i.e., $D^a B_a = 0$.
We shall henceforth refer to such worldsheets as BPS worldsheets, and they correspond to the classical trajectories of the equivalent 2d gauged sigma model with target ${\mathfrak A}_3$.

\bigskip\noindent\textit{BPS Worldsheets with  Boundaries Corresponding to $\theta$-deformed Hitchin Configurations on $M_3$}
\vspace*{0.5em}

Recall from the previous subsection that at $\theta = \pi /2$,  the endpoints of the solitons correspond to flat $G_\C$ connections on $M_3$. If there are `$k$' such configurations $\{ \mathcal{A}_1(\pi/2), \dots, \mathcal{A}_k(\pi/2) \} $, it would mean that a $\pi/2$-deformed soliton at $\tau = \pm \infty$ can be further denoted as $\gamma^{IJ}_{\pm}(t, \pi/2)$, where $I, J = 1, \dots, k$ indicates that its left and right endpoints $\gamma^I(- \infty, \pi/2)$ and $\gamma^J( + \infty, \pi/2)$ would correspond to $\mathcal{A}_I(\pi/2)$ and $\mathcal{A}_J(\pi/2)$, respectively. Again, since the physical theory is symmetric under a variation of $\theta$, this would be true at any $\theta$. In other words, we can also denote any $\theta$-deformed soliton at $\tau=\pm\infty$ as $\gamma^{IJ}_{\pm}(t, \theta)$, where its left and right endpoints $\gamma^I(- \infty, \theta)$ and $\gamma^J( + \infty, \theta)$  would correspond to $\mathcal{A}_I(\theta)$ and $\mathcal{A}_J(\theta)$, respectively, where the $\mathcal{A}_{*}(\theta)$'s are $k$ number of $\theta$-deformed $G$ Hitchin configurations on $M_3$.

\begin{figure}
    \centering
    \begin{tikzpicture}
        \coordinate (lt) at (0,4) [label=left: \footnotesize] {};
        \coordinate (rt) at (4,4) [label=right: \footnotesize {$\gamma^J(+\infty, \theta)$}] {}
        edge node[pos=0.5, above ,name=top-soliton] {$\gamma^{IJ}_+(t, \theta)$} (lt) {};
        \coordinate (lb) at (0,0) [label=left: \footnotesize {$\gamma^I(-\infty, \theta)$}] {}
        edge node[pos=0.5, left, name=left-soliton] {$ \mathcal{A}_I(\theta)$} (lt) {};
        \coordinate (rb) at (4,0) [label=right: \footnotesize {$\gamma^J(+\infty, \theta)$}] {}
        edge node[pos=0.5, below ,name=bot-soliton] {$\gamma^{IJ}_-(t, \theta)$} (lb) {}
        edge node[pos=0.5, right, name=right-soliton] {$\mathcal{A}_J(\theta)$} (rt) {};;
        \draw (lb) -- (lt);
        \draw (rb) -- (rt);
        \coordinate (co) at (5,0);
        \coordinate (cx) at (5.5,0);
        \node at (cx) [right=2pt of cx] {$t$};
        \coordinate (cy) at (5,0.5);
        \node at (cy) [above=2pt of cy] {$\tau$};
        \draw[->] (co) -- (cx);
        \draw[->] (co) -- (cy);
    \end{tikzpicture}
    \caption{BPS worldsheet with solitons $\gamma^{IJ}_\pm$ and boundaries corresponding to $\mathcal{A}_I(\theta)$ and $\mathcal{A}_J(\theta)$.}
    \label{fig:mu1 map}
\end{figure}
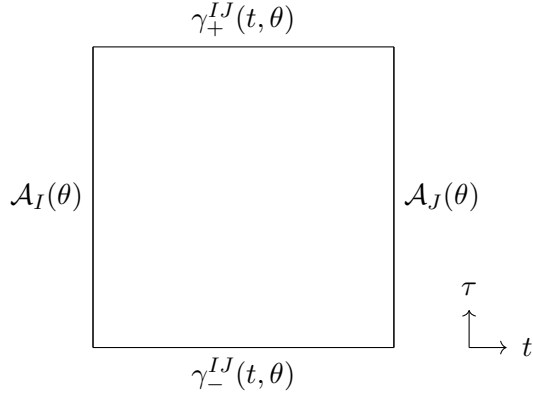

In short, since the $\mathcal{A}_{*}(\theta)$'s are $\tau$-independent and therefore, have the same values for all $\tau$, we have BPS worldsheets of the kind shown in Fig.~\ref{fig:mu1 map}.

\bigskip\noindent\textit{The 2d Model on $\R^2$ and an Open String Theory in $\mathfrak A_3$}
\vspace*{0.5em}

Hence, one can understand the 2d gauged sigma model on $\R^2$ to define an open string theory in $\mathfrak A_3$, with \textit{effective} worldsheet and boundaries shown in Fig.~\ref{fig:mu1 map}, where $\tau$ and $t$ are the temporal and spatial directions, respectively.

\subsection{A Gauged LG Model, Soliton String Theory, the 5d-HW Partition Function, and an FS Type \texorpdfstring{$A_\infty$}{A-infinity}-category of Hitchin Configurations on \texorpdfstring{$M_3$}{M3}}
\label{subsection: Fukaya-Seidel category of M3}

\bigskip\noindent\textit{The 2d Model as a Gauged LG Model}
\vspace*{0.5em}

Notice that we can also express the first two lines of \eqref{m3 x r2 bps without gauge fields} as
\begin{equation}
  \label{m3 x r2 bps without gauge fields complexified}
  \frac{D\phi^a}{D\tau}
  + i\frac{D\phi^{a}}{Dt}
  = - \frac{1}{2} \epsilon^{abc} \bar{\mathcal{F}}_{bc} e^{i\theta}
  \, ,
\end{equation}
where $\phi^a = A^a + iB^a$ can be interpreted as a holomorphic coordinate on the space $\mathscr A_3$ of connections of a complexified gauge group $G_\C$ on $M_3$, and $\bar{\mathcal{F}} = d\bar\phi + \bar\phi \wedge \bar\phi$. That is, the BPS equations of the 2d gauged sigma model with target $\mathfrak{A}_3$ which govern the trajectory of the BPS worldsheet are effectively \eqref{m3 x r2 bps without gauge fields complexified} (plus $F_{\tau t} = 0$).

In turn, this means that we can express the 2d gauged sigma model action in~\eqref{2d action on r2} as
\begin{equation}
  \label{2d LG}
  \begin{aligned}
    S_{\text{LG}, \mathscr{A}_3}
    &= \frac{1}{e^2} \int d\tau dt  \, \bigg(
      \left| D_\tau \phi^a
      + i D_t \phi^a
      + \frac{1}{2} \epsilon^{abc} \bar{\mathcal{F}}_{bc} e^{i\theta} \right|^2
      + |F_{\tau t}|^2
      + \dots
      \bigg)
      \\
    &= \frac{1}{e^2} \int d\tau dt  \, \bigg(
      \left| D_\mu \phi^a \right|^2
      + \left|\frac{\partial W}{\partial \phi^a}\right|^2
      + |F_{\tau t}|^2
      + \dots
      \bigg)
      \, .
  \end{aligned}
\end{equation}
In other words, our 2d gauged sigma model with target $\mathfrak{A}_3$ can also be interpreted as a gauged LG model with target $\mathscr{A}_3$ with holomorphic superpotential $W(\phi)$!

As the gradient vector field of $W$ is $\mathcal F$, $W$ must therefore be a \textit{Chern-Simons function of $\phi$}. Furthermore, $W$ is \textit{complex-valued} (since $\phi$ is complex).

By setting $\phi^{\prime a} = 0$ and $A_\tau, A_t \to 0$ in  \eqref{m3 x r2 bps without gauge fields complexified}, we can read off the  LG soliton equation corresponding to $\gamma^{IJ}_{\pm}(t, \theta)$ (that re-expresses \eqref{soliton equations in 2d model}) as
\begin{equation}
    \label{soliton equations in 2d model complexified}
    \frac{d\phi^{a}}{dt}
    = g^{a \bar b}_{\mathscr{A}_3} \left(
      \frac{i\zeta}{2} \frac{\partial W}{\partial \phi^b}
    \right)^*\, ,
\end{equation}
where $\zeta = e^{- i \theta }$ and $g^{a \bar b}_{\mathscr{A}_3}$ is the metric
on $\mathscr{A}_3$.

By setting $\dot\phi^{a} = 0$ in \eqref{soliton equations in 2d model complexified}, we can read off the LG soliton endpoint equation corresponding to $\gamma^{I,J}(\mp \infty, \theta)$ (that re-expresses \eqref{soliton equations endpoints in 2d model}) as
\begin{equation}
    \label{soliton equations endpoints in 2d model complexified}
    g^{a \bar b}_{\mathscr{A}_3} \left(
      \frac{i\zeta}{2} \frac{\partial W}{\partial \phi^b}
    \right)^* = 0 \, .
\end{equation}

Recall from the end of the previous subsection that we only consider certain $G$ and $M_3$ such that the endpoints of the LG solitons are isolated and non-degenerate. Thus, from their definition in \eqref{soliton equations endpoints in 2d model complexified} which tells us that they correspond to critical points of $W$, we conclude that $W$ can be regarded as a complex Morse function in $\mathscr A_3$.

Furthermore, it is a standard fact that the LG soliton defined by \eqref{soliton equations in 2d model complexified} maps to a straight line segment $[W^I (\theta), W^J (\theta)]$ in the complex $W$-plane that starts and ends at the critical values $W^I (\theta) = W(\gamma^{I}(-\infty, \theta))$ and $W^J (\theta) = W(\gamma^{J}(+\infty, \theta))$, where its slope depends on $\theta$ (via $\zeta$). This fact will be relevant shortly. We shall also assume that $\text{Re} \, W^I (\theta) < \text{Re} \, W^J (\theta)$.

\bigskip\noindent\textit{The LG Model as an LG SQM}
\vspace*{0.5em}

Last but not least, from \eqref{2d LG}, we find that the LG SQM action (that re-expresses  \eqref{sqm action m3 x R2}) will be given by\footnote{%
  As explained in footnote~\ref{footnote:2d model integrate off gauge fields}, $A_\tau$ is non-dynamical in the 1d SQM and has thus been integrated out to furnish the Christoffel connection for the fermion kinetic terms, and the extra term in the action containing $A_t$ is omitted because the 1d SQM localizes onto the trivial condition of it vanishing.}
\begin{equation}
    \label{sqm action 2d LG}
    \begin{aligned}
        S_{\text{LG SQM}, \mathcal{M}(\R, \mathscr{A}_3)}
        = \frac{1}{e^2} \int d\tau \, \bigg(
            & \left\vert \frac{d\phi^\alpha}{d\tau} + g^{\alpha \beta}_{\mathcal{M}(\R, \mathscr{A}_3)} \frac{\partial h}{\partial \phi^\beta} \right\vert^2
            + \dots \bigg) \, ,
    \end{aligned}
\end{equation}
where $\phi^\alpha = A^\alpha + i B^\alpha$ can be interpreted as a holomorphic coordinate of $\mathcal{M}(\R, \mathscr A_3)$, $g^{\alpha \beta}_{\mathcal{M}(\R, \mathscr{A}_3)}$ is the metric on $\mathcal{M}(\R, \mathscr A_3)$, and $h (\phi)$ is a \textit{real-valued} potential in $\mathcal{M}(\R, \mathscr A_3)$.

Again, the LG SQM will localize onto configurations that set the LHS and RHS of the expression within the squared term in \eqref{sqm action 2d LG} \textit{simultaneously} to zero. In other words, it would localize onto the $\tau$-invariant critical points of $h(\phi)$ that correspond to the LG solitons defined by \eqref{soliton equations in 2d model complexified}.

For our consideration of $G$ and $M_3$, the LG solitons, like their endpoints, would be isolated and non-degenerate. Thus, $h$ can be regarded as a real-valued Morse functional in $\mathcal{M}(\R, \mathscr A_3)$.

\bigskip\noindent\textit{Morphisms from $\mathcal{A}_I(\theta)$ to $\mathcal{A}_J(\theta)$ as Floer Homology Classes of Intersecting Thimbles}
\vspace*{0.5em}

It will now be useful to describe the LG soliton solutions via the notion of a thimble, which can be thought of as submanifolds of a certain fiber space over the complex $W$-plane. Specifically, solutions satisfying
\begin{equation}
  \lim_{t \to -\infty} \gamma_\pm (t, \theta)  = \gamma^I (-\infty, \theta)
\end{equation}
are known as left thimbles, and those that satisfy
\begin{equation}
  \lim_{t \to +\infty} \gamma_\pm (t, \theta)  = \gamma^J (+\infty, \theta)
\end{equation}
are known as right thimbles. Clearly, a soliton solution $\gamma^{IJ}_\pm(t, \theta)$  must simultaneously be in a left and right thimble, and is thus represented by a transversal intersection of the two in the fiber space over the line segment $[W^I (\theta), W^J (\theta)]$.\footnote{This intersection is guaranteed at some $\theta$, which we can freely tune as the physical theory is symmetric under its variation.} Denote the set of all such intersections by $S_{IJ}$. Then, for each LG soliton pair $\gamma^{IJ}_\pm (t, \theta)$ whose left and right endpoints correspond to $\mathcal{A}_I(\theta)$ and $\mathcal{A}_J(\theta)$ on a BPS worldsheet as shown in Fig.~\ref{fig:mu1 map}, we have a pair of intersection points $p^{IJ}_\pm (\theta) \in S_{IJ}$.

As in earlier sections, the LG SQM in $\mathcal{M}(\R, \mathscr A_3)$ with action \eqref{sqm action 2d LG} will physically realize an LG Floer homology, where the chains of the LG Floer complex will be generated by LG solitons which we can identify with $p^{*  *}_\pm(\theta)$, and the differential will be realized by the flow lines governed by the gradient flow equation satisfied by $\tau$-varying configurations which set the expression within the squared term in \eqref{sqm action 2d LG} to zero. In particular, the LG SQM partition function will be given by
\begin{equation}\label{LG SQM partition function}
   {  \mathcal{Z}_{\text{LG SQM}, \mathcal{M}(\R, \mathscr{A}_3)}(G) = \sum_{I\neq J = 1}^k \, \sum_{p^{IJ}_\pm (\theta) \in S_{IJ}}  \text{HF}^{G}_{d_p}(p^{IJ}_\pm (\theta))} \, ,
\end{equation}
where the contribution $\text{HF}^{G}_{d_p}$  can be identified with a class in a Floer homology generated by intersection points of thimbles which represent solitons whose endpoints correspond to $\theta$-deformed $G$ Hitchin configurations on $M_3$, and $d_p$ is its degree measured by the number of outgoing flow lines from the critical points of $h$ in $\mathcal{M}(\R, \mathscr A_3)$ that can be identified as $p^{IJ}_\pm (\theta)$.

Notice that we have omitted the `$\theta$' label on the LHS of \eqref{LG SQM partition function}, as we recall that the physical theory is actually equivalent for all values of $\theta$.

At any rate, $\mathcal{Z}_{\text{LG SQM}, \mathcal{M}(\R, \mathscr{A}_3)}(G)$ in \eqref{LG SQM partition function} is a sum of LG solitons with endpoints defined by \eqref{soliton equations in 2d model complexified} with \eqref{soliton equations endpoints in 2d model complexified}, or equivalently, \eqref{soliton equations in 2d model} with \eqref{soliton equations endpoints in 2d model}, respectively. In other words, we can write
\begin{equation} \label{CF = HF}
  {  \text{CF}(\mathcal{A}_I (\theta), \mathcal{A}_J (\theta))_\pm = \text{HF}^{G}_{d_p}(p^{IJ}_\pm (\theta))} \, ,
\end{equation}
where  $\text{CF}(\mathcal{A}_I (\theta), \mathcal{A}_J (\theta))_\pm$ is a vector representing a $\gamma^{IJ}_\pm(t, \theta)$ soliton defined by \eqref{soliton equations in 2d model} whose start and endpoints correspond to  $\mathcal{A}_I(\theta)$ and $\mathcal{A}_J(\theta)$ (as seen in Fig.~\ref{fig:mu1 map}), such that $\text{Re} \, W^I (\theta) < \text{Re} \, W^J (\theta)$.

Notice that a $\gamma^{IJ}_\pm(t, \theta)$ soliton can also be regarded as a morphism $\text{Hom}(\mathcal{A}_I (\theta), \mathcal{A}_J (\theta))_\pm$ from $\mathcal{A}_I(\theta)$ to $\mathcal{A}_J(\theta)$. Since there is a one-to-one identification between $\text{Hom}(\mathcal{A}_I (\theta), \mathcal{A}_J (\theta))_\pm$  and $\text{CF}(\mathcal{A}_I (\theta), \mathcal{A}_J (\theta))_\pm$, we will also have the following one-to-one identification
\begin{equation} \label{Haydys Hom = CF}
  \boxed{ \text{Hom}(\mathcal{A}_I, \mathcal{A}_J)_\pm  \Longleftrightarrow \text{HF}^{G}_{d_p}(p^{IJ}_\pm)}
\end{equation}
where if $I = J$, the RHS is proportional to the identity class, and if $I \leftrightarrow J$, the RHS is zero (as the soliton only goes from left to right in Fig.~\ref{fig:mu1 map}). Here, one can also interpret the $p^{IJ}_\pm$'s to correspond to solutions of \eqref{HW 5d} at $\tau = \pm \infty$ where $A_\tau, A_t \to 0$, that at $t = \pm\infty$, are given by $G$ Hitchin configurations on $M_3$ satisfying \eqref{endpoints HW}, whence they coincide with the definition of `$\mathcal{B}_\mp$' in~\cite[$\S$5]{haydys2010fukaya}.

Again, we have omitted the `$\theta$' label in \eqref{Haydys Hom = CF}, as we recall that the physical theory is actually equivalent for all values of $\theta$.

\bigskip\noindent\textit{Soliton String Theory from the LG Model}
\vspace*{0.5em}

Just like the 2d gauged sigma model with target $\mathfrak{A}_3$, the equivalent gauged LG model will define an open string theory in $\mathscr{A}_3$ with effective worldsheet and boundaries shown in Fig.~\ref{fig:mu1 map}, where $\tau$ and $t$ are the temporal and spatial directions, respectively.

The dynamics of this open string in $\mathscr A_3$ will be governed by \eqref{m3 x r2 bps without gauge fields complexified}, where $\phi^a$ and ${\bar\phi}^{\bar a}$ are scalars on the worldsheet that correspond to the holomorphic and antiholomorphic coordinates of $\mathscr A_3$. At some arbitrary time instant $\tau = T$ whence ${d\phi^a / d\tau} = 0$ in \eqref{m3 x r2 bps without gauge fields complexified}, we see that the dynamics of the $\phi^a$ fields along the spatial $t$-direction will be governed by the soliton equation ${d\phi^{a}}/ {dt} = - i [A_\tau, \phi^a] - [A_t, \phi^a] + g^{ab} ({i\zeta} {\partial W / \partial \phi^b})^*/2$.

Thus, just as a topological A-model can be interpreted as an instanton string theory whose corresponding dynamics of the $\phi^a$ fields along the spatial $t$-direction  will be governed by the instanton equation $d\phi^a / dt = 0$, our LG model can be interpreted as a \textit{soliton} string theory.

\bigskip\noindent\textit{The Normalized 5d-HW Partiton Function, Soliton String Scattering, and Maps of an $A_\infty$-structure}
\vspace*{0.5em}

The spectrum of HW theory is given by the $\mathcal{Q}$-cohomology of operators. In particular, its normalized 5d partition function will be a sum over the free-field correlation functions of these operators.\footnote{Recall from footnote~\ref{balanced TQFT} that HW theory is a balanced TQFT, whence the normalized 5d partition function can be computed by bringing down interaction terms to absorb fermion pair zero-modes in the path integral measure. These interaction terms can be regarded as operators of the free-field theory that are necessarily in the $\mathcal{Q}$-cohomology (as the nonvanishing partition function ought to remain $\mathcal{Q}$-invariant), and their contribution to the partition function can be understood as their free-field correlation functions.} Because HW theory is semi-classical, these correlation functions will correspond to tree-level scattering only.

\begin{figure}
    \centering
    \begin{tikzpicture}[declare function={
        lenX(\legLength,\leftAngle,\rightAngle) = \legLength * cos((\leftAngle + \rightAngle)/2);
        lenY(\legLength,\leftAngle,\rightAngle) = \legLength * sin((\leftAngle + \rightAngle)/2);
        lenLX(\segAngle,\leftAngle,\rightAngle) = -2 * tan(\segAngle/2) * sin((\leftAngle + \rightAngle)/2);
        lenLY(\segAngle,\leftAngle,\rightAngle) = 2 * tan(\segAngle/2) * cos((\leftAngle + \rightAngle)/2);
        }]
    \def \NumSeg {8}                                
    \def \Rad {1}                                   
    \def \Leg {1.5}                                 
    \def \SegAngle {180/\NumSeg}                    
    \def \TpRtAngle {{(\NumSeg - 1)*\SegAngle/2}}   
    \def \TpLtAngle {{(\NumSeg + 1)*\SegAngle/2}}   
    \def \BaLtAngle {{(\NumSeg + 1)*\SegAngle}}     
    \def \BaRtAngle {{(\NumSeg + 2)*\SegAngle}}     
    \def \BbLtAngle {{(\NumSeg + 3)*\SegAngle}}     
    \def \BbRtAngle {{(\NumSeg + 4)*\SegAngle}}     
    \def \BcLtAngle {{(2 * \NumSeg - 2)*\SegAngle}} 
    \def \BcRtAngle {(2 * \NumSeg - 1)*\SegAngle}   
    \draw ([shift=({\BcRtAngle-360}:\Rad)]3,3) arc ({\BcRtAngle-360}:\TpRtAngle:\Rad);
    \draw ([shift=(\TpLtAngle:\Rad)]3,3) arc (\TpLtAngle:\BaLtAngle:\Rad);
    \draw ([shift=(\BaRtAngle:\Rad)]3,3) arc (\BaRtAngle:\BbLtAngle:\Rad);
    \draw[dashed] ([shift=(\BbRtAngle:\Rad)]3,3) arc (\BbRtAngle:
    \BcLtAngle:\Rad);
    \draw ([shift=(\TpRtAngle:\Rad)]3cm,3cm)
        -- node[right] {\footnotesize $\mathcal{A}_{I_{d+1}}$} ++({lenX(\Leg,\TpLtAngle,\TpRtAngle)},{lenY(\Leg,\TpLtAngle,\TpRtAngle)})
        -- node[above] {$+$} ++({lenLX(\SegAngle,\TpLtAngle,\TpRtAngle)},{lenLY(\SegAngle,\TpLtAngle,\TpRtAngle)})
        -- node[left] {\footnotesize $\mathcal{A}_{I_1}$} ++(-{lenX(\Leg,\TpLtAngle,\TpRtAngle)},-{lenY(\Leg,\TpLtAngle,\TpRtAngle)});
    \draw ([shift=(\BaLtAngle:\Rad)]3,3)
        -- node[near end, above left] {\footnotesize $\mathcal{A}_{I_1}$} ++({lenX(\Leg,\BaLtAngle,\BaRtAngle)},{lenY(\Leg,\BaLtAngle,\BaRtAngle)})
        -- node[left] {$-$} ++({lenLX(\SegAngle,\BaLtAngle,\BaRtAngle)},{lenLY(\SegAngle,\BaLtAngle,\BaRtAngle)})
        -- node[near start, below] {\footnotesize $\mathcal{A}_{I_2}$} ++(-{lenX(\Leg,\BaLtAngle,\BaRtAngle)},-{lenY(\Leg,\BaLtAngle,\BaRtAngle)});
    \draw ([shift=(\BbLtAngle:\Rad)]3,3)
        -- node[near end, left=1pt] {\footnotesize $\mathcal{A}_{I_2}$} ++({lenX(\Leg,\BbLtAngle,\BbRtAngle)},{lenY(\Leg,\BbLtAngle,\BbRtAngle)})
        -- node[below] {$-$} ++({lenLX(\SegAngle,\BbLtAngle,\BbRtAngle)},{lenLY(\SegAngle,\BbLtAngle,\BbRtAngle)})
        -- node[near start, right] {\footnotesize $\mathcal{A}_{I_3}$} ++(-{lenX(\Leg,\BbLtAngle,\BbRtAngle)},-{lenY(\Leg,\BbLtAngle,\BbRtAngle)});
    \draw ([shift=(\BcLtAngle:\Rad)]3,3)
        -- node[near end, below] {\footnotesize $\mathcal{A}_{I_d}$} ++({lenX(\Leg,\BcLtAngle,\BcRtAngle)},{lenY(\Leg,\BcLtAngle,\BcRtAngle)})
        -- node[right] {$-$} ++({lenLX(\SegAngle,\BcLtAngle,\BcRtAngle)},{lenLY(\SegAngle,\BcLtAngle,\BcRtAngle)})
        -- node[near start, above right] {\footnotesize $\mathcal{A}_{I_{d+1}}$} ++(-{lenX(\Leg,\BcLtAngle,\BcRtAngle)},-{lenY(\Leg,\BcLtAngle,\BcRtAngle)});
    \end{tikzpicture}
    \caption{Tree-level scattering BPS worldsheet of incoming ($-$) and outgoing ($+$)  LG soliton strings.}
    \label{fig:mud maps}
\end{figure}
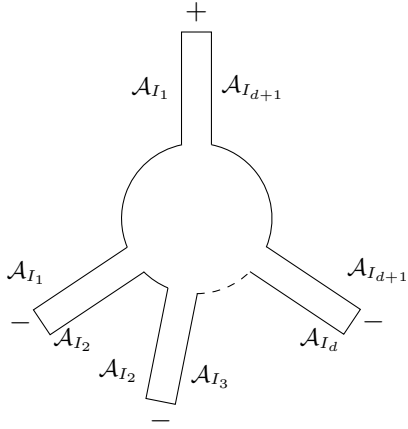

From the equivalent SQM and gauged LG perspective, the $\mathcal{Q}$-cohomology will be spanned by the LG soliton strings defined by \eqref{soliton equations in 2d model complexified}. In turn, this means that the normalized 5d-HW partition function can also be regarded as a sum over tree-level scattering amplitudes of these LG soliton strings. The BPS worldsheet underlying such a tree-level scattering amplitude is shown in Fig.~\ref{fig:mud maps}.\footnote{Here, we have exploited the topological and hence conformal invariance of the soliton string theory to replace the outgoing LG soliton strings with their vertex operators on the disc, then used their coordinate-independent operator products to reduce them to a single vertex operator, before finally translating it back as a single outgoing LG soliton string.}

In other words, we can express the normalized 5d-HW partition function as
\begin{equation}
\label{HW partition function A-infty}
  {  \widetilde{\mathcal{Z}}_{\text{HW}, M_3 \times \R^2} (G) = \sum_d \mu^d, \qquad  d= 1, 2, 3, \dots, k-1}
\end{equation}
where each
\begin{equation}
  \label{Haydys mu maps}
  \boxed{
    \mu^d: \bigotimes_{i = 1}^d
    \text{Hom}(\mathcal{A}_{I_i}, \mathcal{A}_{I_{i + 1}})_-
    \longto
    \text{Hom}(\mathcal{A}_{I_1}, \mathcal{A}_{I_{d + 1}})_+
  }
\end{equation}
is a scattering amplitude of $d$ incoming LG soliton strings $\text{Hom}(\mathcal{A}_{I_1}, \mathcal{A}_{I_{2}})_-, \dots, \text{Hom}(\mathcal{A}_{I_d}, \mathcal{A}_{I_{d+1}}) _-$ and a single outgoing LG soliton string $\text{Hom}(\mathcal{A}_{I_1}, \mathcal{A}_{I_{d+1}})_+$ with left and right boundaries as labeled, whose underlying worldsheet shown in Fig.~\ref{fig:mud maps} can be regarded as a disc with $d+1$ vertex operators at the boundary. That is, $\mu^d$ counts pseudoholomorphic discs with $d+1$ punctures at the boundary that are mapped to $\mathscr{A}_3$ according to the BPS worldsheet equation \eqref{m3 x r2 bps without gauge fields complexified}.

In turn, this means that $\mu^d$ counts the moduli of solutions to \eqref{m3 x r2 bps 2} with $d+1$ boundary conditions that can be described as follows. First, note that we can regard $\R^2$ as the effective worldsheet in Fig.~\ref{fig:mud maps} that we shall denote as $\Omega$, so $M_5$ can be interpreted as a trivial $M_3$ fibration over $\Omega$. Then, at the $d+1$ soliton strings on $\Omega$ where $\tau = \pm \infty$, \eqref{m3 x r2 bps 2} will become \eqref{HW 5d} with $A_\tau, A_t \to 0$, and over the soliton string boundaries on $\Omega$ where $t = \pm\infty$, \eqref{m3 x r2 bps 2} will become \eqref{endpoints HW} which define $G$ Hitchin configurations on $M_3$. Hence, they coincide with the definition of the maps `$\mu^d$' in~\cite[$\S5$]{haydys2010fukaya}.

A relevant fact at this juncture, is that the collection of maps $\mu^d$ is also known to define an $A_\infty$-structure.

\bigskip\noindent\textit{A Fukaya-Seidel Type $A_\infty$-category of Hitchin Configurations on $M_3$}
\vspace*{0.5em}

Altogether, this means that the normalized 5d partition function of HW theory on $M_5 = M_3 \times \R^2$, as expressed in \eqref{HW partition function A-infty}, manifests a Fukaya-Seidel type $A_\infty$-category defined by the maps \eqref{Haydys mu maps} and the identification \eqref{Haydys Hom = CF}, where the $k$ objects $\{ \mathcal{A}_{1}, \dots, \mathcal{A}_k \}$ correspond to $G$ Hitchin configurations on $M_3$!

\bigskip\noindent\textit{A Physical Proof of Haydys' Mathematical Conjecture and Realization of Wang's Mathematical Construction}
\vspace*{0.5em}

Such a Fukaya-Seidel $A_\infty$-category of three-manifolds
was first conjectured to exist by Haydys in~\cite[\S5]{haydys2010fukaya}, and later rigorously constructed by Wang in~\cite{wang2022complgradien}. Thus, we have furnished a purely physical proof of Haydys' mathematical conjecture and realization of Wang's mathematical construction.

\bigskip\noindent\textit{A Physical Proof and Generalization of Abouzaid-Manolescu's Mathematical Conjecture}
\vspace*{0.5em}

Also, in the frame where $\theta = \pi/ 2$, we saw in the previous subsection that the $k$ objects $\{ \mathcal{A}_{1}, \dots, \mathcal{A}_k \}$ would correspond to irreducible flat $G_\C$ connections on $M_3$. These will generate a $G_\C$-instanton Floer homology of $M_3$ which, via~\cite[$\S$5]{ong2022vafa}, can be identified with $\text{HP}^*(M_3)$, a hypercohomology of a perverse sheaf of vanishing cycles in the moduli space  of irreducible flat $G_\C$ connections on $M_3$ constructed in~\cite{abouzaid2017sheaftheor}. Thus, our results mean that $\text{HP}^*(M_3)$ can be categorified to give a Fukaya-Seidel $A_\infty$-category of $M_3$. As Abouzaid-Manolescu conjectured in $\S$9.2 of \textit{loc}.~\textit{cit}. that $\text{HP}^*(M_3)$ can be categorified to give an $A_\infty$-category of $M_3$, we have also furnished a purely physical proof and generalization (when $\theta \neq \pi/2$) of Abouzaid-Manolescu's mathematical conjecture.

\subsection{An Atiyah-Floer Type Correspondence for the FS Type \texorpdfstring{$A_\infty$}{A-infinity}-category of Hitchin Configurations on \texorpdfstring{$M_3$}{M3}}

\bigskip\noindent\textit{Intersecting Thimbles as Intersecting Branes}
\vspace*{0.5em}

Let $M_1 = \R$ in $\S$\ref{Heegaard Split of M5}. Then, by the same arguments that made use of \eqref{interval to loop space floer classes} to lead us to \eqref{Atiyah-Floer partition equality S1}, we can make use of \eqref{interval to path space floer classes} to get
\begin{equation}
\label{Atiyah-Floer partition equality R}
{ \sum_k  \text{HF}^{\text{HW}^\theta}_{d_k}(M_3 \times \R, G) =  \sum_v \text{HSF}^{\text{Int}}_{d_v}(\mathcal{M}\big(\R, \mathcal{M}^{G,\Sigma}_{\text{H}, \theta}\big) , \mathscr{P}_0, \mathscr{P}_1)}\,,
\end{equation}
where $\text{HW}^\theta$ refers to HW theory on $M_3 \times \R^2$ with the $\R^2$-plane rotated by an angle $\theta$, as per our formulation in this section thus far.

The LHS of \eqref{Atiyah-Floer partition equality R} is just a $\theta$-generalization of \eqref{5d partition function floer}, i.e., we can also express \eqref{Atiyah-Floer partition equality R} as
\begin{equation}
\label{Z-theta = HSF-theta}
{ \mathcal{Z}_{\text{HW}^\theta,M_3 \times \R^2}(G)  =  \sum_v \text{HSF}^{\text{Int}}_{d_v}(\mathcal{M}\big(\R, \mathcal{M}^{G,\Sigma}_{\text{H}, \theta}\big) , \mathscr{P}_0, \mathscr{P}_1)}\,.
\end{equation}

In turn, the LHS of \eqref{Z-theta = HSF-theta} is given by the LHS of \eqref{LG SQM partition function}, whence we can write
\begin{equation}
\label{Z-theta-SQM = HSF-theta}
{  \sum_{I\neq J = 1}^k \, \sum_{p^{IJ}_\pm (\theta) \in S_{IJ}}  \text{HF}^{G}_{d_p}(p^{IJ}_\pm (\theta)) =  \sum_v \text{HSF}^{\text{Int}}_{d_v}(\mathcal{M}\big(\R, \mathcal{M}^{G,\Sigma}_{\text{H}, \theta}\big) , \mathscr{P}_0, \mathscr{P}_1)}\,,
\end{equation}
which implies
\begin{equation}
{\text{HF}^{G}_{d_p}(p^{IJ}_\pm (\theta)) \cong     \text{HSF}^{\text{Int}}_{d_v}(\mathcal{M}\big(\R, \mathcal{M}^{G,\Sigma}_{\text{H}, \theta}\big) , \mathscr{P}_0, \mathscr{P}_1) }\,.
\end{equation}
That is, we have a correspondence between a gauge-theoretic Floer homology generated by intersecting thimbles and a symplectic Floer homology generated by intersecting branes!

\bigskip\noindent\textit{An Atiyah-Floer Type Correspondence for the FS \texorpdfstring{$A_\infty$}{A-infinity}-category of Hitchin Configurations on \texorpdfstring{$M_3$}{M3}}
\vspace*{0.5em}

Via \eqref{Haydys Hom = CF}, we would have the one-to-one identification
\begin{equation} \label{Haydys Hom = CF Atiyah-Floer}
  \boxed{
    \text{Hom}(\mathcal{A}_I(\theta), \mathcal{A}_J(\theta))_\pm
    \Longleftrightarrow
    \text{HSF}^{\text{Int}}_{d_v}(\mathcal{M}\big(\R, \mathcal{M}^{G,\Sigma}_{\text{H}, \theta}\big) , \mathscr{P}_0, \mathscr{P}_1)
  }
\end{equation}

This identification is indeed a consistent one as follows. When $\theta = \pi/2$, $\mathcal{A}_{I,J}(\pi/2)$ would correspond to flat $G_\C$ connections on $M_3$, while $\mathscr{P}_{0,1}(\pi/2)$ would be isotropic-coisotropic intersecting branes that when restricted to $\mathcal{M}^{G,\Sigma}_{\text{H}, \pi/2} = \mathcal{M}^{G_\C,\Sigma}_{\text{flat}}$, would have intersection points that correspond to flat $G_\C$ connections on $\Sigma$ that can be extended to all of $M_3 = M_3' \bigcup_\Sigma M_3''$. When $\theta = \pi$, $\mathcal{A}_{I,J}(\pi)$ would correspond to $G$ Higgs pairs on $M_3$, while $\mathscr{P}_{0,1}(\pi)$ would be isotropic-coisotropic intersecting branes that when restricted to $\mathcal{M}^{G,\Sigma}_{\text{H}, \pi} = \mathcal{M}^{G,\Sigma}_{\text{Higgs}}$, would have intersection points that correspond to $G$ Higgs pairs on $\Sigma$ that can be extended to all of $M_3 = M_3' \bigcup_\Sigma M_3''$.

Thus, we now have a Fukaya-Seidel type $A_\infty$-category defined by the maps \eqref{Haydys mu maps} and identification \eqref{Haydys Hom = CF Atiyah-Floer}, where the $k$ objects $\{ \mathcal{A}_{1}, \dots, \mathcal{A}_k \}$ which correspond to $G$ Hitchin configurations on $M_3$ would be related to intersecting branes whose intersection points correspond to $G$ Hitchin configurations on $\Sigma$ that can be extended to all of $M_3 = M_3' \bigcup_\Sigma M_3''$.

In other words, we have a \emph{novel} Atiyah-Floer type correspondence for the Fukaya-Seidel $A_\infty$-category of $G$ Hitchin configurations on $M_3$!

\bigskip\noindent\textit{The Soliton as a Hom-category}
\vspace*{0.5em}

At $\theta = \pi/2$, an $\mathcal{A}_{M}(\pi/2)$ corresponding to a flat $G_\C$ connection on $M_3$ will generate a $G_\C$-instanton Floer homology of $M_3$. Therefore, via the Atiyah-Floer correspondence in~\eqref{VW partition equality HW_4}, we can identify $\mathcal{A}_M(\pi/2)$ with a class in an intersection Floer homology generated by intersections of Lagrangian branes $L_{M, 0}', L_{M, 1}'$ of types $(\text{A}, \text{B}, \text{A})$ in the complex structure $(I, J, K)$ of $\mathcal{M}^{G_\C, \Sigma}_{\text{flat}} = \mathcal{M}^{G, \Sigma}_{\text{H}, \pi/2}$, i.e.,
\begin{equation}
  \mathcal{A}_M(\pi/2)
  \Longleftrightarrow
  \text{HSF}^{\text{Int}}_*( \mathcal{M}^{G, \Sigma}_{\text{H}, \pi/2}, L_{M, 0}', L_{M, 1}')
  \, .
\end{equation}

At $\theta = \pi$, an $\mathcal{A}_{M}(\pi)$ corresponding to a $G$ Hitchin configuration on $M_3$ will generate a 4d-HW Floer homology of $M_3$. Therefore, via the 4d-HW Atiyah-Floer correspondence of~\eqref{Atiyah-Floer HW_4}, we can identify $\mathcal{A}_M(\pi)$ with a class in an intersection Floer homology generated by intersections of Lagrangian branes $L_{M, 0}, L_{M, 1}$ of types $(\text{A}, \text{A}, \text{B})$ in the complex structure $(I, J, K)$ of $\mathcal{M}^{G, \Sigma}_{\text{Higgs}} = \mathcal{M}^{G, \Sigma}_{\text{H}, \pi}$, i.e.,
\begin{equation}
  \mathcal{A}_M(\pi)
  \Longleftrightarrow
  \text{HSF}^{\text{Int}}_*( \mathcal{M}^{G, \Sigma}_{\text{H}, \pi}, L_{M, 0}, L_{M, 1})
  \, .
\end{equation}

This therefore means that an $\mathcal{A}_{M}(\theta)$ corresponding to a $\theta$-deformed $G$ Hitchin configuration on $M_3$ can be identified with a class in an intersection Floer homology generated by intersections of $\theta$-deformed Lagrangian branes $L_{M, 0}(\theta), L_{M, 1}(\theta)$ of $\mathcal{M}^{G, \Sigma}_{\text{H}, \theta}$, i.e.,
\begin{equation}
  \label{theta G Hitchin configs as HSF}
  \mathcal{A}_M(\theta)
  \Longleftrightarrow
  \text{HSF}^{\text{Int}}_*( \mathcal{M}^{G, \Sigma}_{\text{H}, \theta}, L_{M, 0}(\theta), L_{M, 1}(\theta))
  \, .
\end{equation}

Note that the classes on the RHS of \eqref{theta G Hitchin configs as HSF} correspond to open string states of a 2d A$_{\theta}$-model with branes $L_{M, 0}(\theta), L_{M, 1}(\theta)$, whence we can interpret them as $\text{Hom}(L_{M, 0}(\theta), L_{M, 1}(\theta))$, i.e.,
\begin{equation}
  \label{theta G Hitchin configs as Homs}
  \text{HSF}^{\text{Int}}_*( \mathcal{M}^{G, \Sigma}_{\text{H}, \theta}, L_{M, 0}(\theta), L_{M, 1}(\theta))
  \Longleftrightarrow
  \text{Hom}(L_{M, 0}(\theta), L_{M, 1}(\theta))
  \, .
\end{equation}
Then, the LHS of~\eqref{Haydys Hom = CF} can be identified as
\begin{equation}
  \label{FS-cat hom as Hom-cat}
  \boxed{
    \text{Hom}(\mathcal{A}_I(\theta), \mathcal{A}_J(\theta))_\pm
    \Longleftrightarrow
    \text{Hom}\left(
      \text{Hom}(L_{I, 0}(\theta), L_{I, 1}(\theta)),
      \text{Hom}(L_{J, 0}(\theta), L_{J, 1}(\theta))
    \right)_\pm
  }
\end{equation}

In other words, the morphisms defining a Fukaya-Seidel $A_\infty$-category of $G$ Hitchin configurations on $M_3$ can be identified as a Hom-category with objects themselves being morphisms between Lagrangian branes (submanifolds) of $\mathcal{M}^{G, \Sigma}_{\text{H}, \theta}$.

\bigskip\noindent\textit{A Physical Proof and Generalization of Bousseau's Mathematical Conjecture}
\vspace*{0.5em}

In the frame where $\theta = \pi /2$, the objects of the Hom-category are morphisms between Lagrangian submanifolds of $\mathcal{M}^{G_\C, \Sigma}_{\text{flat}}$ corresponding to flat $G_\C$ connections on $M_3$ generating a $G_\C$-instanton Floer homology, whilst the $k$ objects $\{\mathcal{A}_1(\pi/2), \dots, \mathcal{A}_k(\pi/2)\}$ of the Fukaya-Seidel $A_\infty$-category correspond to flat $G_\C$ connections on $M_3$. Thus, our results mean that there is a correspondence between (i) a Hom-category of morphisms between Lagrangian submanifolds of $\mathcal{M}^{G_\C, \Sigma}_{\text{flat}}$, (ii) a Fukaya-Seidel $A_\infty$-category of flat $G_\C$ connections on $M_3$, and (iii) a $G_\C$-instanton Floer homology of $M_3$. Such a correspondence has been conjectured by Bousseau in \cite[$\S$2.8]{bousseau-2024-holom-floer}. Thus, in arriving at \eqref{FS-cat hom as Hom-cat}, we have furnished a purely physical proof and generalization (when $\theta \neq \pi/2$) of Bousseau's mathematical conjecture!

\bigskip\noindent\textit{Intersecting Branes as a Hom-category}
\vspace*{0.5em}

Finally, by applying~\eqref{FS-cat hom as Hom-cat} to~\eqref{Haydys Hom = CF Atiyah-Floer}, we would have the one-to-one identification
\begin{equation}
  \label{CF Atiyah-Floer = Hom-cat}
  \boxed{
    \text{HSF}^{\text{Int}}_{d_v}(\mathcal{M}\big(\R, \mathcal{M}^{G,\Sigma}_{\text{H}, \theta}\big) , \mathscr{P}_0, \mathscr{P}_1)
    \Longleftrightarrow
    \text{Hom}\left(
      \text{Hom}(L_{I, 0}(\theta), L_{I, 1}(\theta)),
      \text{Hom}(L_{J, 0}(\theta), L_{J, 1}(\theta))
    \right)_\pm
  }
\end{equation}
between an intersection Floer homology and a Hom-category of morphisms!

This identification is indeed a consistent one as follows. The LHS of~\eqref{CF Atiyah-Floer = Hom-cat} actually corresponds to open membrane states of a 3d sigma model on $\R \times I \times \R$, which in turn can be understood as morphisms between the open string states of two 2d sigma models on $\R \times I$ with branes $L_{*, 0}(\theta)$ and $L_{*, 1}(\theta)$. From $\S$\ref{subsection: 7d-hw atiyah-floer correspondence} and the generalization to general $\theta$ above, such open string states of 2d sigma models are given by the intersection Floer homology classes generated by intersections of Lagrangian branes of $\mathcal{M}^{G, \Sigma}_{\text{H}, \theta}$. That is to say, we have
\begin{equation}
  \label{HSF of 3d model = Hom of HSF of 2d model}
  \begin{gathered}
    \text{HSF}^{\text{Int}}_*(\mathcal{M}\big(\R, \mathcal{M}^{G,\Sigma}_{\text{H}, \theta}\big) , \mathscr{P}_0, \mathscr{P}_1)
    \\
    \cong
    \text{Hom}\left(
      \text{HSF}^{\text{Int}}_*(\mathcal{M}^{G,\Sigma}_{\text{H}, \theta}, L_{A,0}(\theta), L_{A,1}(\theta)),
      \text{HSF}^{\text{Int}}_*(\mathcal{M}^{G,\Sigma}_{\text{H}, \theta}, L_{B,0}(\theta), L_{B,1}(\theta))
    \right)
    \, .
  \end{gathered}
\end{equation}
Then, via~\eqref{theta G Hitchin configs as Homs}, the bottom of~\eqref{HSF of 3d model = Hom of HSF of 2d model} would become the RHS of~\eqref{CF Atiyah-Floer = Hom-cat}, concluding the consistency check.

\vspace{0.4cm}

\printbibliography




\end{document}